\documentclass[aps,prd,groupedaddress,preprintnumbers,%
  showpacs,showkeys,floatfix,amssymb,amsfonts,twocolumn]{revtex4-1}  
\usepackage{hyperref}
\usepackage[usenames]{color}
\usepackage{todonotes}
\usepackage{bbold}
\usepackage{amsmath}
\usepackage{multirow}
\usepackage{graphicx}
\usepackage{xfrac}

\begin{document}
\title{Nucleon electromagnetic form factors using lattice simulations at the physical point}
\preprint{DESY 17-085}
\author{
  C.~Alexandrou$^{1,2}$,
  M.~Constantinou$^{3}$,
  K.~Hadjiyiannakou$^{1}$,
  K.~Jansen$^{4}$,
  Ch.~Kallidonis$^{1}$, 
  G.~Koutsou$^{1}$, and
  A.~Vaquero Aviles-Casco$^5$
}
\affiliation{
  $^1$Computation-based Science and Technology Research Center,
  The Cyprus Institute, 20 Kavafi Str., Nicosia 2121, Cyprus \\
  $^2$Department of Physics, University of Cyprus,  P.O. Box 20537,  1678 Nicosia, Cyprus\\
  $^3$Department of Physics, Temple University, 1925 N. 12th Street, Philadelphia, PA 19122-1801, USA\\
  $^4$NIC, DESY, Platanenallee 6,  D-15738 Zeuthen,  Germany\\
  $^5$Department of Physics and Astronomy, University of Utah, Salt Lake City, UT 84112, USA\\
}

\begin{abstract}
  We present results for the nucleon electromagnetic form factors
  using an ensemble of maximally twisted mass clover-improved fermions
  with pion mass of about 130 MeV. We use multiple sink-source
  separations and three analysis methods to probe ground-state
  dominance.  We evaluate both the connected and disconnected
  contributions to the nucleon matrix elements.  We find that the
  disconnected quark loop contributions to the isoscalar matrix
  elements are small giving an upper bound of up to 2\% of the
  connected and smaller than its statistical error. We present results
  for the isovector and isoscalar electric and magnetic Sachs form
  factors and the corresponding proton and neutron form factors. By
  fitting the momentum dependence of the form factors to a dipole form
  or to the z-expansion we extract the nucleon electric and magnetic
  radii, as well as, the magnetic moment. We compare our results to
  experiment as well as to other recent lattice QCD calculations.
\end{abstract}
\maketitle

\section{Introduction}
Electromagnetic form factors probe the internal structure of hadrons
mapping their charge and magnetic distributions. The slope of the
electric and magnetic form factors at zero momentum yields the
electric and magnetic root mean square radius, while the value of the
form factors at zero momentum give its electric charge and magnetic
moment. Extensive electron scattering experiments have been carried
out since the fifties for the precise determination of the nucleon
form factors, including recent experiments at Jefferson Lab, MIT-Bates
and Mainz. For a recent review on electron elastic scattering
experiments, see Ref.~\cite{Punjabi:2015bba}. The proton radius can
also be obtained spectroscopically, namely via the Lamb shifts of the
hydrogen atom and of muonic hydrogen~\cite{Pohl:2010zza} and via
transition frequencies of electronic and muonic deuterium. In these
measurements, including a recent experiment using muonic
deuterium~\cite{Pohl1:2016xoo}, discrepancies are observed in the
resulting proton radius between hydrogen and deuterium and their
corresponding muonic equivalents. Whether new physics is responsible
for this discrepancy, or errors in the theoretical or experimental
analyses, a first principles calculation of the electromagnetic form
factors of the nucleon can provide valuable insight. Although nucleon
electromagnetic form factors have been extensively studied in lattice
QCD, most of these studies have been carried out at higher than
physical pion masses, requiring extrapolations to the physical point,
which for the case of baryons carry a large systematic uncertainty.

In this paper, we calculate the electromagnetic form factors of the
nucleon using an ensemble of two degenerate light quarks
(N$_\textrm{f}=2$) tuned to reproduce a pion mass of about 130~MeV, in
a volume with $m_\pi L\simeq$3~\cite{Abdel-Rehim:2015pwa}. We use the
twisted mass fermion action with clover
improvement~\cite{Frezzotti:2003ni,Frezzotti:2000nk}. We employ
$\mathcal{O}(10^5)$ measurements to reduce the statistical errors, and
multiple sink-source separations to study excited state effects using
three different analyses. We extract the momentum dependence of the
electric and magnetic Sachs form factors for both isovector and
isoscalar combinations, i.e. for both the difference ($p-n$) and sum
($p+n$) of proton and neutron form factors. For the latter we compute
the computationally demanding disconnected contributions and find them
to be smaller than the statistical errors of the connected
contributions. To fit the momentum dependence we use both a dipole
form as well as the z-expansion~\cite{Hill:2010yb}. From these fits we
extract the electric and magnetic radii, as well as, the magnetic
moments of the proton, the neutron and the isovector and isoscalar
combinations. For the electric root mean squared (rms) radius of the
proton we find $\sqrt{\langle r_E^2\rangle_p}=0.767(25)(21)$~fm where
the first error is statistical and the second a systematic due to
excited states. Although this value is closer to the value of
0.84087(39)~fm extracted from muonic hydrogen~\cite{Pohl1:2016xoo}, a
more complete analysis of systematic errors using multiple ensembles
is required to assess accurately all lattice artifacts.

The remainder of this paper is organized as follows: in
Section~\ref{sec:setup} we provide details of the lattice set-up for
this calculation and in Section~\ref{sec:results} we present our
results. In Section~\ref{sec:others} we compare our results with other
lattice calculations and in Section~\ref{sec:conclusions} we summarize
our findings and conclude.

\section{Setup and lattice parameters}
\label{sec:setup}
\subsection{Electromagnetic form factors}
The electromagnetic form factors are extracted from the
electromagnetic nucleon matrix element given by
\begin{align}
  \langle  &N(p', s')|\mathcal{O}^{V}_\mu|N(p,s)\rangle = \nonumber\\
  &\sqrt{\frac{m^2_N}{E_N(\vec{p}')E_N(\vec{p})}}\bar{u}_N(p',s')\Lambda^{V}_\mu(q^2)u_N(p,s)
  \label{eq:matrix element}
\end{align}
with $N(p,s)$ the nucleon state of momentum $p$ and spin $s$,
$E_N(\vec{p}) = p_0$ its energy and $m_N$ its mass, $\vec
q=\vec{p}\,^\prime-\vec p$, the spatial momentum transfer from initial
($\vec p$) to final ($\vec{p}\,'$) momentum, $u_N$ the nucleon spinor
and $\mathcal{O}^{V}$ the vector current.  In the isospin limit, where
an exchange between up and down quarks ($u\leftrightarrow d$) and
between proton and neutron ($p \leftrightarrow n$) is a symmetry, the
isovector matrix element can be related to the difference between
proton and neutron form factors as follows:
\begin{align}
  \langle p | \frac{2}{3}\bar{u}\gamma_\mu u - \frac{1}{3}\bar{d}\gamma_\mu d | p \rangle -   \langle n | \frac{2}{3}\bar{u}\gamma_\mu u - \frac{1}{3}\bar{d}\gamma_\mu d | n \rangle\nonumber\\\xrightarrow[p\leftrightarrow n]{u\leftrightarrow d} \langle p | \bar{u}\gamma_\mu u - \bar{d}\gamma_\mu d | p \rangle.
  \label{eq:isovector}
\end{align}
Similarly, for the isoscalar combination we have
\begin{align}
  \langle p | \frac{2}{3}\bar{u}\gamma_\mu u - \frac{1}{3}\bar{d}\gamma_\mu d | p \rangle + \langle n | \frac{2}{3}\bar{u}\gamma_\mu u - \frac{1}{3}\bar{d}\gamma_\mu d | n \rangle\nonumber\\\xrightarrow[p\leftrightarrow n]{u\leftrightarrow d} \frac{1}{3}\langle p | \bar{u}\gamma_\mu u + \bar{d}\gamma_\mu d | p \rangle.
  \label{eq:isoscalar}
\end{align}
We will use these relations to compare our lattice results, obtained
for the isovector and isoscalar combinations, with the experimental
data for the proton and neutron matrix elements.

We use the symmetrized lattice conserved vector current,
$\mathcal{O}^{V}_\mu$ = $\frac{1}{2}[j_\mu(x)+j_\mu(x-\hat{\mu})]$,
with
\begin{align}
j_\mu(x) =
\frac{1}{2}[&\bar{\psi}(x+\hat{\mu})U^\dagger_\mu(x)(1+\gamma_\mu)\tau_a\psi(x)
  \nonumber\\ -&\bar{\psi}(x)U_\mu(x)(1-\gamma_\mu)\tau_a\psi(x+\hat{\mu})],
\end{align}
where $\bar{\psi}=(\bar{u}, \bar{d})$ and $\tau_a$ acts in flavor
space. We consider $\tau_a=\tau_3$, the third Pauli matrix, for the
isovector case, and $\tau_a=\sfrac{\mathbb{1}}{3}$ for the isoscalar case.
$\hat{\mu}$ is the unit vector in direction $\mu$ and $U_\mu(x)$ is
the gauge link connecting site $x$ with $x+\hat{\mu}$. Using the
conserved lattice current means that no renormalization of the vector
operator is required. 

The matrix element of the vector current can be decomposed in terms of
the Dirac $F_1$ and Pauli $F_2$ form factors as
\begin{align}
  \Lambda_\mu^V(q^2) = \gamma_\mu
  F_1(q^2)+\frac{i\sigma_{\mu\nu}q^\nu}{2m_N}F_2(q^2).
\end{align}
$F_1$ and $F_2$ can also be expressed in terms of the nucleon electric
$G_E$ and magnetic $G_M$ Sachs form factors via the relations
\begin{align}
  G_E(q^2) =& F_1(q^2)+\frac{q^2}{(2m_N)^2}F_2(q^2),\,\textrm{and}\nonumber\\
  G_M(q^2) =& F_1(q^2)+F_2(q^2).
  \label{eq:f1f2}
\end{align}

\subsection{Lattice extraction of form factors}

\begin{figure}
  \includegraphics[width=1\linewidth]{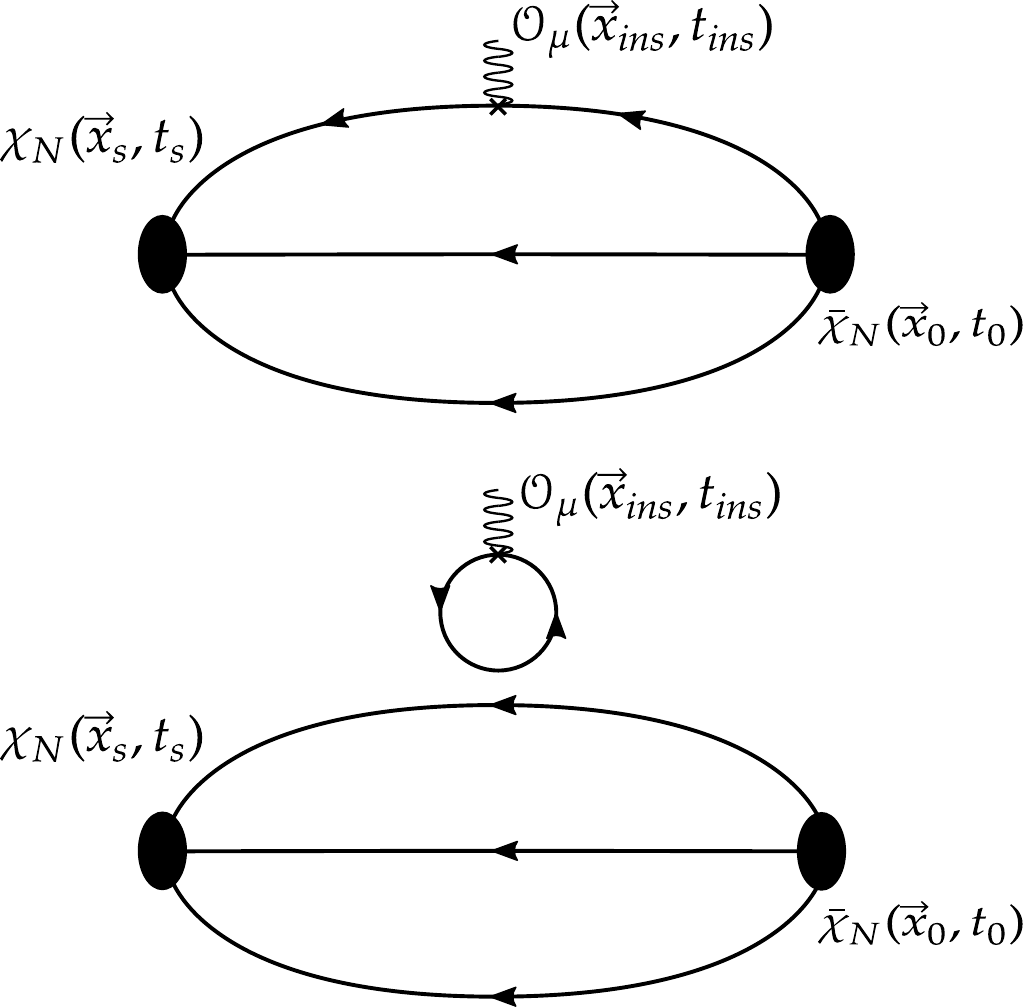}
  \caption{Three-point nucleon correlation function with source at
    $x_0$, sink at $x_s$ and current insertion $\mathcal{O}_\mu$ at
    $x_{\rm ins}$. The connected contribution is shown in the upper
    panel and the disconnected contribution in the lower panel.
  }
  \label{fig:thrp}
\end{figure}

On the lattice, after Wick rotation to Euclidean time, extraction of
matrix elements requires the calculation of a three-point correlation
function shown schematically in Fig.~\ref{fig:thrp}. For simplicity we will
take $x_0=(\vec{0}, 0)$ from here on. We use sequential inversions
through the sink, fixing the sink momentum $\vec{p}\,^\prime$ to zero, which
constrains $\vec{p}=-\vec{q}$:
\begin{widetext}
\begin{align}
  G_\mu(\Gamma;\vec{q};t_s,t_{\rm ins}) = &
  \sum_{\vec{x}_s\vec{x}_{\rm ins}}e^{-i\vec q .\vec{x}_{\rm
      ins}}\Gamma^{\alpha\beta}\langle
  \bar{\chi}^\beta_N(\vec{x}_s;t_s) | \mathcal{O}^\mu(\vec{x}_{\rm
    ins};t_{\rm ins}) | \chi^\alpha_N(\vec{0};0)\rangle\nonumber
  \\ \xrightarrow[t_{\rm ins} \rightarrow \infty]{t_{\rm s}-t_{\rm
      ins}\rightarrow\infty} & \sum_{ss'}\Gamma^{\alpha\beta}\langle
  \chi_N^\beta | N(0,s') \rangle \langle N(p,s)|
  \bar{\chi}_N^\alpha\rangle\langle
  N(0,s')|\mathcal{O}_\mu(q)|N(p,s)\rangle e^{-E_N(\vec{p})t_{\rm
      ins}} e^{-m_N(t_{\rm s}-t_{\rm ins})},
  \label{eq:three-point}
\end{align}
\end{widetext}
where $\Gamma$ is a matrix acting on Dirac indices $\alpha$ and
$\beta$ and $\chi_N$ is the standard nucleon interpolating operator
given by
\begin{equation}
  \chi_N^\alpha(\vec{x},t)=\epsilon^{abc}u^a_\alpha(x)[u^{b\intercal}(x)C\gamma_5d^c(x)].
\end{equation}
with $C=\gamma_0\gamma_2$ the charge conjugation matrix.  In the
second line of Eq.~(\ref{eq:three-point}) we have inserted twice a
complete set of states with the quantum numbers of the nucleon, of
which, after assuming large time separations, only the nucleon
survives with higher energy states being exponentially suppressed.  We
use Gaussian smeared point-sources~\cite{Alexandrou:1992ti,
  Gusken:1989ad} to increase the overlap with the nucleon state with
APE smearing applied to the gauge links, with the same parameters as
in Ref.~\cite{Abdel-Rehim:2015owa}, tuned so as to yield a rms radius
of about 0.5~fm. These are the same parameters as in
Ref.~\cite{Abdel-Rehim:2016won}, namely $(N_G,\alpha_G) = (50, 4)$ for
the Gaussian smearing and $(N_{\rm APE},\alpha_{\rm APE}) = (50, 0.5)$
for the APE smearing.

We construct an optimized ratio dividing $G_\mu$ by a combination of
two-point functions. The optimized ratio $R_\mu$ is given by
\begin{align}
  R_\mu(\Gamma;\vec{q};t_s;t_{\rm ins}) =& \frac{G_{\mu}(\Gamma;\vec{q};t_s;t_{\rm ins})}{G(\vec{0};t_s)}\times\nonumber\\
  &\Biggl[
     \frac{G(\vec{0};t_s)G(\vec{q};t_s-t_{\rm ins})G(\vec{0};t_{\rm ins})}
          {G(\vec{q};t_s)G(\vec{0};t_s-t_{\rm ins})G(\vec{q};t_{\rm ins})}
   \Biggr ]^{\frac{1}{2}}
  \label{eq:ratio}
\end{align}
with the two-point function given by
\begin{align}
  G(\vec{p};t)=\sum_{\vec{x}}
e^{-i\vec{p}\vec{x}}\Gamma^{\alpha\beta}_0\langle
\bar{\chi}^\beta_N(\vec{x};t) | \chi^\alpha_N(\vec{0};0)\rangle.
\label{eq:two-point}
\end{align}
$\Gamma_0$ is the unpolarized projector,
$\Gamma_0=\frac{1+\gamma_0}{4}$. After taking the large time limit,
unknown overlaps and energy exponentials cancel in the ratio, leading to the time-independent quantity $\Pi_\mu(\Gamma;\vec{q})$, defined via:
\begin{align}
R_\mu(\Gamma;\vec{q};t_s;t_{\rm ins})\xrightarrow[t_{\rm
    ins}\rightarrow\infty]{t_s-t_{\rm ins}\rightarrow\infty}\Pi_\mu(\Gamma;\vec{q}).
\end{align}
Having $\Pi_\mu(\Gamma;\vec{q})$, different combinations of current
insertion directions ($\mu$) and nucleon polarizations determined by
$\Gamma$ yield different expressions for the form
factors~\cite{Alexandrou:2006ru,Alexandrou:2011db}. Namely, we have
\begin{align}
  \Pi_0(\Gamma_0;\vec{q}) =&
  \mathcal{C}\frac{E_N+m_N}{2m_N}G_E(Q^2), \nonumber \\
  \Pi_i(\Gamma_0;\vec{q}) =&
  \mathcal{C}\frac{q_i}{2m_N}G_E(Q^2),\nonumber\\ \Pi_i(\Gamma_k;\vec{q})
  =& \mathcal{C}\frac{\epsilon_{ijk}q_j}{2m_N}G_M(Q^2),\label{eq:ffs}
\end{align}
where $Q^2=-q^2$, is the Euclidean momentum transfer squared,
$\mathcal{C}=\sqrt{\frac{2m_N^2}{E_N(E_N+m_N)}}$, and the polarized
projector is given by $\Gamma_k=i\gamma_5\gamma_k\Gamma_0$, and
$i,k=1,2,3$.

In what follows we will use three methods to extract $\Pi_\mu$ from
lattice data:

\noindent i) \textit{Plateau method.} We seek to identify a range of
values of $t_{\rm ins}$ where the ratio $R_\mu$ is time-independent
(plateau region). We fit, within this window, $R_\mu$ to a constant
and use multiple $t_s$ values. Excited states are considered
suppressed when our result does not change with $t_s$.

\noindent ii) \textit{Two-state fit method.}  We fit the time
dependence of the three- and two-point functions keeping contributions
up to the first excited state. Namely, we truncate the two-point
function of Eq.~(\ref{eq:two-point}) keeping only the ground and first
excited states to obtain
\begin{equation}
  G(\vec{p};t) = c_0(\vec{p}) e^{-E(\vec{p})t} [1 + c_1(\vec{p}) e^{-\Delta E_1(\vec{p})t}+ {\cal O}(e^{-\Delta E_2(\vec{p})t})].
\end{equation}
Similarly, the three-point function of Eq.~(\ref{eq:three-point}) becomes
\begin{align}
  G_\mu(\Gamma;\vec{q};t_s,t_{\rm ins}) =& a^\mu_{00}(\Gamma;\vec{q})e^{-m (t_s-t_{\rm ins})} e^{-E(\vec{q}) t_{\rm ins}}\times \nonumber\\
  \bigg [1+&a^\mu_{01}(\Gamma;\vec{q}) e^{-\Delta E_1(\vec{q}) t_{\rm ins}}\nonumber\\
    +&a^\mu_{10}(\Gamma;\vec{q})e^{-\Delta m_1 (t_s-t_{\rm ins})} \nonumber\\
    +&a^\mu_{11}(\Gamma;\vec{q})e^{-\Delta m_1 (t_s-t_{\rm ins})} e^{-\Delta E_1(\vec{q}) t_{\rm ins}}\nonumber\\
    +&\mathcal{O}[\min(e^{-\Delta m_2 (t_s-t_{\rm ins})}, e^{-\Delta E_2(\vec{q})t_{\rm ins}})]\bigg],
  \label{eq:two state}
\end{align}
where $\Delta E_k(\vec{p})=E_k(\vec{p})-E(\vec p)$ is the energy
difference between the $k^{\rm th}$ nucleon excited state and the
ground state at momentum $\vec{p}$ and $m=E(\vec{0})$ and $\Delta
m_k=\Delta E_k(\vec{0})$. The desired ground state matrix element is
given by
\begin{equation}
  \Pi_\mu(\Gamma;\vec{q}) = \frac{a^\mu_{00}(\Gamma;\vec{q})}{\sqrt{c_0(\vec{0})c_0(\vec{q})}}.
\end{equation}
In practice, we fit simultaneously the three-point function and the
finite and zero momentum two-point functions in a twelve parameter fit
to determine $m$, $E(\vec{q})$, $\Delta m_1$, $\Delta E_1(\vec{q})$,
$c_0(\vec{q})$, $c_0(\vec{0})$, $c_1(\vec{q})$, $c_1(\vec{0})$,
$a^\mu_{00}(\Gamma;\vec{q})$, $a^\mu_{01}(\Gamma;\vec{q})$,
$a^\mu_{10}(\Gamma;\vec{q})$ and $a^\mu_{11}(\Gamma;\vec{q})$.
The two-point function is evaluated using the maximum statistics available at time separation $t_s/a=18$. 

\noindent iii) \textit{Summation method}. We sum the ratio of
Eq.~(\ref{eq:ratio}) over the insertion time-slices. From the
expansion up to first excited state of Eq.~(\ref{eq:two state}) one
sees that a geometric sum arises, which yields:
\begin{equation}
  \sum_{t_{\rm ins}=a}^{t_s-a}R_\mu(\Gamma;\vec{q};t_s;t_{\rm
    ins})\xrightarrow{t_s\rightarrow\infty}c+\Pi_{\mu}(\Gamma;\vec{q})t_s+\mathcal{O}(t_se^{-\Delta
    m_1t_s}).
  \label{eq:summation}
\end{equation}
The summed ratio is then fitted to a linear form and the slope is
taken as the desired matrix element. We note that, in quoting final
results, we do not use the values extracted from summation method
However, it does provide an additional  consistency check 
for the plateau values.

\subsection{Lattice setup}
The simulation parameters of the ensemble we use are tabulated in
Table~\ref{table:sim}. We use an N$_\textrm{f}=2$ ensemble of twisted
mass fermion configurations with clover improvement with quarks tuned
to maximal twist, yielding a pion mass of about 130~MeV. The
lattice volume is 48$^3\times$96 and the lattice spacing is determined
at $a=$0.0938$(3)$~fm yielding a physical box length of about
4.5~fm. The value of the lattice spacing is determined using the
nucleon mass, as explained in Ref.~\cite{Alexandrou:2017xwd}.  Details
of the simulation and first results using this ensemble were presented
in Refs.~\cite{Abdel-Rehim:2015owa, Abdel-Rehim:2015pwa}.

\begin{table}
  \caption{Simulation parameters of the ensemble used in this
    calculation, first presented in Ref.~\cite{Abdel-Rehim:2015pwa}.
    The nucleon and pion mass and the lattice spacing have been
    determined in Ref.~\cite{Alexandrou:2017xwd}.}
  \label{table:sim}
  \begin{tabular}{c|r@{=}l}
    \hline\hline
    \multicolumn{3}{c}{$\beta$=2.1, $c_{\rm SW}$=1.57751, $a$=0.0938(3)~fm, $r_0/a$=5.32(5)} \\
    \hline
    \multirow{4}{*}{48$^3\times$96, $L$=4.5~fm} & $\alpha\mu$  & 0.0009       \\
      & $m_\pi$      & 0.1304(4)~GeV\\
      & $m_\pi L$    & 2.98(1)      \\
      & $m_N$        & 0.932(4)~GeV \\
      \hline\hline
  \end{tabular}
\end{table}

The parameters used for the calculation of the correlation functions
are given in Table~\ref{table:statistics}. We use increasing
statistics with increasing sink-source separation so that statistical
errors are kept approximately constant. Furthermore, as will be
discussed in Section~\ref{sec:results}, $G_E(Q^2)$ is found to be more
susceptible to excited states compared to $G_M(Q^2)$, requiring larger
separations for ensuring their suppression.  Therefore, we carry out
sequential inversions for five sink-source separations using the
unpolarized projector $\Gamma_0$, which yields $G_E(Q^2)$ according
Eq.~\ref{eq:ffs}. To obtain $G_M(Q^2)$, we carry out three additional
sequential inversions, one for each polarized projector $\Gamma_k$,
$k=1,2,3$, for each of the three smallest separations.

\begin{table}
  \caption{Parameters of the calculation of the form factors. The
    first column shows the sink-source separations used, the second
    column the sink projectors and the last column the total
    statistics ($N_{\rm st}$) obtained using $N_{\rm cnf}$
    configurations times $N_{\rm src}$ source-positions per
    configuration.}
  \label{table:statistics}
  \begin{tabular}{ccr@{ = }r}
    \hline\hline
    $t_s$ [$a$] & Proj. & $N_{\rm cnf}\cdot N_{\rm src}$&$N_{\rm st}$ \\
    \hline
    10,12,14 & $\Gamma_0$, $\Gamma_k$ & 578$\cdot$16& 9248 \\
    16 & $\Gamma_0$ & 530$\cdot$88& 46640 \\
    18 & $\Gamma_0$ & 725$\cdot$88& 63800 \\
    \hline\hline
  \end{tabular}
\end{table}

\section{Results}
\label{sec:results}
\subsection{Analysis}
\subsubsection{Isovector contributions}
We use the three methods, described in the previous section, to
analyze the contribution due to the excited states and extract the
desired nucleon matrix element.

\begin{figure}
  \includegraphics[width=\linewidth]{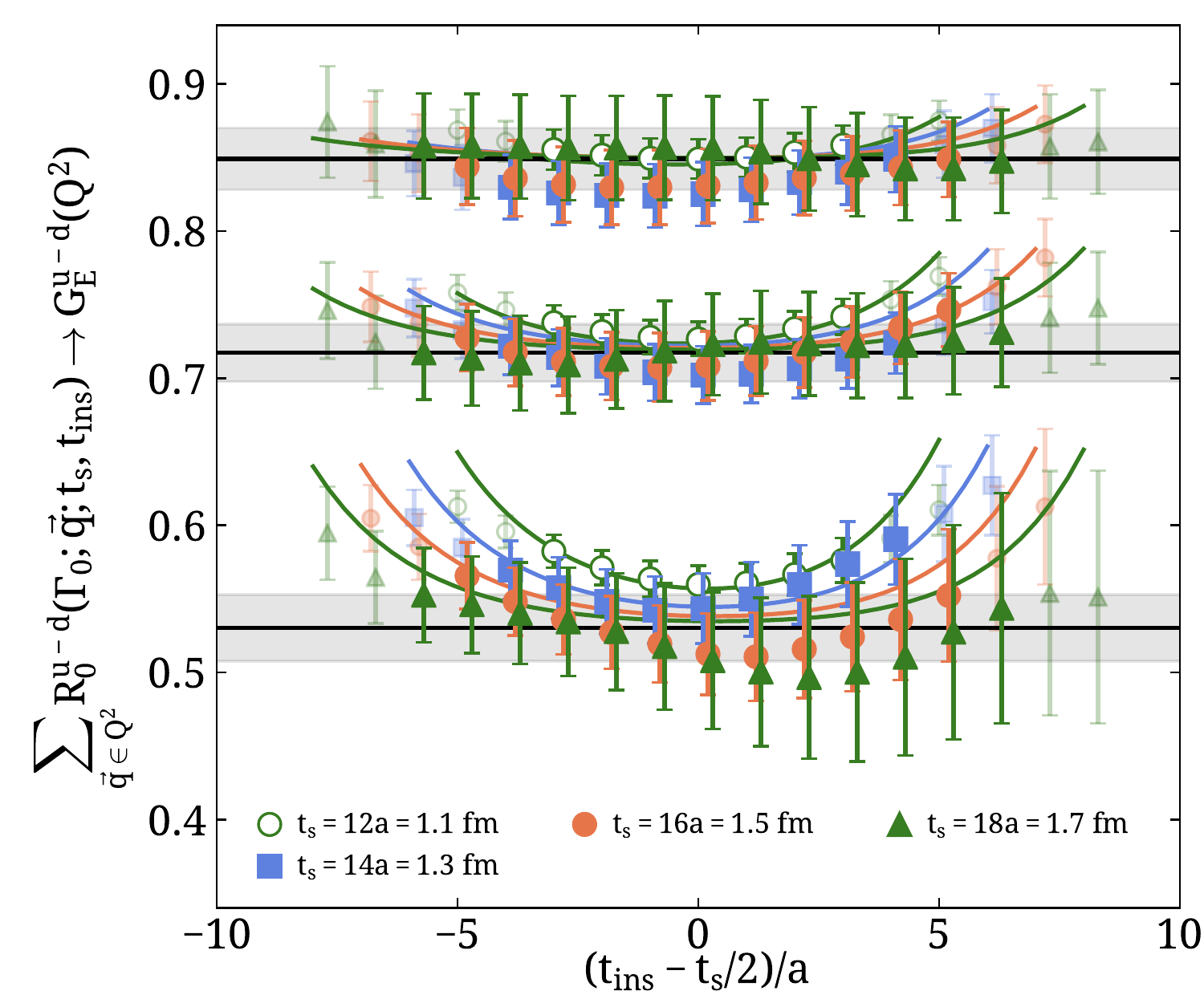}
  \caption{Ratio yielding the isovector electric Sachs form factor. We
    show results for three representative $Q^2$ values, namely the
    first, second and fourth non-zero $Q^2$ values from top to bottom,
    for $t_s=12a$ (open circles), $t_s=14a$ (filled squares),
    $t_s=16a$ (filled circles) and $t_s=18a$ (filled triangles). The
    curves are the results from the two-state fits, with the fainter
    points excluded from the fit. The band is the form factor value
    extracted using the two-state fit.}
  \label{fig:gev two-state}
\end{figure}

\begin{figure}
  \includegraphics[width=\linewidth]{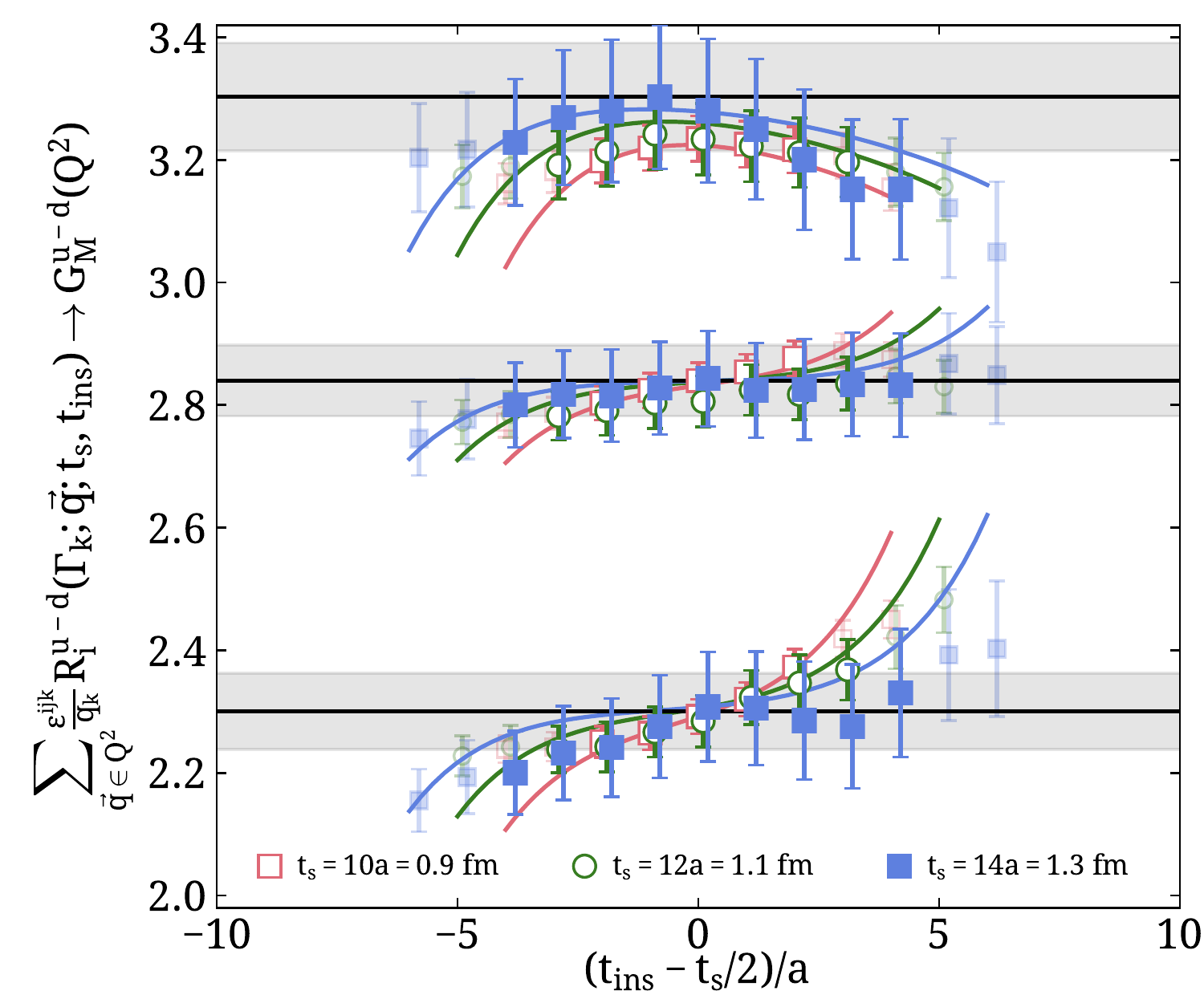}
  \caption{Ratio yielding the isovector magnetic Sachs form factor. We
    show results for three representative $Q^2$ values, namely the
    first, second and fourth non-zero $Q^2$ values from top to bottom,
    for $t_s=10a$ (open squares), $t_s=12a$ (open circles), and
    $t_s=14a$ (filled squares). The curves are the results from the
    two-state fits, with the fainter points excluded from the fit. The
    band is the form factor value extracted using the two-state fit.}
  \label{fig:gmv two-state}
\end{figure}

We demonstrate the quality of our data and two-state fits in
Figs.~\ref{fig:gev two-state} and~\ref{fig:gmv two-state} for the
isovector contributions to $G_E(Q^2)$ and $G_M(Q^2)$ respectively, for
three momentum transfers, namely the first, second and fourth non-zero
$Q^2$ values of our setup. In these figures we show the ratio after
the appropriate combinations of Eq.~(\ref{eq:ffs}) are taken to yield
either $G^{u-d}_E(Q^2)$ or $G^{u-d}_M(Q^2)$.  We indeed observe larger
excited state contamination in the case of $G^{u-d}_E(Q^2)$, which is
the reason for considering larger values of $t_s$ for this case.  We note that for fitting the plateau and
summation methods, the ratios of Eq.~(\ref{eq:ratio}) are constructed
with two- and three-point functions with the same source positions
and gauge configurations. For the
two-state fit, as already mentioned, we use the two-point correlation function at the
maximum statistics available, namely 725 configurations times 88
source positions, as indicated in Table~\ref{table:statistics}. These
are the ratios shown in Figs.~\ref{fig:gev two-state} and~\ref{fig:gmv two-state}, which differ from those used for the plateau fits.

\begin{figure}
  \includegraphics[width=\linewidth]{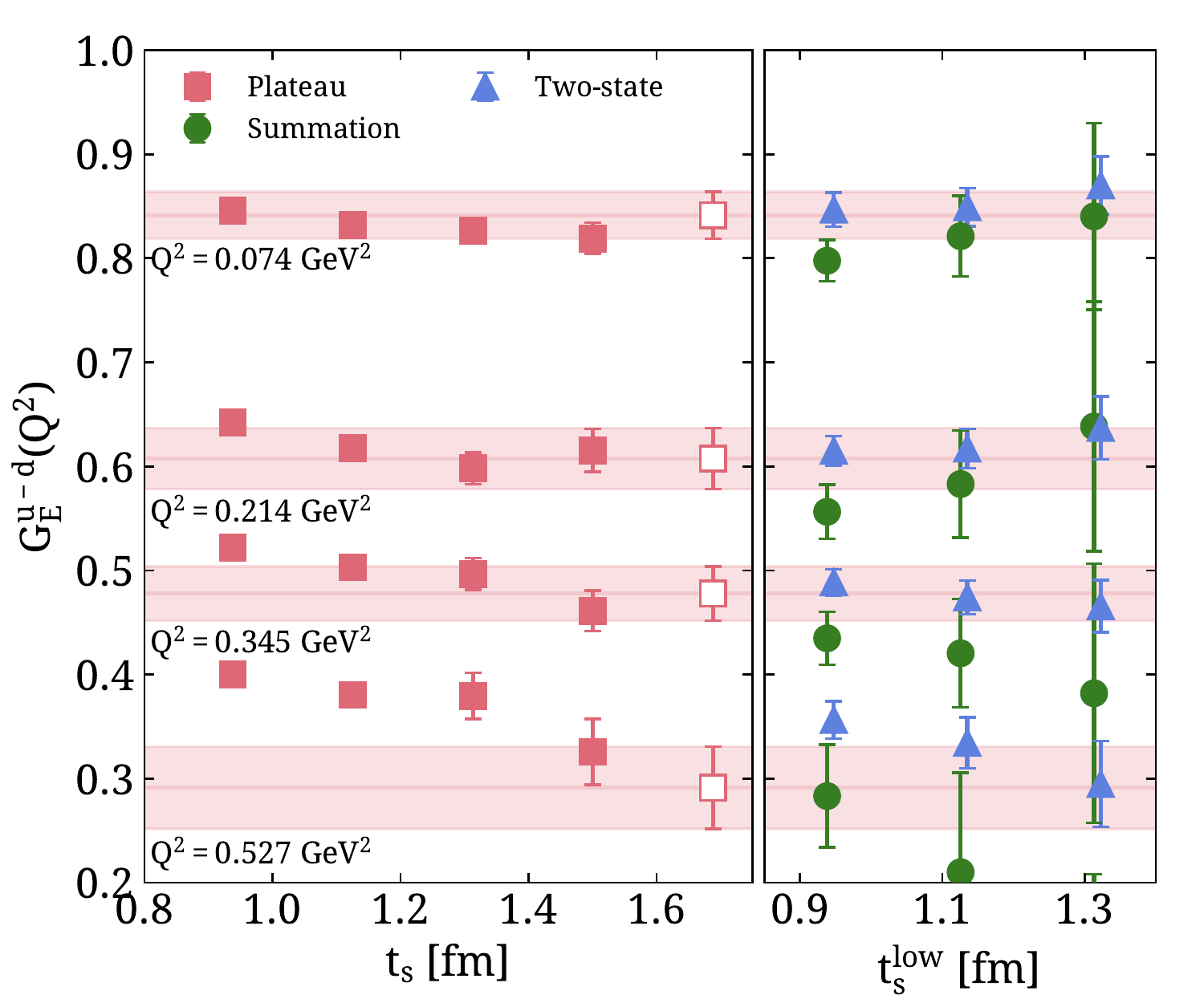}
  \caption{Isovector electric form factor, for four non-zero $Q^2$
    values, extracted from the plateau method (squares), the summation
    method (circles) and the two-state fit method (triangles). The
    plateau method results are plotted as a function of the
    sink-source separation while the summation and two-state fit
    results are plotted as a function of $t_s^{\rm low}$, i.e. of the
    smallest sink-source separation included in the fit, with
    $t_s^{\rm high}$ kept fixed at $t_s=18a=1.7$~fm. The open square
    and band shows the selected value and its statistical error used
    to obtain our final results.}
  \label{fig:gev fits}
\end{figure}

\begin{figure}
  \includegraphics[width=\linewidth]{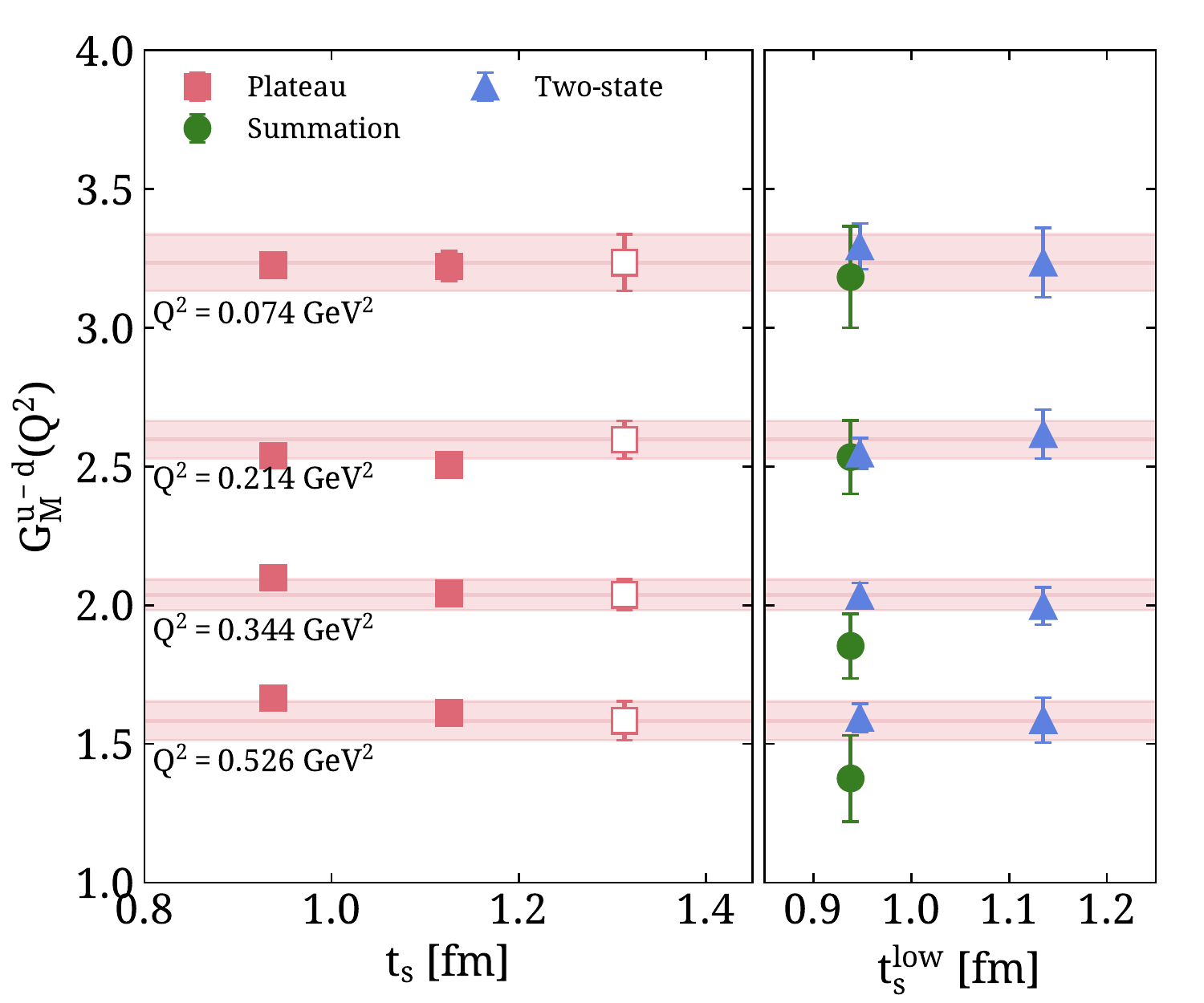}
  \caption{Isovector magnetic form factor. The notation is the same as
    that in Fig.~\ref{fig:gev fits}. For the summation and two-state
    fit methods, the largest sink-source separation included in the
    fit is kept fixed at $t^{\rm high}_s=14a=1.3$~fm.}
  \label{fig:gmv fits}
\end{figure}

The investigation of excited states is facilitated further by
Figs.~\ref{fig:gev fits} and~\ref{fig:gmv fits}. These plots indicate
that excited state contributions are present in $G^{u-d}_E(Q^2)$ for
the first three sink-source separations of $t_s/a=10$, $12$ and~$14$
in particular for larger momentum transfer. For the two larger
sink-source separations we see convergence of the results extracted
from the plateau method, which are in agreement with those from the
summation method and the two-state fits when the lower fit range is
$t_s^{\rm low}=12a=1.1$~fm. For $G^{u-d}_M(Q^2)$, all results from
the three sink-source separations are in agreement and consistent with
the summation and two-state fit methods within their errors. The
values obtained at $t_s=18a=1.7$~fm for the case of $G^{u-d}_E(Q^2)$
and $t_s=14a=1.3$~fm for the case of $G^{u-d}_M(Q^2)$ are shown in
Figs.~\ref{fig:gev fits} and~\ref{fig:gmv fits} with the open symbols
and associated error band that demonstrates consistency with the
values extracted using the summation and two-state fit methods.

\begin{figure}
  \includegraphics[width=\linewidth]{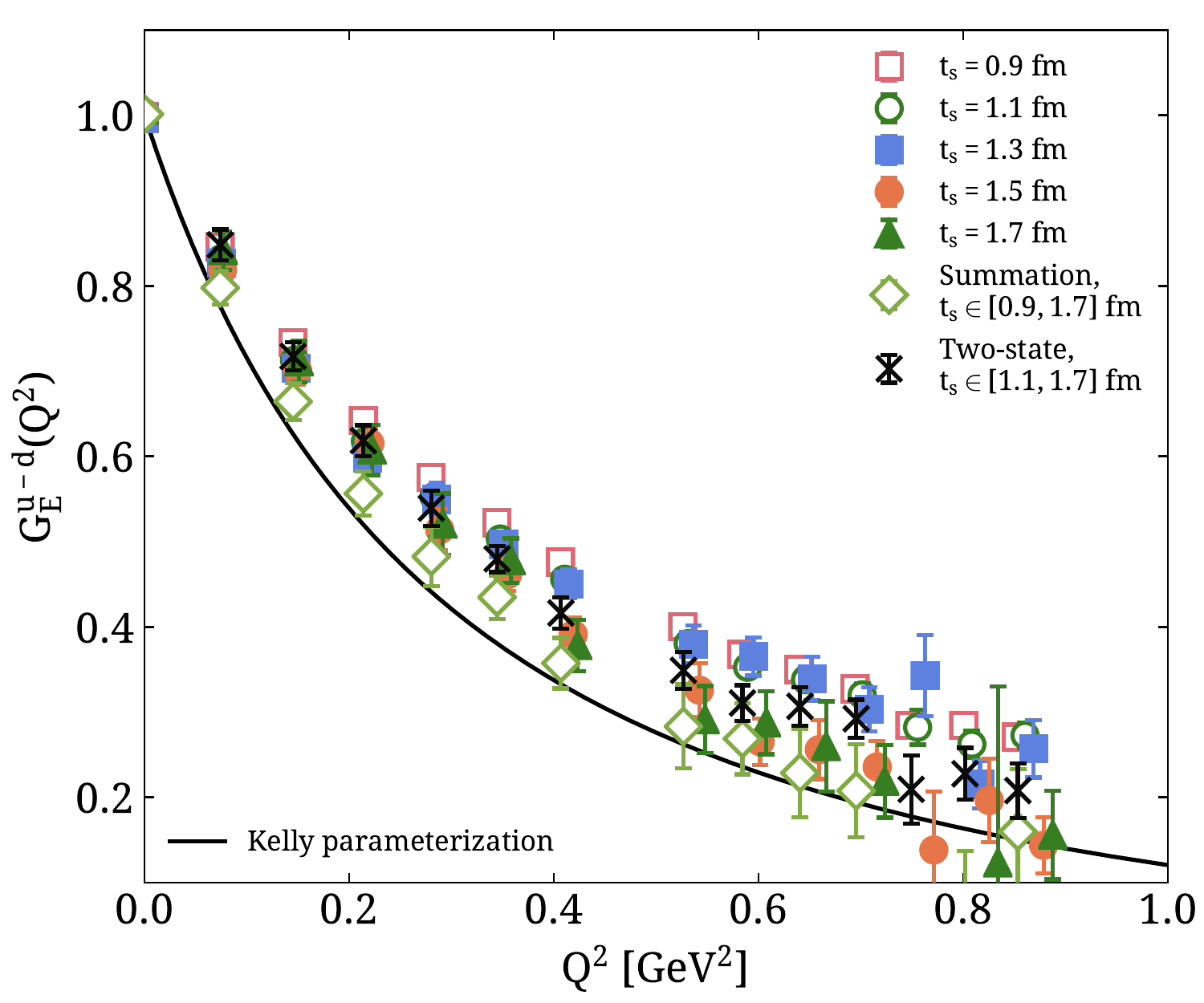}
  \caption{Isovector electric Sachs form factor as a function of the
    momentum transfer squared ($Q^2$). Symbols for the plateau method
    follow the notation of Figs.~\ref{fig:gev two-state}
    and~\ref{fig:gmv two-state}. Results from the summation method are
    shown with open diamonds and for the two-state fit method with the
    crosses. The solid line shows $G_E^p(Q^2)-G_E^n(Q^2)$ using
    Kelly's parameterization of the experimental
    data~\cite{Kelly:2004hm} with parameters taken from Alberico
    \textit{et al.}~\cite{Alberico:2008sz}.}
  \label{fig:gev all ts}
\end{figure}

Our results for the isovector electric Sachs form factor extracted
using all available $t_s$ values and from the summation and two-state
fit methods are shown in Fig.~\ref{fig:gev all ts}. On the same plot
we show the curve obtained from a parameterization of experimental
data for $G^p_E(Q^2)$ and $G_E^n(Q^2)$ according to
Ref.~\cite{Kelly:2004hm}, using the parameters obtained in
Ref.~\cite{Alberico:2008sz}, and taking the isovector combination
$G^p_E(Q^2)-G_E^n(Q^2)$. We see that as the sink-source separation is
increased, our results tend towards the experimental curve.  The
results from the two-state fit method using $t_s^{\rm low}$=$1.1$~fm is
consistent with those extracted from the plateau for $t_s=1.7$~fm for
all $Q^2$ values. Results extracted using the summation method are
consistent within their large errors to those obtained from fitting
the plateau for $t_s=1.7$~fm.

\begin{figure}[!h]
    \includegraphics[width=\linewidth]{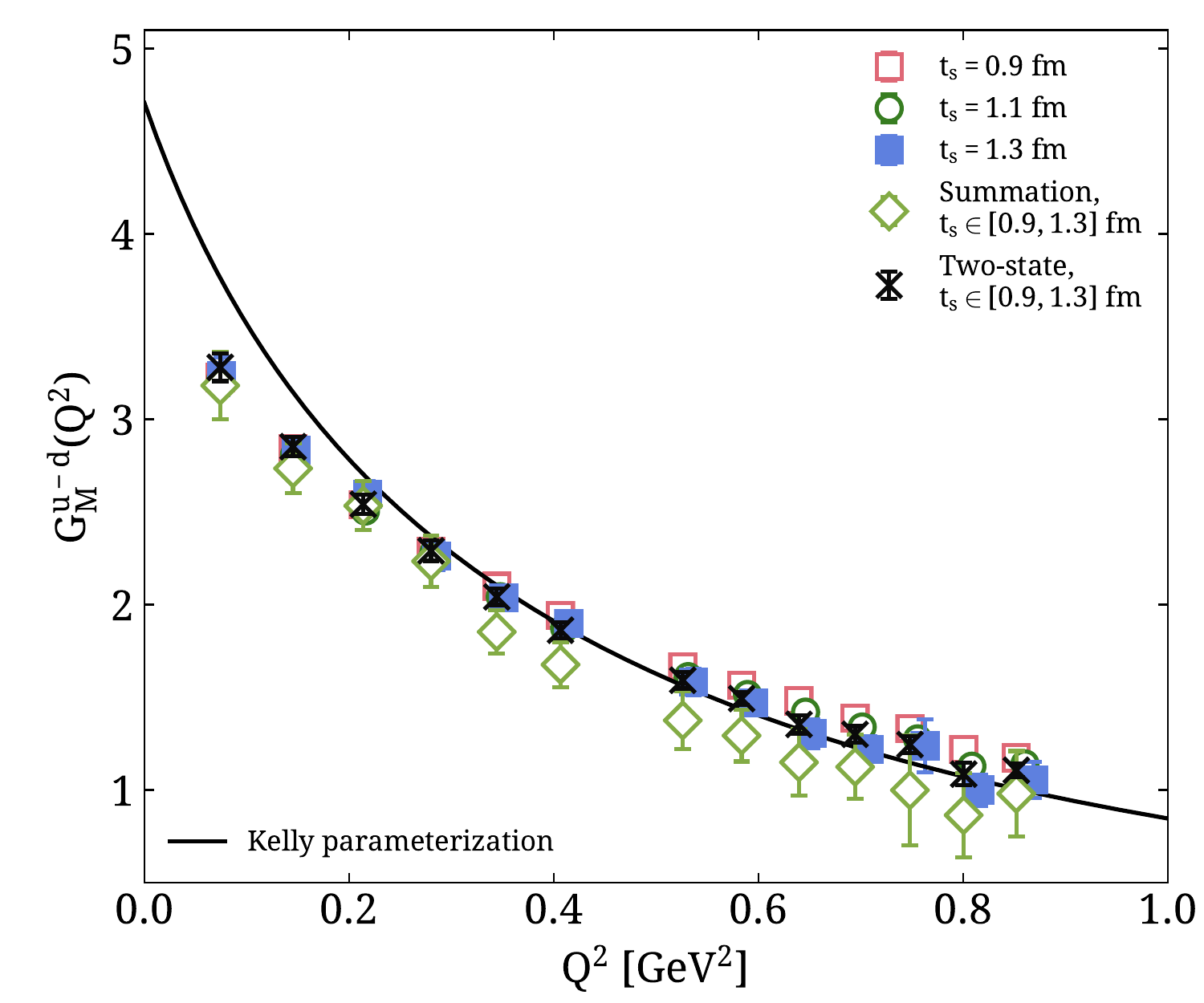}
    \caption{Isovector magnetic Sachs form factor as a function of the
      momentum transfer squared. The notation is the same as that of
      Fig.~\ref{fig:gev all ts}.}
    \label{fig:gmv all ts}
\end{figure}

In Fig.~\ref{fig:gmv all ts} we show our results for the isovector
magnetic form factor. We observe that excited state effects are milder
than in the case of $G^{u-d}_E(Q^2)$, corroborating the conclusion
drawn by observing Fig.~\ref{fig:gmv fits}. We also see agreement with
the experimental curve for $Q^2$ values larger than $\sim$0.2~GeV$^2$.
However, our lattice results underestimate the experimental ones at
the two lowest $Q^2$ values. Excited state effects are seen to be
small for this quantity, and thus they are unlikely to be the cause of
this discrepancy given the consistency of our results at three
separations, as well as with those extracted using the summation and
the two-state fit method. This small discrepancy could be due to
suppressed pion cloud effects, due to the finite volume, that could be
more significant at low momentum transfer. For example, a study of the
magnetic dipole form factor $G_{M1}$ in the $N\rightarrow \Delta$
transition using the Sato-Lee model predicts larger pion cloud
contributions at low momentum transfer~\cite{Sato:2003rq}. Lattice QCD
computations also observe a discrepancy at lower $Q^2$ for $G_{M1}$
when compared to experiment~\cite{Alexandrou:2010uk}. Analysis on a
larger volume is ongoing to investigate volume effects not only in
$G_M(Q^2)$ but also for other nucleon matrix elements and the results
will be reported in subsequent publications. Our results for the form
factors at all sink-source separations and using the summation and
two-state fit methods are included in Appendix~\ref{sec:appendix
  results} in Tables~\ref{table:results gev} to~\ref{table:results
  gms}. Preliminary results for the isovector electromagnetic form
factors have been presented for this ensemble in
Refs.~\cite{Alexandrou:2017msl,Abdel-Rehim:2015jna}.

\subsubsection{Isoscalar contributions}
We perform a similar analysis for the isoscalar contributions, denoted
by $G^{u+d}_E(Q^2)$ and $G^{u+d}_M(Q^2)$. As mentioned, we use the
combination $\sfrac{(u+d)}{3}$ in the matrix element for the isoscalar
such that it yields $G_{E,M}^{u+d}(Q^2) = G^p_{E,M}(Q^2) +
G^n_{E,M}(Q^2)$.  Having also the isovector combination
$G_{E,M}^{u-d}(Q^2)=G^p_{E,M}(Q^2)-G^n_{E,M}(Q^2)$ the individual
proton and neutron form factors can be extracted. While isovector
matrix elements receive no disconnected contributions since they
cancel in the isospin limit, the isoscalar form factors do include
disconnected fermion loops, shown schematically in
Fig.~\ref{fig:thrp}. These disconnected contributions are included for
the first time here at the physical point to obtain the isoscalar form
factors.

\begin{figure}
  \includegraphics[width=\linewidth]{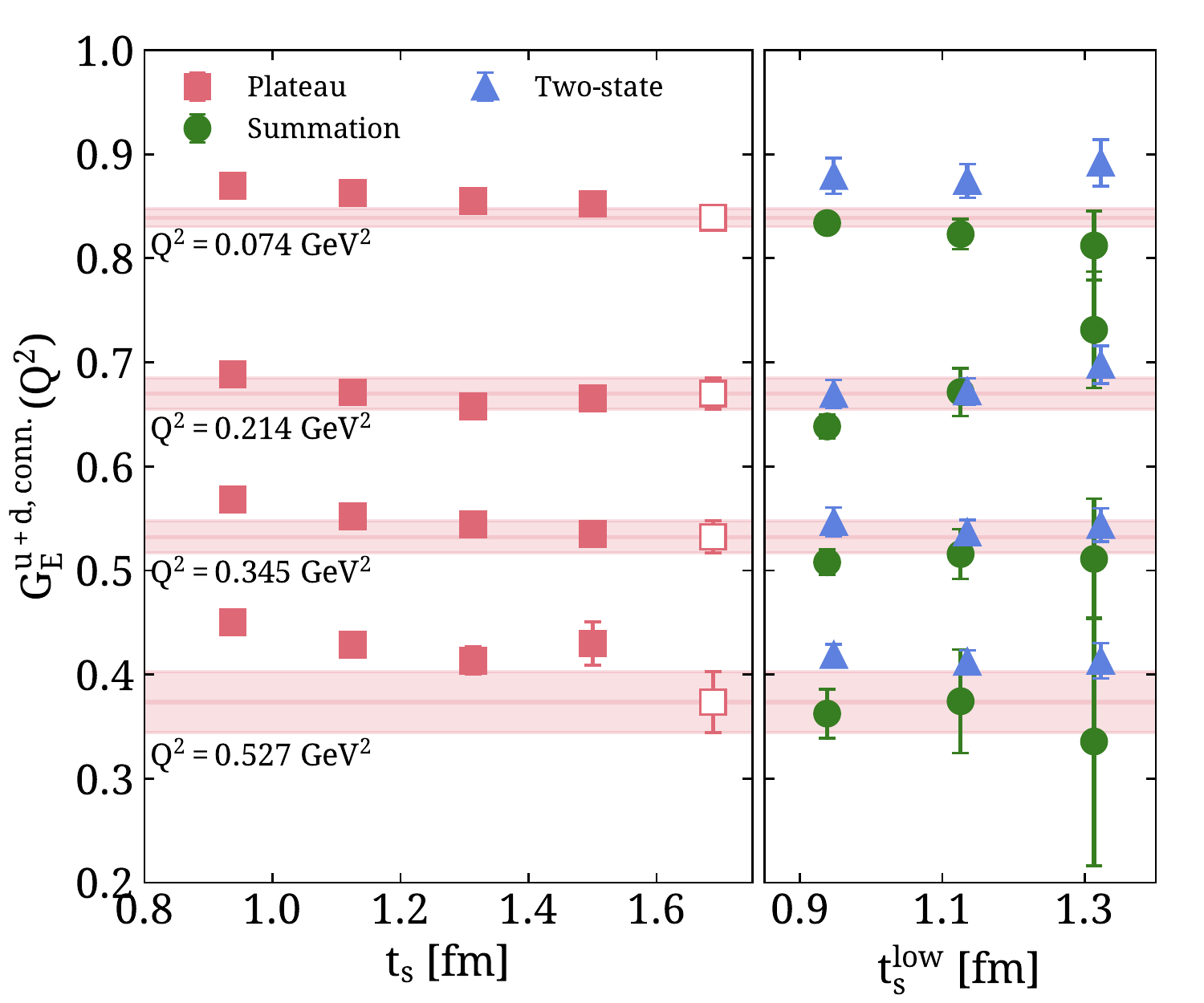}
  \caption{Connected contribution to the $G^{u+d}_E(Q^2)$ form factor,
    for four non-zero $Q^2$ values, extracted from the plateau method
    (squares), the summation method (filled circles) and the two-state
    fit method (filled triangles). The notation is the same as in
    Fig.~\ref{fig:gev fits}.}
  \label{fig:ges fits}
\end{figure}

\begin{figure}
  \includegraphics[width=\linewidth]{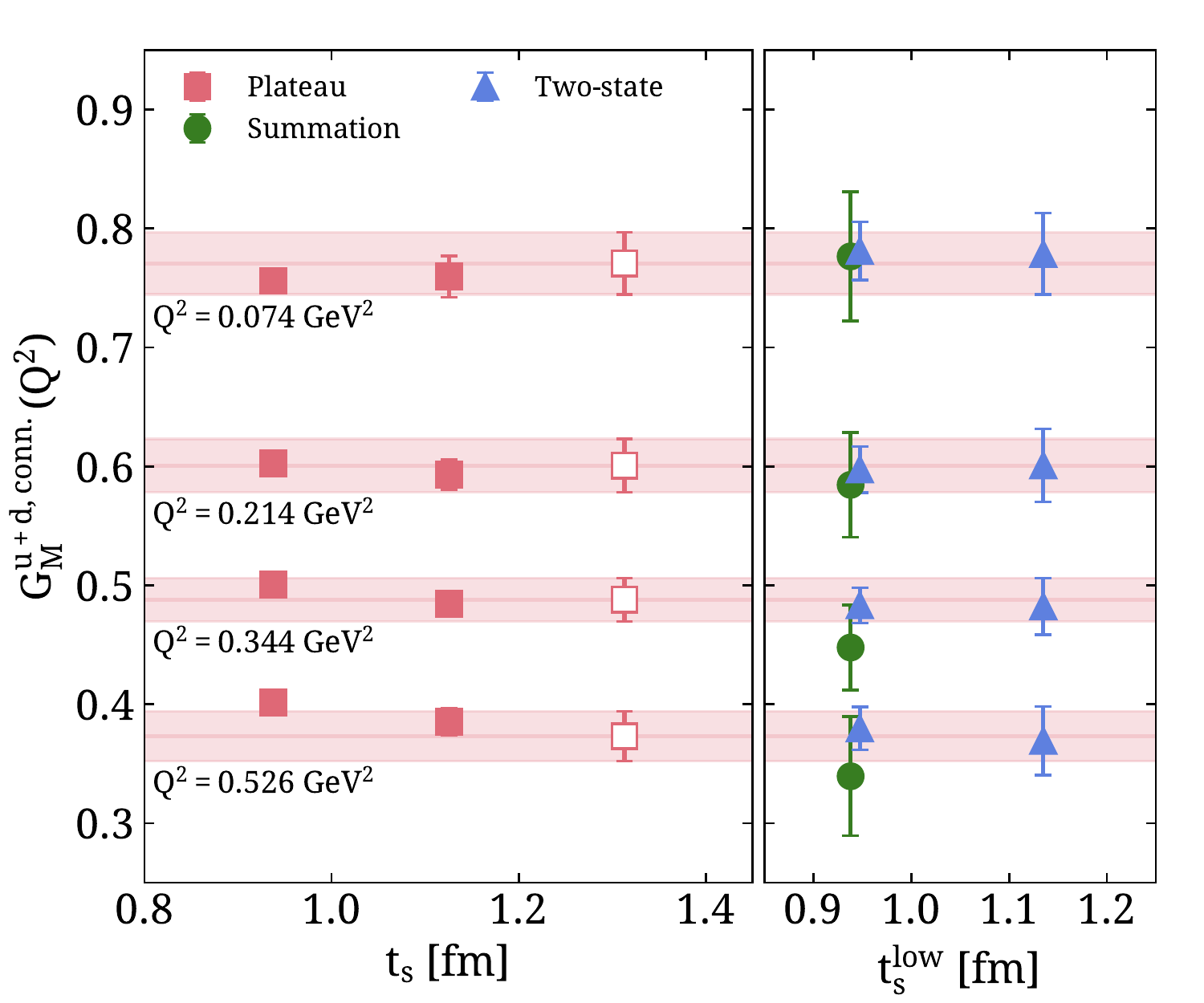}
  \caption{Connected contribution to the $G^{u+d}_M(Q^2)$ form factor.
    The notation is the same as in Fig.~\ref{fig:gmv fits}.  }
  \label{fig:gms fits}
\end{figure}

The connected isoscalar three-point function is computed using the
same procedure as in the isovector case. We show results for the
connected contribution to $G_E^{u+d}(Q^2)$ and $G_M^{u+d}(Q^2)$ in
Figs.~\ref{fig:ges fits} and~\ref{fig:gms fits} respectively. These
results are for the same momentum transfer values as used in
Figs.~\ref{fig:gev fits} and~\ref{fig:gmv fits}. In the case of the
isoscalar electric form factor, we observe contributions due to
excited states that are similar to those observed for the isovector
case.  Namely, we find that a separation of about $t_s$=1.7~fm is
required for their suppression. For the isoscalar magnetic form
factor, we observe that the values extracted from fitting the plateau
at time separations $t_s=1.1$~fm and $t_s=1.3$~fm are consistent and
also in agreement with the values extracted using the two-state fit
and summation methods.

\begin{figure}
    \includegraphics[width=\linewidth]{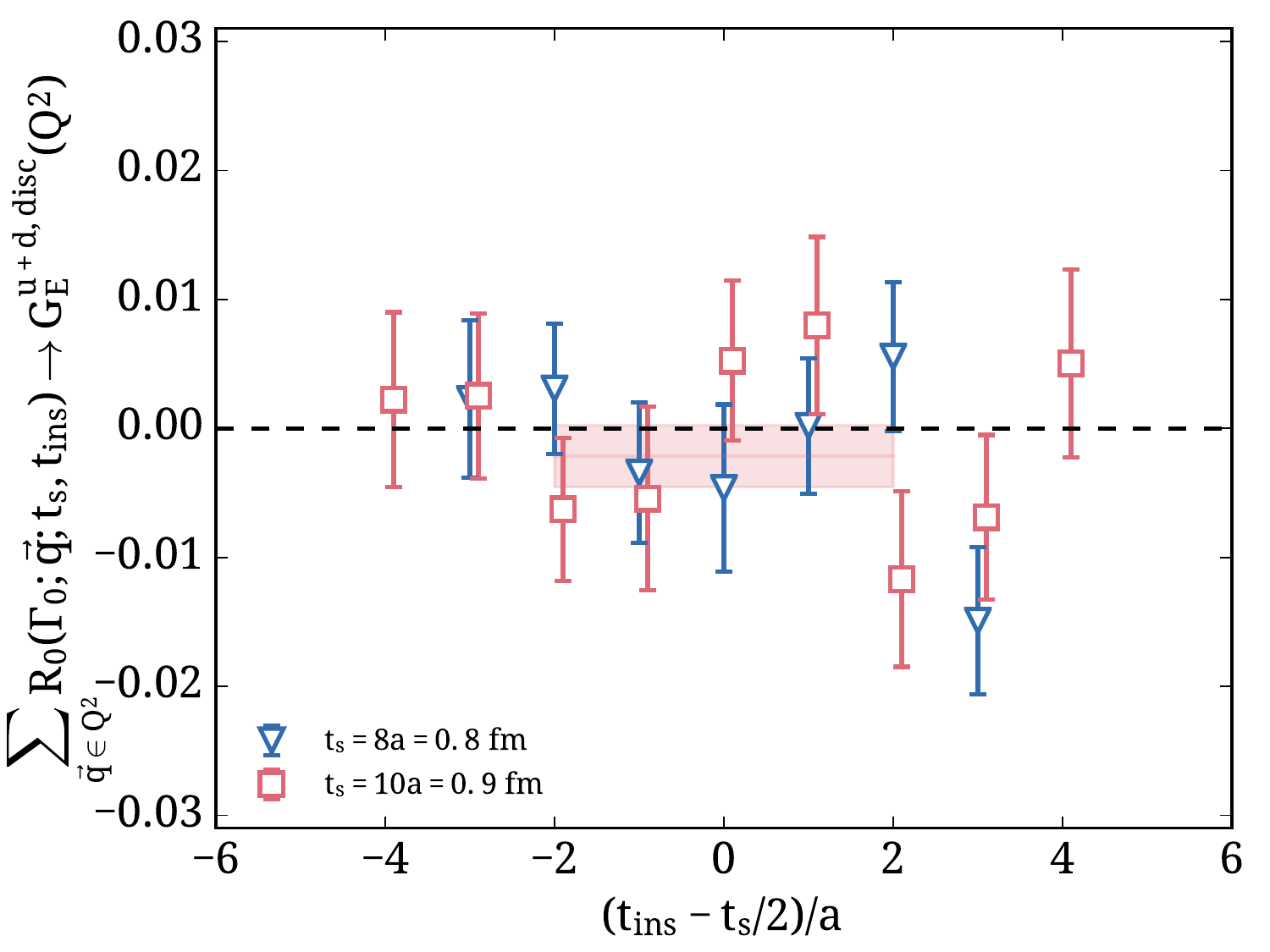}\\
    \includegraphics[width=\linewidth]{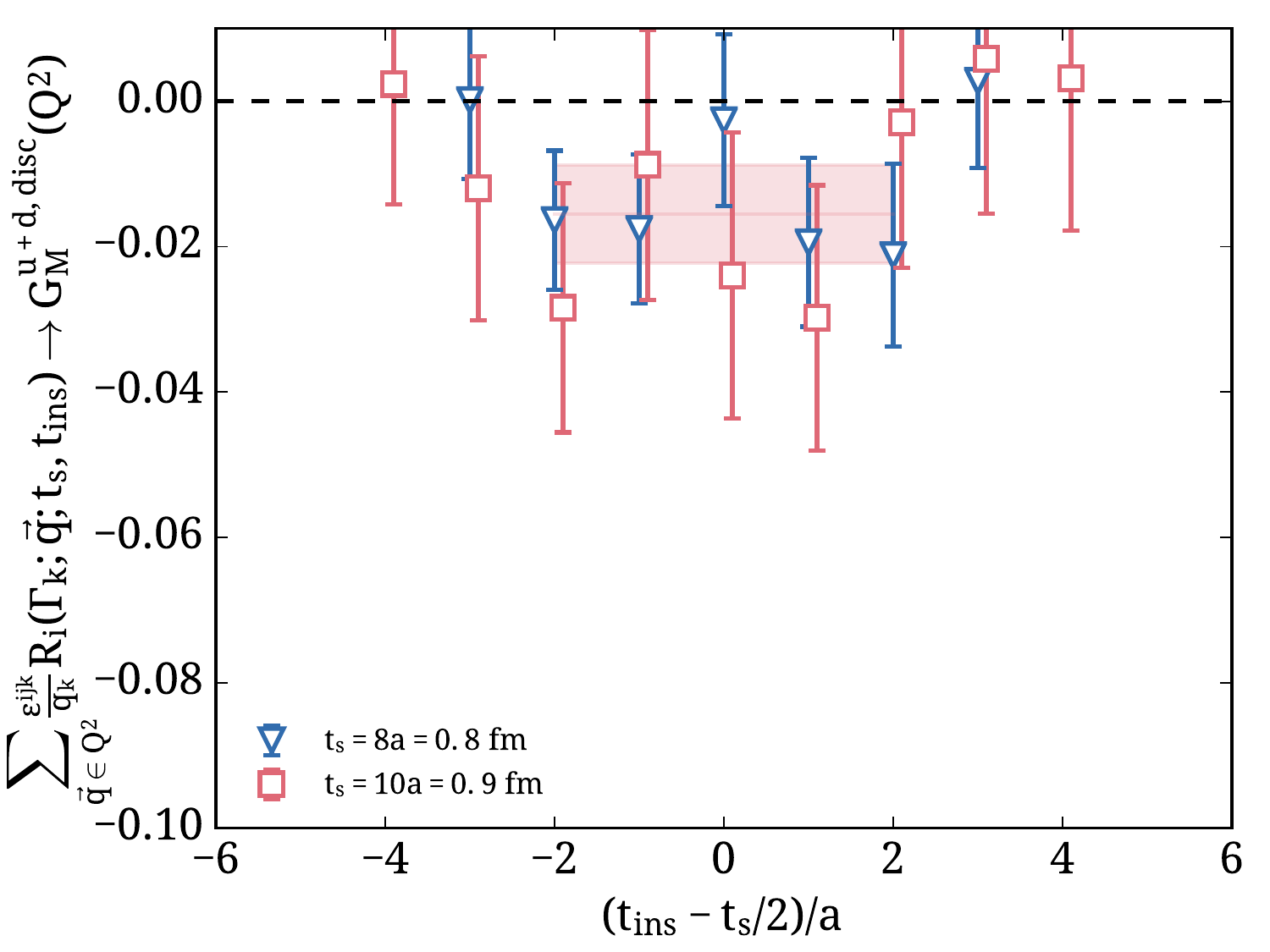}
    \caption{Disconnected contribution to the electric (upper panel)
      and magnetic (lower panel) isoscalar Sachs form factors for
      sink-source separation $t_s=8a=0.75$~fm (inverted triangles) and
      $t_s=10a=0.94$~fm (squares) for the first non-zero momentum
      transfer of $Q^2=0.074$~GeV$^2$. The horizontal bands show the
      values obtained after fitting with the plateau method to the
      results at $t_s=10a=0.94$~fm.}
    \label{fig:gem disc}
\end{figure}

The disconnected diagrams of the electromagnetic form factors are
particularly susceptible to statistical fluctuations, even at larger
pion masses of 370~MeV as reported in Refs.~\cite{Abdel-Rehim:2013wlz}
and~\cite{Alexandrou:2012zz}. Here we show results, at the physical
pion mass, for the disconnected contribution to $G_E^{u+d}(Q^2)$ and
$G_M^{u+d}(Q^2)$ in Fig.~\ref{fig:gem disc} for the first non-zero
momentum transfer. The results are obtained using the same ensemble
used for the connected contributions, detailed in
Table~\ref{table:sim}, using 2120 configurations, with two-point
functions computed on 100 randomly chosen source positions per
configuration. 2250 stochastic noise vectors are used for estimating
the fermion loop. Averaging the proton and neutron two-point functions
and the forward and backwards propagating nucleons yields a total of
8$\cdot 10^5$ statistics. More details of this calculation are
presented in Ref.~\cite{Alexandrou:2017hac}, where results for the
axial form factors are shown.

In the case of the electric form factor, we obtain
$G^{u+d,\textrm{disc.}}_E(Q^2=0.074~\textrm{GeV}^2) = -0.002(3)$,
which is consistent with zero and about 0.2\% of the value of the
connected contribution and four times smaller than its statistical
error. For the magnetic form factor, fitting to the plateau we obtain
$G^{u+d,\textrm{disc.}}_M(Q^2=0.074~\textrm{GeV}^2) = -0.016(7)$ which
is 2\% of the value of the connected contribution at this $Q^2$ and
half the value of the statistical error. These values are consistent
with a dedicated study of the disconnected contributions using an
ensemble of clover fermions with pion mass of
317~MeV~\cite{Green:2015wqa} and a recent result at the physical point
presented in Ref.~\cite{Sufian:2017osl}. There it was shown that
$G^{u+d,\textrm{disc.}}_M(Q^2)$ is negative and largest in magnitude
at $Q^2=0$ while $G^{u+d,\textrm{disc.}}_E(Q^2)$ is largest at around
$Q^2=0.4$~GeV$^2$. In our case, at our largest momentum transfer, we
find $G^{u+d,\textrm{disc.}}_E(Q^2=0.280~\textrm{GeV}^2) =
-0.0056(40)$, which is 1\% of the value of the connected contribution
at this momentum transfer and smaller than the associated statistical
error. Investigation of methods for increasing the precision at the
physical point is ongoing, with preliminary results presented in
Ref.~\cite{Abdel-Rehim:2016pjw} for the ensemble used here, and will
be reported in a separate work.

\begin{figure}
  \includegraphics[width=\linewidth]{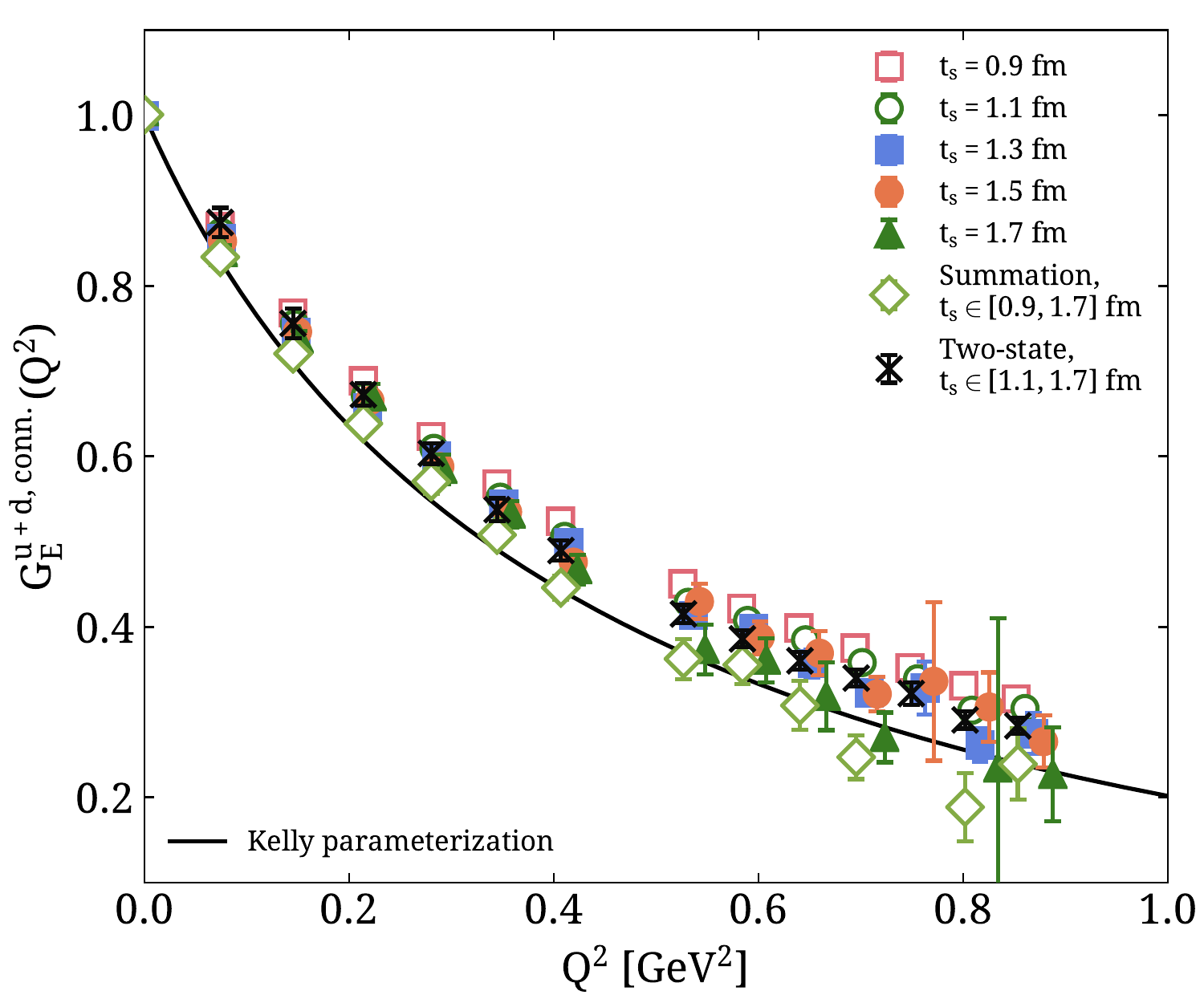}
  \caption{Connected contribution to the isoscalar electric Sachs form
    factor as a function of the momentum transfer, using the notation
    of Fig.~\ref{fig:ges all ts}. The solid line shows
    $G_E^p(Q^2)+G_E^n(Q^2)$ using the Kelly parameterization of
    experimental data from Ref.~\cite{Kelly:2004hm} with parameters
    taken from Alberico \textit{et al.}~\cite{Alberico:2008sz}.}
  \label{fig:ges all ts}
\end{figure}

\begin{figure}[!h]
    \includegraphics[width=\linewidth]{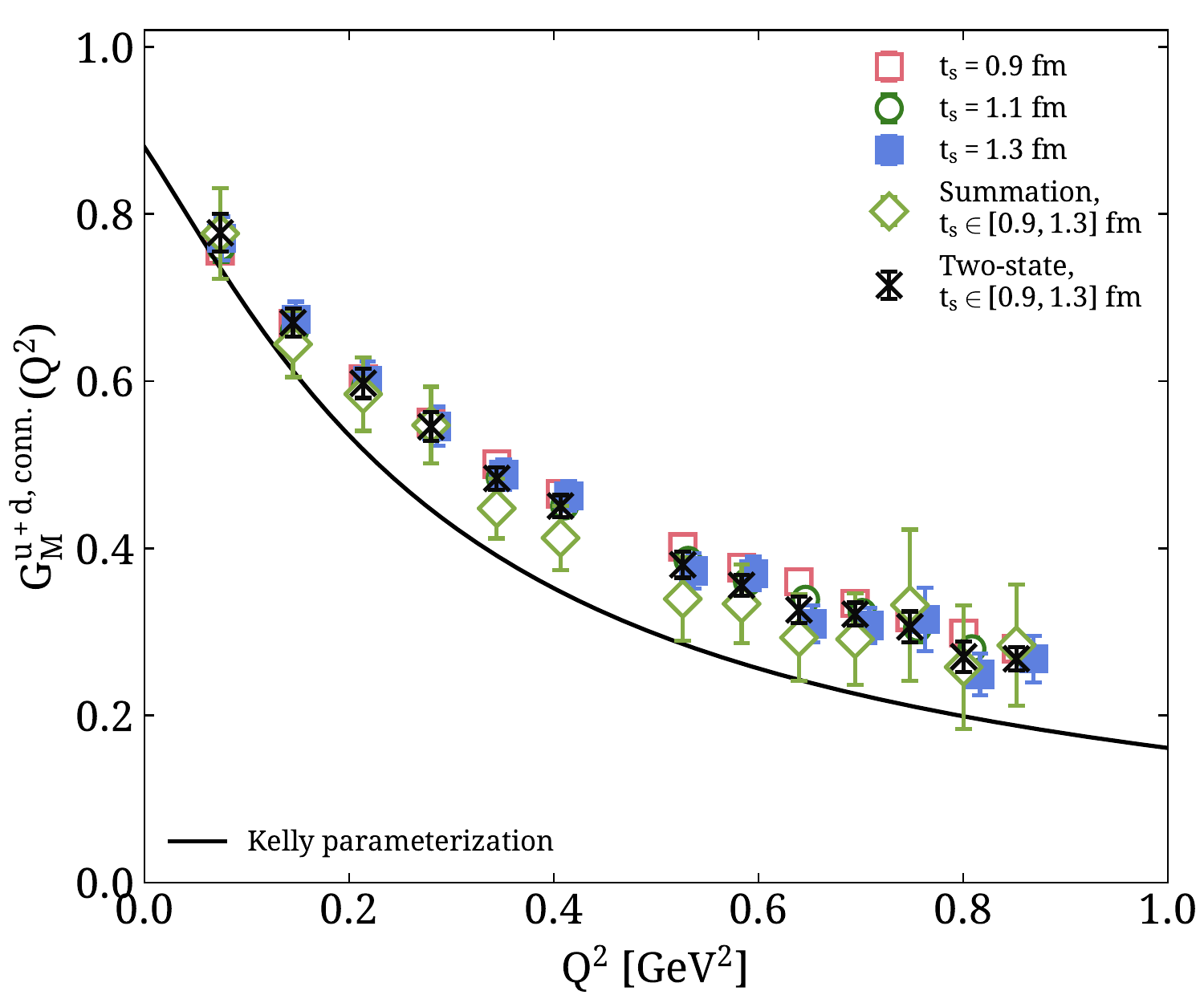}
    \caption{Connected contribution to the isoscalar magnetic Sachs
      form factor as a function of the momentum transfer. The notation
      is the same as in Fig.~\ref{fig:ges all ts}.}
    \label{fig:gms all ts}
\end{figure}

We show our results for the connected contribution to the isoscalar
electric and magnetic form factors in Figs.~\ref{fig:ges all ts}
and~\ref{fig:gms all ts} extracted from the plateau method for all
available sink-source separations, and from the summation and the
two-state fit methods. The isoscalar electric form factor tends to
decrease as the sink-source separation increases approaching the
experimental parameterization. This may indicate residual excited
state effects, that need to be further investigated by going to larger
time separations. For the isoscalar magnetic form factor, we observe a
weaker dependence on $t_s$ pointing to less severe excited state
effects.

\subsection{$Q^2$-dependence of the form factors}
\subsubsection{Isovector and isoscalar form factors}

\begin{figure}
    \includegraphics[width=\linewidth]{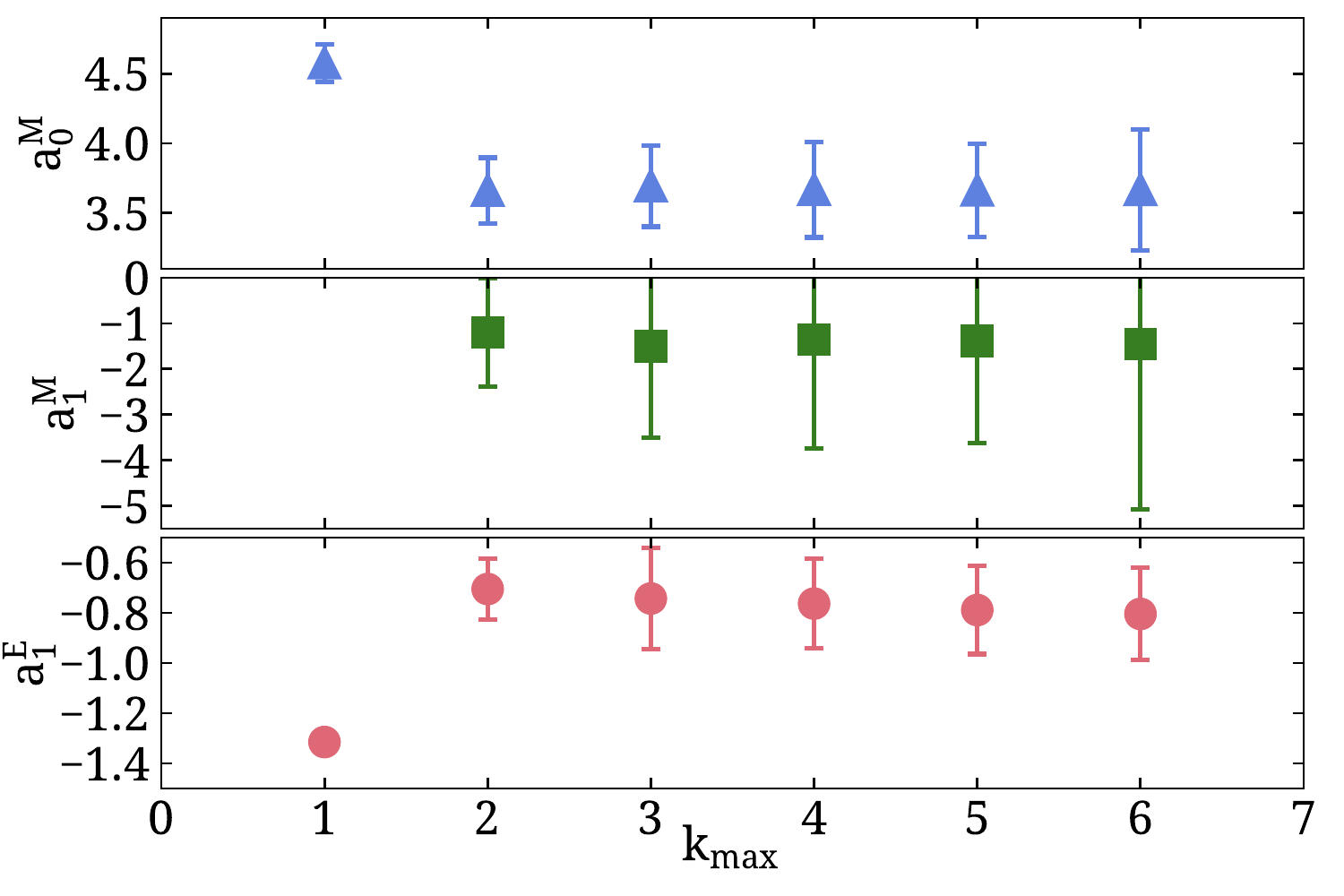}
    \caption{Results from fitting using the z-expansion as a function
      of $k_{\rm max}$ for $a_1^E$ (lower panel), $a_1^M$ (center
      panel) and $a_0^M$ (top panel) of Eq.~\ref{eq:zexp fit}.}
    \label{fig:zexp fit kmax}
\end{figure}

We fit $G_E(Q^2)$ and $G_M(Q^2)$ to both a dipole Ansatz and the
z-expansion form. The truncated z-expansion is expected to model
better the low-$Q^2$~\cite{Hill:2010yb} dependence of the form
factors, while the dipole form is motivated by vector-meson pole
contributions to the form factors~\cite{Perdrisat:2006hj}. For the
case of the dipole fits, we use
\begin{equation}
  G_i(Q^2) = \frac{G_i(0)}{(1+\frac{Q^2}{M_i^2})^2},
  \label{eq:dipole fit}
\end{equation}
with $i=E,\,M$, allowing both $G_M(0)$ and $M_M$ to vary for the case
of magnetic form factor, while constraining $G_E(0)=1$ for the case of
the electric form factor. For the z-expansion, we use the
form~\cite{Hill:2010yb}
\begin{align}
  G_i(Q^2) &= \sum_{k=0}^{k_{\rm max}} a_k^{i}z^k,\,\textrm{where}\nonumber\\
  z&=
  \frac{\sqrt{t_{\rm cut}+Q^2} - \sqrt{t_{\rm cut}}}{\sqrt{t_{\rm cut}+Q^2} + \sqrt{t_{\rm cut}}}
  \label{eq:zexp fit}
\end{align}
and take $t_{\rm cut}=4m_\pi^2$. For both isovector and isoscalar
$G_E(Q^2)$ we fix $a^E_0=1$ while for $G_M(Q^2)$ we allow all
parameters to vary. We use Gaussian priors for $a^i_k$ for $k\ge2$
with width $w=5\max(|a_0^i|, |a_1^i|)$ as proposed in
Ref.~\cite{Epstein:2014zua}. We observe larger errors when fitting
with the z-expansion compared to the dipole form. In
Fig.~\ref{fig:zexp fit kmax} we show $a^M_0$ and $a^M_1$ from fits to
the magnetic isovector form factor and $a^E_1$ from fits to the
electric as a function of $k_\textrm{max}$ and observe no significant
change in the fitted parameters beyond $k_{\rm max}\ge2$. We also note
that the resulting values for $a^{i}_k$ for $k\ge2$ obtained are well
within the Gaussian priors, i.e. $|a^{i}_k|\ll 5\max(|a_0^i|,
|a_1^i|)$. We therefore quote results using $k_{\rm max}=2$ from here
on.

\begin{figure}
  \includegraphics[width=\linewidth]{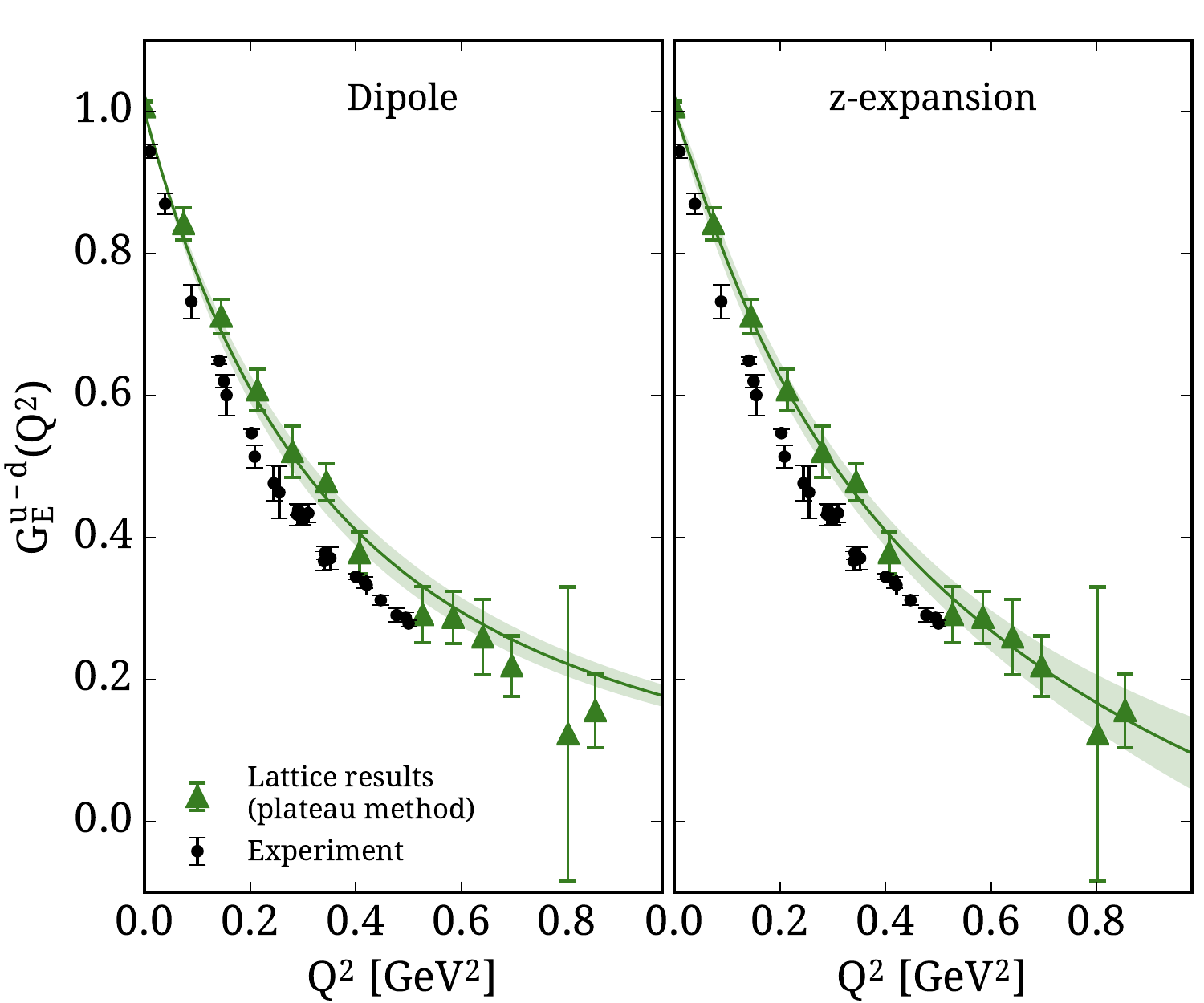}
  \caption{Isovector electric Sachs form factor as a function of the
    momentum transfer extracted from the plateau method at
    $t_s=18a=1.7$~fm (triangles).  We show fits using the dipole form
    (left) and the z-expansion (right). The black points are obtained
    using experimental data for $G_E^p(Q^2)$ from
    Ref.~\cite{Bernauer:2013tpr} and for $G_E^n(Q^2)$ from
    Refs.~\cite{Golak:2000nt, Becker:1999tw, Eden:1994ji,
      Meyerhoff:1994ev, Passchier:1999cj, Warren:2003ma, Zhu:2001md,
      Plaster:2005cx, Madey:2003av, Rohe:1999sh, Bermuth:2003qh,
      Glazier:2004ny, Herberg:1999ud, Schiavilla:2001qe,
      Ostrick:1999xa}.}
  \label{fig:gev ff fit}
\end{figure}

\begin{figure}
    \includegraphics[width=\linewidth]{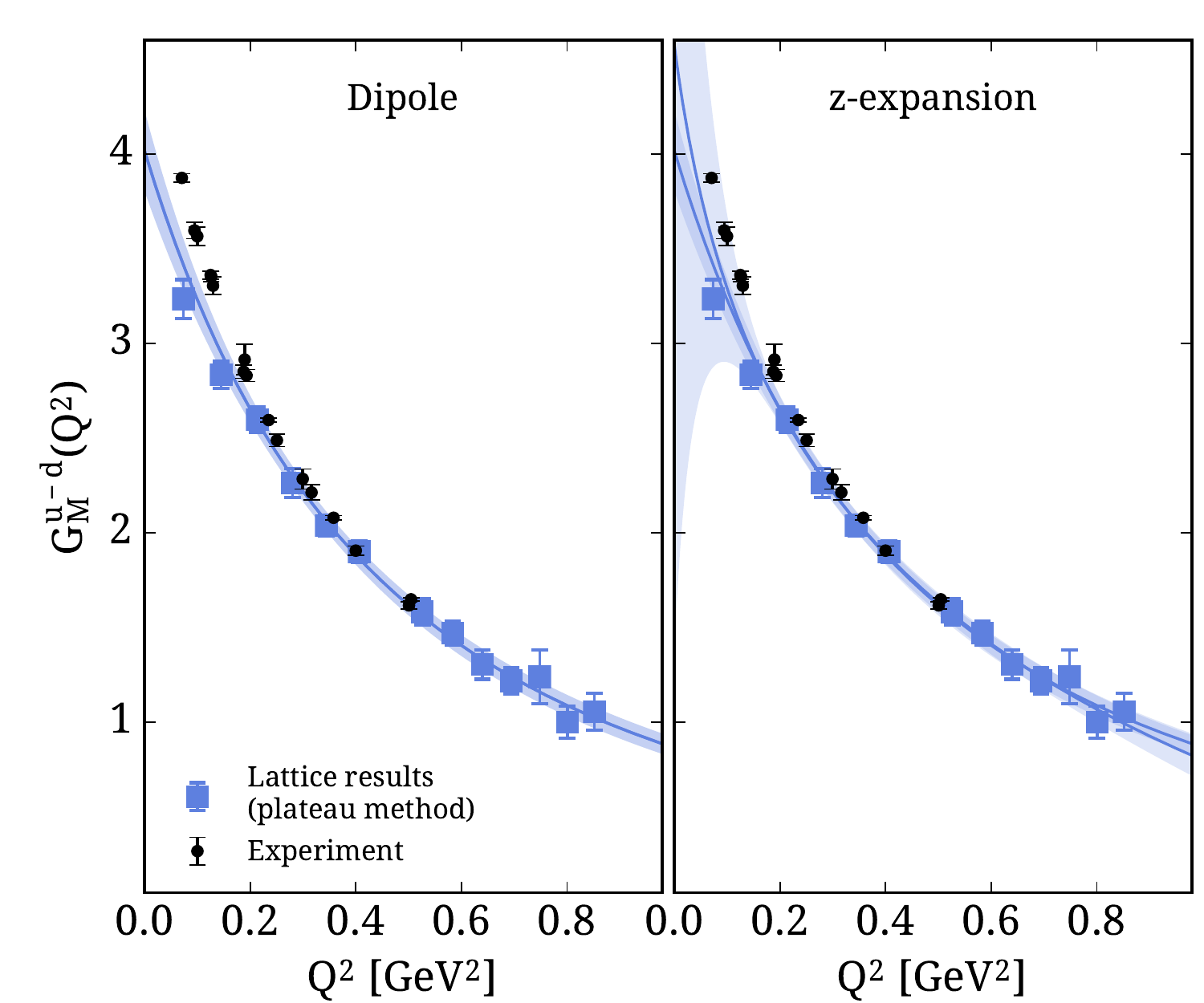}
    \caption{Isovector magnetic Sachs form factor as a function of the
      momentum transfer extracted from the plateau method at
      $t_s=14a=1.3$~fm (squares).  We show fits using the dipole form
      (left) and the z-expansion (right). The smaller error band
      corresponds to fitting to all $Q^2$ values, while the larger
      band is obtained after omitting the two smallest values.  The
      black points are obtained using experimental data for
      $G_M^p(Q^2)$ from Ref.~\cite{Bernauer:2013tpr} and for
      $G_M^n(Q^2)$ from Refs.~\cite{Anderson:2006jp, Gao:1994ud,
        Anklin:1994ae, Anklin:1998ae, Kubon:2001rj, Alarcon:2007zza}.}
    \label{fig:gmv ff fit}
\end{figure}

\begin{figure}
    \includegraphics[width=\linewidth]{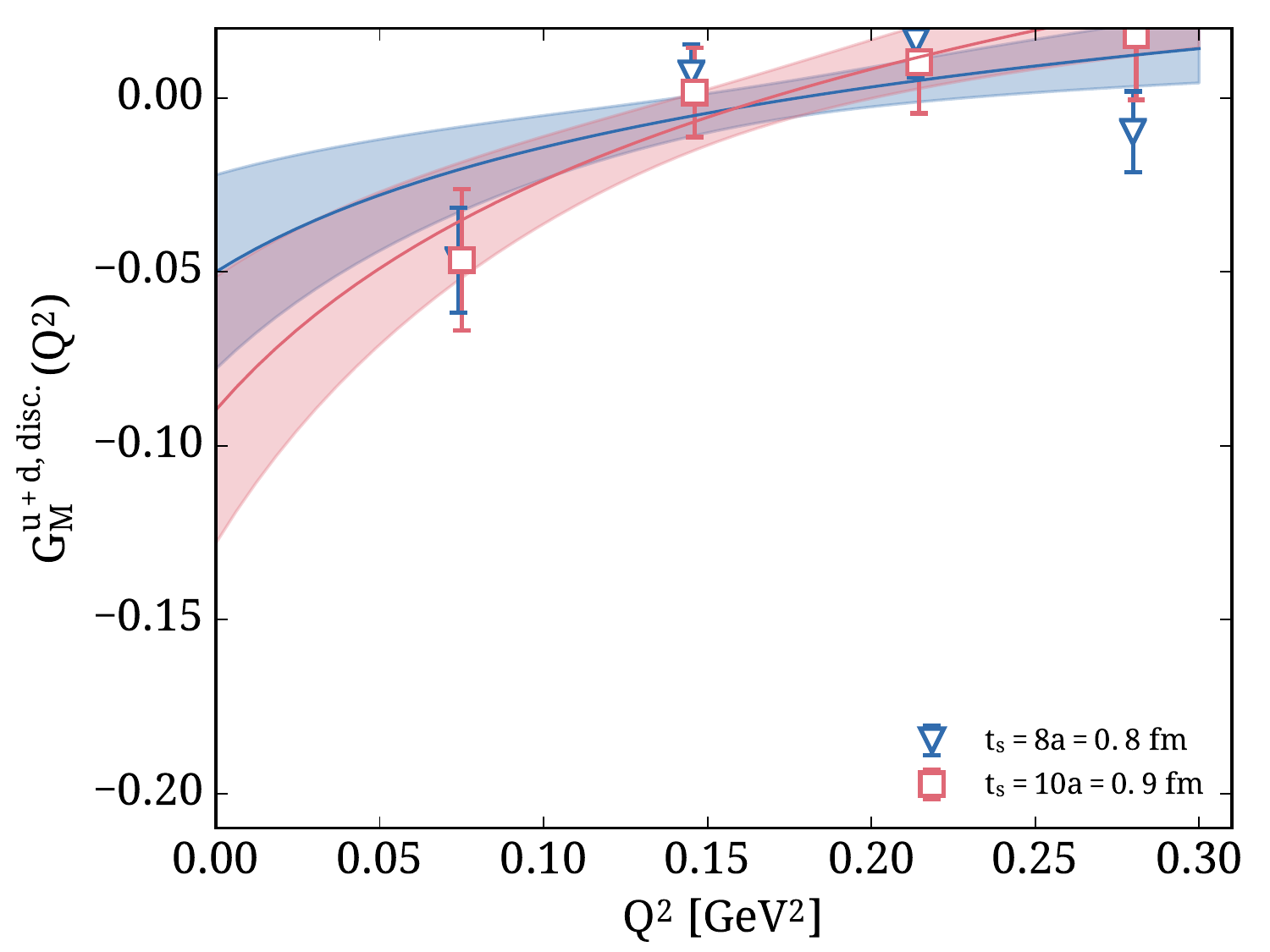}
    \caption{Disconnected contribution to the isoscalar magnetic Sachs
      form factor as a function of the momentum transfer for
      $t_s=8a=0.7$~fm (inverted triangles) and $t_s=10a=0.9$~fm
      (squares).  The bands show fits to the z-expansion form with
      $k_{\rm max}=1$.}
    \label{fig:gms disc ff fit}
\end{figure}

\begin{figure}
  \includegraphics[width=\linewidth]{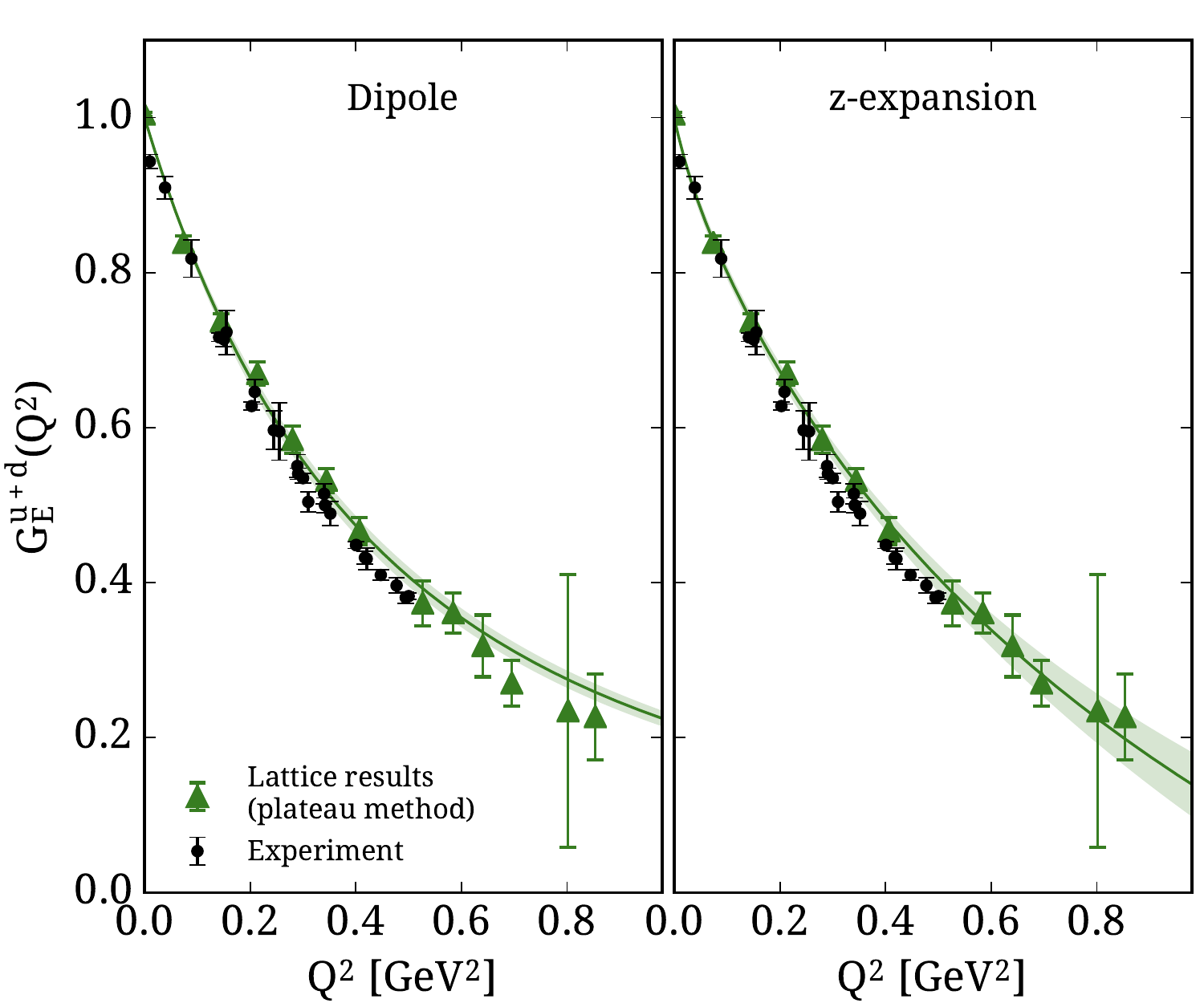}
  \caption{Isoscalar electric Sachs form factor with fits to the
    dipole form (left) and to the z-expansion (right). We show with
    triangles the sum of connected and disconnected contributions,
    with the plateau result for $t_s=18a=1.7$~fm for the connected and
    for $t_s=10a=0.9$~fm for the disconnected. The black points show
    experiment using the same data as for Fig.~\ref{fig:gev ff fit}. }
  \label{fig:ges ff fit}
\end{figure}

\begin{figure}
    \includegraphics[width=\linewidth]{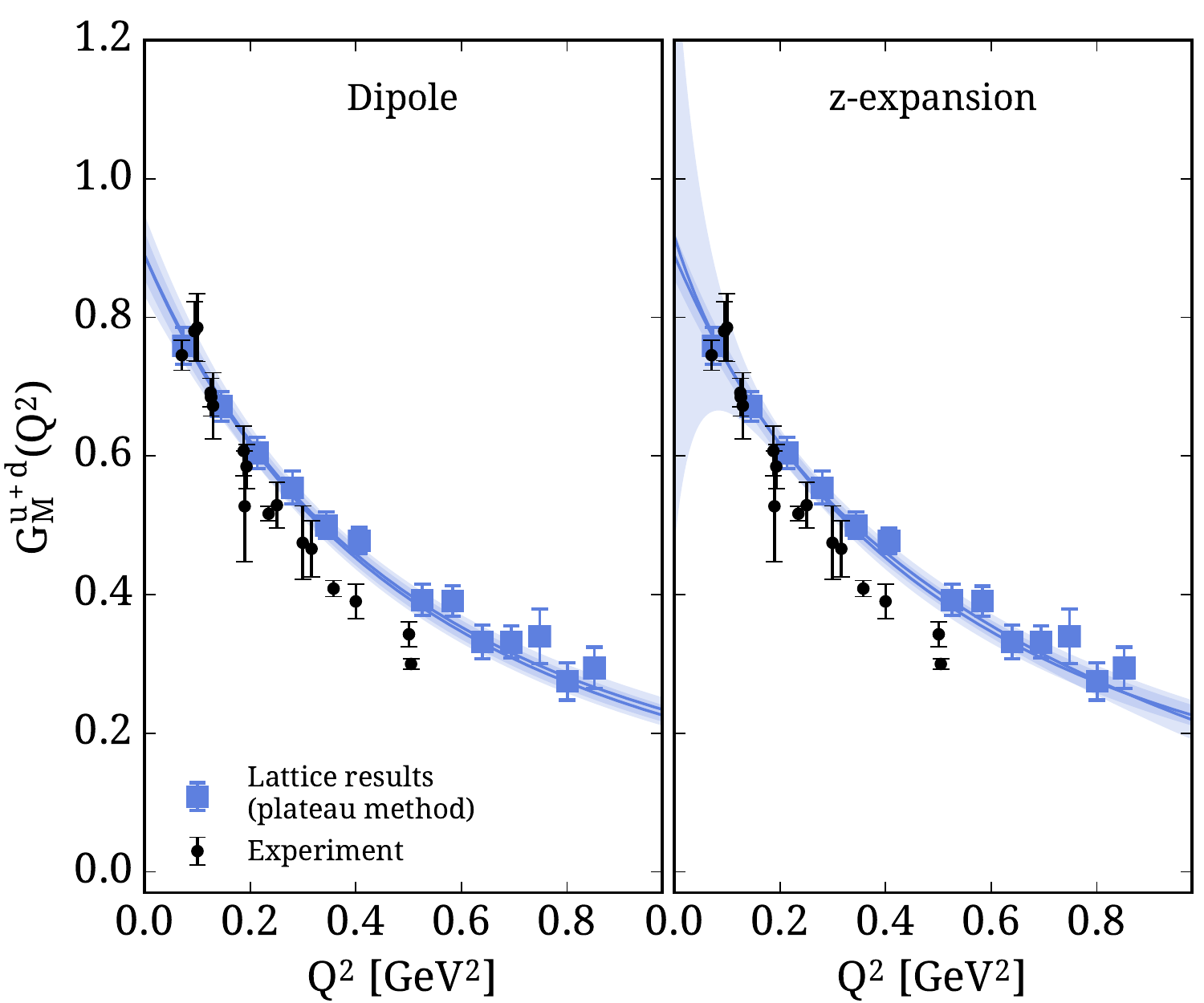}
    \caption{Isoscalar magnetic Sachs form factor with fits to the
      dipole form (left) and to the z-expansion (right). We show with
      triangles the sum of connected and disconnected contributions,
      with the plateau result for $t_s=14a=1.3$~fm for the connected
      and for $t_s=10a=0.9$~fm for the disconnected. The black points
      show experiment using the same data as for Fig.~\ref{fig:gmv ff
        fit}.}
    \label{fig:gms ff fit}
\end{figure}

Fits to the $Q^2$ dependence of $G^{u-d}_E(Q^2)$ are shown in
Fig.~\ref{fig:gev ff fit} using the values extracted from the plateau
at $t_s=18a=1.7$~fm. The line and error band are the result of fitting
to either dipole or the z-expansion for all available $Q^2$ values.
Both dipole and z-expansion form describe the lattice QCD results
well. In this plot we also show results from experiment, using data
for $G_E^p(Q^2)$ obtained from Ref.~\cite{Bernauer:2013tpr} and data
for $G_E^n(Q^2)$ from Refs.~\cite{Golak:2000nt, Becker:1999tw,
  Eden:1994ji, Meyerhoff:1994ev, Passchier:1999cj, Warren:2003ma,
  Zhu:2001md, Plaster:2005cx, Madey:2003av, Rohe:1999sh,
  Bermuth:2003qh, Glazier:2004ny, Herberg:1999ud, Schiavilla:2001qe,
  Ostrick:1999xa}. To subtract the two form factors and obtain the
isovector combination, we linearly interpolate the more accurate
experimental data of $G_E^p(Q^2)$ to the $Q^2$ values for which
$G_E^n(Q^2)$ is available.

For both dipole and z-expansion fit, the resulting curve lies about
one standard deviation above the experimental data. This small
discrepancy may be due to small residual excited state effects, which
would require significant increase of statistics at larger sink-source
separations to identify. Having only performed the calculation using
one ensemble we cannot check directly for finite volume and cut-off
effects. However, in a previous study employing $N_\textrm{f}=2$
twisted mass fermions at heavier than physical pion masses and three
values of the lattice spacing, we found no detectable cut-off effects
in these quantities for a lattice spacing similar to the one used
here~\cite{Alexandrou:2011db}. We have also performed a volume
assessment using the aforementioned heavier mass twisted mass
ensembles with $m_\pi L$ values ranging from 3.27 to 5.28. Namely, we
found no volume dependence within our statistical accuracy between two
ensembles with $m_\pi L=3.27$ and $m_\pi L$=4.28 respectively and
similar pion mass of $m_\pi\simeq$300 MeV. We plan to carry out a high
accuracy analysis of the volume dependence at the physical point on a
lattice size of $64^3\times 128$ keeping the other parameters fixed in
a forthcoming publication.

The same analysis carried out for $G^{u-d}_E(Q^2)$ is also performed
for $G^{u-d}_M(Q^2)$ in Fig.~\ref{fig:gmv ff fit}, where we use the
result from fitting to the plateau at the largest sink-source
separation available, namely $t_s=14a=1.3$~fm. As for the case of
$G_E^{u-d}(Q^2)$, both the dipole Ansatz and z-expansion describe well
the lattice QCD data. The plots show two bands, one when including all
$Q^2$ values, resulting in the smaller error band, and one in which
the first two $Q^2$ values are omitted, resulting in the larger
band. The experimental data shown are obtained using $G_M^p(Q^2)$ from
the same experiment as for $G_E^p(Q^2)$ shown in Fig.~\ref{fig:gev ff fit},
namely Ref.~\cite{Bernauer:2013tpr}, and $G_M^n(Q^2)$ from
Refs.~\cite{Anderson:2006jp, Gao:1994ud, Anklin:1994ae, Anklin:1998ae,
  Kubon:2001rj, Alarcon:2007zza}.

In both dipole and z-expansion fits of $G_M^{u-d}(Q^2)$ we find that
the $Q^2$ dependence is consistent with experiment after
$Q^2\simeq0.2$~GeV$^2$. We suspect that the deviation at the two
smallest $Q^2$ values is due to finite volume effects. As already
mentioned, we plan to further investigate this using an ensemble of
$N_\textrm{f}=2$ twisted mass fermions on a larger volume of
$64^3\times 128$. As can be seen, discarding the two lowest $Q^2$
values results in a larger error for $G^{u-d}_M(0)$, in particular in
the case of the z-expansion.

We show the momentum dependence of the disconnected contribution to
$G_M^{u+d}(Q^2)$ in Fig.~\ref{fig:gms disc ff fit}. The large errors
do not permit as thorough analysis as for the connected
contribution. Since the disconnected isoscalar contributions do not
follow a dipole form, and in the absence of any theoretically
motivated form for the disconnected contributions, we use a
z-expansion fit with $k_{\rm max}=2$, fixing $a_0=0$ for
$G_E^{u+d,\textrm{disc.}}(Q^2)$ and with $k_{\rm max}=1$, allowing
both $a_0$ and $a_1$ to vary. For the case of
$G_E^{u+d,\textrm{disc.}}(Q^2)$ we find results consistent with
zero. For the magnetic case, the disconnected contribution decreases
the form factor by at most 3\% at $Q^2=0$.

We add connected and disconnected contributions to obtain the
isoscalar contributions shown in Figs.~\ref{fig:ges ff fit}
and~\ref{fig:gms ff fit}. There are small discrepancies between our
lattice data and experiment at larger $Q^2$ values. Whether these are
due to volume effects or other lattice artifacts will be investigated
in a follow-up study.

The slope of the form factors at $Q^2$=0 is related to the isovector
electric and magnetic radius as follows
\begin{equation}
  \frac{\partial}{\partial Q^2}G_i(Q^2)|_{Q^2=0}=-\frac{1}{6}G_i(0)\langle r_{i}^2\rangle,
  \label{eq:ff deriv}
\end{equation}
with $i=E,M$ for the electric and magnetic form factors respectively.
For the z-expansion, this is given by
\begin{equation}
  \langle r_{i}^2\rangle = -\frac{6}{4 t_{\rm cut}} \frac{a^i_1}{a^i_0}
\end{equation} 
  and for the dipole fit
\begin{equation}
  \langle r_{i}^2\rangle = \frac{12}{M_i^2}.
\end{equation}
Furthermore, the nucleon magnetic moment is defined as $\mu=G_M(0)$
and is obtained directly from the fitted parameter in both cases. As
for the form factors, we will denote the isovector radii and magnetic
moment with the $u-d$ superscript and for the isoscalar with $u+d$. We
tabulate our results for the isovector radii and magnetic moment from
both dipole and z-expansion fits in Tables~\ref{table:fit params gev}
and~\ref{table:fit params gmv}, and from fits to the isoscalar form
factors in Tables~\ref{table:fit params ges} and~\ref{table:fit params
  gms}. For the isoscalar results shown in Tables~\ref{table:fit
  params ges} and~\ref{table:fit params gms}, we show two results for
each case, namely the result of fitting only the connected
contribution in the first column of each case and the total
contribution, by combining connected and disconnected, in the second
column.

\begin{table}
  \caption{Results for the isovector electric charge radius of the
    nucleon ($\langle r_E^2\rangle^{u-d}$) from fits to
    $G^{u-d}_E(Q^2)$. In the first column we show $t_s$ for the
    plateau method and the $t_s$ fit range for the summation and
    two-state fit methods.}
  \label{table:fit params gev}
  \begin{tabular}{cr@{.}lcr@{.}lc}
    \hline\hline
    \multirow{2}{*}{$t_s$ [fm]} &  \multicolumn{2}{c}{dipole} & \multicolumn{2}{c}{z-expansion}\\
    & \multicolumn{2}{c}{$\langle r_E^2\rangle^{u-d}$ [fm$^2$]} & $\frac{\chi^2}{\rm d.o.f}$
    & \multicolumn{2}{c}{$\langle r_E^2\rangle^{u-d}$ [fm$^2$]} & $\frac{\chi^2}{\rm d.o.f}$ \\
    \hline
    \multicolumn{7}{c}{Plateau}\\
               0.94  &            0&523(08) & 2.0 &            0&562(19) & 1.2 \\
               1.13  &            0&562(14) & 1.9 &            0&677(37) & 0.7 \\
               1.31  &            0&580(26) & 1.2 &            0&718(75) & 0.7 \\
               1.50  &            0&666(33) & 0.9 &           0&61(10) & 0.3 \\
               1.69  &            0&653(48) & 0.6 &           0&52(14) & 0.2 \\
    \hline
    \multicolumn{7}{c}{Summation}\\
            0.9-1.7  &            0&744(55) & 0.3 &           0&79(14) & 0.2 \\
    \hline
    \multicolumn{7}{c}{Two-state}\\
            1.1-1.7  &            0&623(33) & 1.0 &           0&56(10) & 0.8 \\
    \hline\hline
  \end{tabular}
\end{table}

\begin{table}
  \caption{Results for the isovector magnetic charge radius of the
    nucleon ($\langle r_M^2\rangle^{u-d}$) and the isovector magnetic
    moment $G_M(0) = \mu^{u-d}$ from fits to $G^{u-d}_M(Q^2)$.
    In the first column we show $t_s$ for the plateau method and the
    $t_s$ fit range for the summation and two-state fit methods. The
    two smallest $Q^2$ values are omitted from the fit.}
  \label{table:fit params gmv}

  \begin{tabular}{cr@{.}lcr@{.}lc}
    \hline\hline
    \multirow{2}{*}{$t_s$ [fm]} &  \multicolumn{2}{c}{dipole} & \multicolumn{2}{c}{z-expansion}\\
    & \multicolumn{2}{c}{$\langle r_M^2\rangle^{u-d}$ [fm$^2$]} & $\frac{\chi^2}{\rm d.o.f}$
    & \multicolumn{2}{c}{$\langle r_M^2\rangle^{u-d}$ [fm$^2$]} & $\frac{\chi^2}{\rm d.o.f}$ \\
    \hline
    \multicolumn{7}{c}{Plateau}\\
               0.94  &            0&404(10) & 0.3 &           0&59(13) & 0.3 \\
               1.13  &            0&434(22) & 0.3 &           0&82(23) & 0.3 \\
               1.31  &            0&536(52) & 0.3 &           0&79(40) & 0.3 \\
    \hline
    \multicolumn{7}{c}{Summation}\\
            0.9-1.3  &           0&68(16) & 0.1 &           1&83(49) & 0.1\\    
    \hline
    \multicolumn{7}{c}{Two-state}\\
            0.9-1.3  &            0&470(31) & 0.3 &           1&15(25) & 0.3\\    
    \hline\hline
  \end{tabular}
  
  \begin{tabular}{cr@{.}lcr@{.}lc}
    \hline\hline
    \multirow{3}{*}{$t_s$ [fm]} &  \multicolumn{2}{c}{dipole} & \multicolumn{2}{c}{z-expansion}\\
    & \multicolumn{2}{c}{$G^{u-d}_M(0)$} & $\frac{\chi^2}{\rm d.o.f}$
    & \multicolumn{2}{c}{$G^{u-d}_M(0)$} & $\frac{\chi^2}{\rm d.o.f}$ \\
    \hline
    \multicolumn{7}{c}{Plateau}\\
               0.94  &            3&548(52) & 0.3 &          3&85(16) & 0.3 \\
               1.13  &            3&595(90) & 0.3 &          4&13(31) & 0.3 \\
               1.31  &            4&02(21) & 0.3 &           4&31(57) & 0.3 \\
    \hline
    \multicolumn{7}{c}{Summation}\\
            0.9-1.3  &           4&32(57) & 0.1 &         6&35(1.35) & 0.1\\
    \hline
    \multicolumn{7}{c}{Two-state}\\
            0.9-1.3  &           3&74(14) & 0.3 &           4&71(42) & 0.3\\
    \hline\hline
  \end{tabular}
\end{table}

\begin{table}
  \caption{Results for the isoscalar electric charge radius of the
    nucleon ($\langle r_E^2\rangle^{u+d}$). In the first column we
    show $t_s$ for the plateau method and the $t_s$ fit range for the
    summation and two-state fit methods. For each $t_s$ and for each
    fit Ansatz, we give the result from fitting to the connected
    contribution in the first column and to the total contribution of
    connected plus disconnected in the second column.}
  \label{table:fit params ges}
  \begin{tabular}{cr@{.}lr@{.}lcr@{.}lr@{.}lc}
    \hline\hline
    \multirow{3}{*}{$t_s$ [fm]} &  \multicolumn{5}{c}{dipole} & \multicolumn{5}{c}{z-expansion}\\
    & \multicolumn{4}{c}{$\langle r_E^2\rangle^{u+d}$ [fm$^2$]} & $\frac{\chi^2}{\rm d.o.f}$
    & \multicolumn{4}{c}{$\langle r_E^2\rangle^{u+d}$ [fm$^2$]} & $\frac{\chi^2}{\rm d.o.f}$ \\
    & \multicolumn{2}{c}{Connected}& \multicolumn{2}{c}{Total}& &\multicolumn{2}{c}{Connected}& \multicolumn{2}{c}{Total}& \\
    \hline
    \multicolumn{11}{c}{Plateau}\\
               0.94  &             0&440(3) &             0&449(49) & 4.5 &             0&418(9) & 0&427(49) & 0.9 \\
               1.13  &             0&469(6) &             0&478(49) & 1.9 &            0&464(17) & 0&474(52) & 0.7\\
               1.31  &            0&494(12) &             0&503(50) & 0.9 &            0&485(34) & 0&495(59) & 0.5\\
               1.50  &            0&502(14) &             0&512(50) & 0.3 &            0&494(41) & 0&503(63) & 0.4\\
               1.69  &            0&527(22) &             0&537(53) & 0.9 &            0&493(60) & 0&503(77) & 0.8\\
               \hline
    \multicolumn{11}{c}{Summation}\\
            0.9-1.7  &            0&565(20) &             0&576(53) & 0.9 &            0&555(54) & 0&564(72) & 0.6\\
    \hline
    \multicolumn{11}{c}{Two-state}\\
            1.1-1.7  &            0&490(16) &             0&499(51) & 0.5 &            0&453(77) & 0&462(91) & 0.7\\
    \hline\hline
  \end{tabular}
\end{table}

\begin{table}
  \caption{Results for the isoscalar magnetic charge radius of the
    nucleon ($\langle r_M^2\rangle^{u+d}$) and the isoscalar magnetic
    moment $G^{u+d}_M(0)$. The notation is as in Table~\ref{table:fit
      params ges}.}
  \label{table:fit params gms}
  \begin{tabular}{cr@{.}lr@{.}lcr@{.}lr@{.}lc}
    \hline\hline
    \multirow{3}{*}{$t_s$ [fm]} &  \multicolumn{5}{c}{dipole} & \multicolumn{5}{c}{z-expansion}\\
    & \multicolumn{4}{c}{$\langle r_M^2\rangle^{u+d}$ [fm$^2$]} & $\frac{\chi^2}{\rm d.o.f}$
    & \multicolumn{4}{c}{$\langle r_M^2\rangle^{u+d}$ [fm$^2$]} & $\frac{\chi^2}{\rm d.o.f}$ \\
    & \multicolumn{2}{c}{Connected}& \multicolumn{2}{c}{Total}& &\multicolumn{2}{c}{Connected}& \multicolumn{2}{c}{Total}& \\
    \hline
    \multicolumn{11}{c}{Plateau}\\
               0.94  &            0&392(13) &             0&302(34) & 0.2 &           0&41(19) & 0&32(20) & 0.2\\
               1.13  &            0&419(29) &             0&329(47) & 0.1 &           0&84(28) & 0&78(32) & 0.1\\
               1.31  &            0&476(59) &             0&394(82) & 0.4 &           0&4(1.0) & 0&4(1.1) & 0.5\\
               \hline
    \multicolumn{11}{c}{Summation}\\    
            0.9-1.3  &           0&50(18) &            0&42(24) & 0.2 &             1&94(92) & 2&0(1.3) & 0.2\\
    \hline
    \multicolumn{11}{c}{Two-state}\\
            0.9-1.3  &            0&439(44) &             0&353(65) & 0.2 &           0&89(47) & 0&83(52) & 0.2\\
            \hline\hline
  \end{tabular}

  \begin{tabular}{cr@{.}lr@{.}lcr@{.}lr@{.}lc}
    \hline\hline
    \multirow{3}{*}{$t_s$ [fm]} &  \multicolumn{5}{c}{dipole} & \multicolumn{5}{c}{z-expansion}\\
    & \multicolumn{4}{c}{$G^{u+d}_M(0)$} & $\frac{\chi^2}{\rm d.o.f}$
    & \multicolumn{4}{c}{$G^{u+d}_M(0)$} & $\frac{\chi^2}{\rm d.o.f}$ \\
    & \multicolumn{2}{c}{Connected}& \multicolumn{2}{c}{Total}& &\multicolumn{2}{c}{Connected}& \multicolumn{2}{c}{Total}& \\
    \hline
    \multicolumn{11}{c}{Plateau}\\
               0.94  &          0&838(16) &           0&808(18) & 0.2 &         0&867(50) & 0&837(50) & 0.2\\
               1.13  &          0&841(29) &           0&811(30) & 0.1 &         0&981(90) & 0&951(90) & 0.1\\
               1.31  &          0&900(59) &           0&870(60) & 0.4 &         0&90(19)  & 0&87(19)  & 0.5\\
               \hline                                                              
    \multicolumn{11}{c}{Summation}\\                                     
            0.9-1.3  &          0&88(16)  &           0&85(16)  & 0.2 &         1&51(45) & 1&48(45) & 0.2\\
            \hline                                                              
            \multicolumn{11}{c}{Two-state}\\
            0.9-1.3  &          0&861(47) &           0&831(48) & 0.2 &           1&01(14) & 0&98(14) & 0.2\\
    \hline\hline
  \end{tabular}
\end{table}

For our final result for the isovector electric charge radius, we use
the central value and statistical error of the result from the plateau
method at $t_s=18a=1.7$~fm using a dipole fit to all $Q^2$ values.  We
also include a systematic error from the difference of the central
values when comparing with the two-state fit method to account for
excited states effects. Similarly, for the magnetic radius and moment,
we take the result from the dipole fits to our largest sink-source
separation, which for this case is $t_s=14a=1.31$~fm and as in the
case of the electric charge radius, we take the difference with the
two-state fit method as an additional systematic error. In this case,
the values at the two lowest momenta are not included in the fit. Our
final values for the isovector radii and isovector nucleon magnetic
moment are:
\begin{align}
  \langle r_E^2\rangle^{u-d} &= 0.653(48)(30)~\textrm{fm}^2, \nonumber\\
  \langle r_M^2\rangle^{u-d} &= 0.536(52)(66)~\textrm{fm}^2,\,\textrm{and} \nonumber\\
  \mu^{u-d}                  &= 4.02(21)(28),
\end{align}
where the first error is statistical and the second error is a
systematic obtained when comparing the plateau method to the two-state
fit method as a measure of excited state effects. For the isoscalar
radii and moment we follow a similar analysis after adding the
disconnected contribution from the plateau method for
$t_s=10a=0.9$~fm. We obtain
\begin{align}
  \langle r_E^2\rangle^{u+d} &= 0.537(53)(38)~\textrm{fm}^2, \nonumber\\
  \langle r_M^2\rangle^{u+d} &= 0.394(82)(42)~\textrm{fm}^2,\,\textrm{and} \nonumber\\
  \mu^{u+d}                  &= 0.870(60)(39).
\end{align}

\subsection{Proton and neutron form factors}
\begin{figure}
    \includegraphics[width=\linewidth]{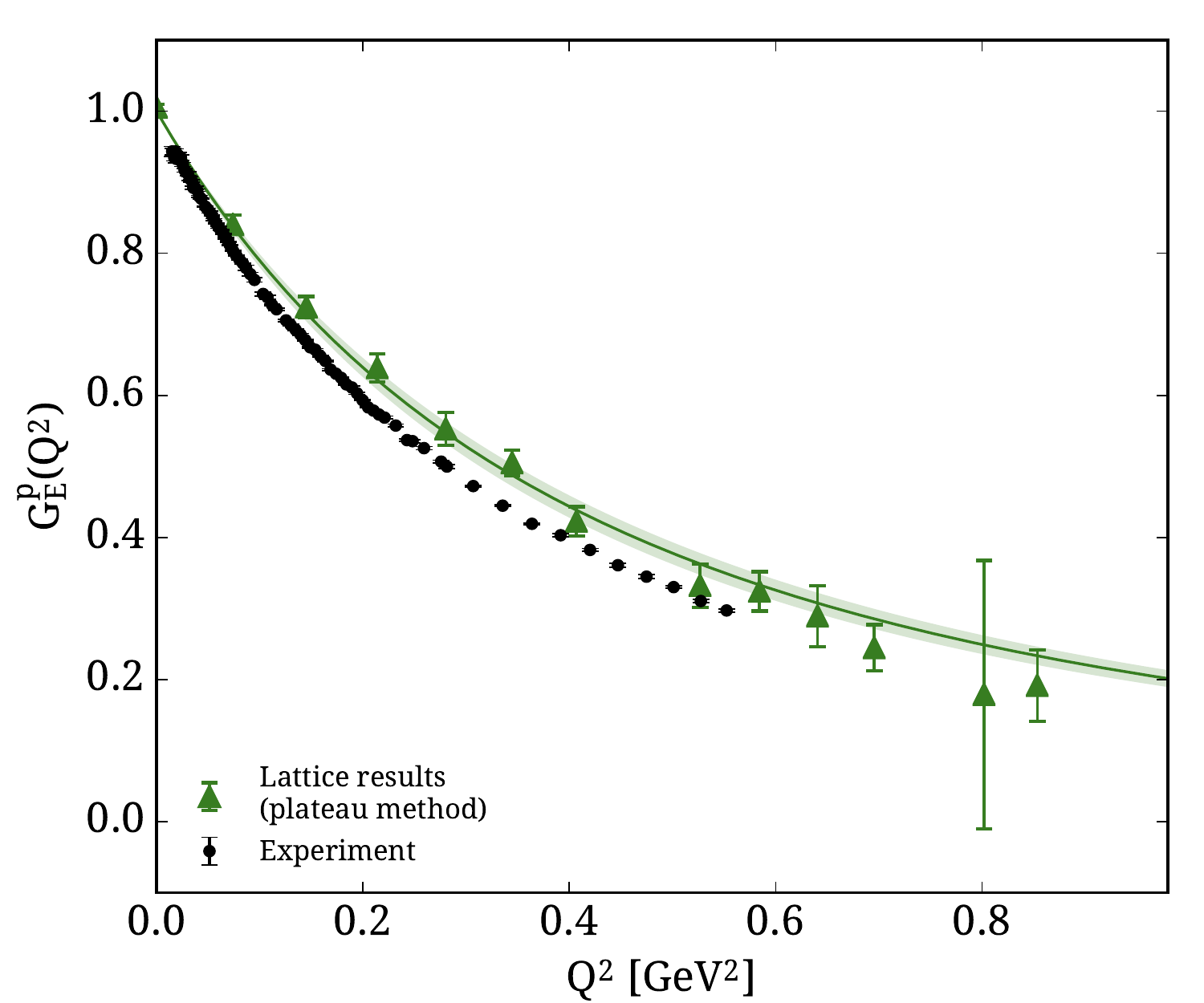}
    \caption{Proton electric Sachs form factor as a function of the
      momentum transfer. We show with triangles the sum of connected
      and disconnected contributions, with the plateau result for
      $t_s=18a=1.7$~fm for the connected and for $t_s=10a=0.9$~fm for
      the disconnected. The band is a fit to the dipole form. The
      black points show experimental data from
      Ref.~\cite{Bernauer:2013tpr}.}
    \label{fig:gep ff fit}
\end{figure}

\begin{figure}
    \includegraphics[width=\linewidth]{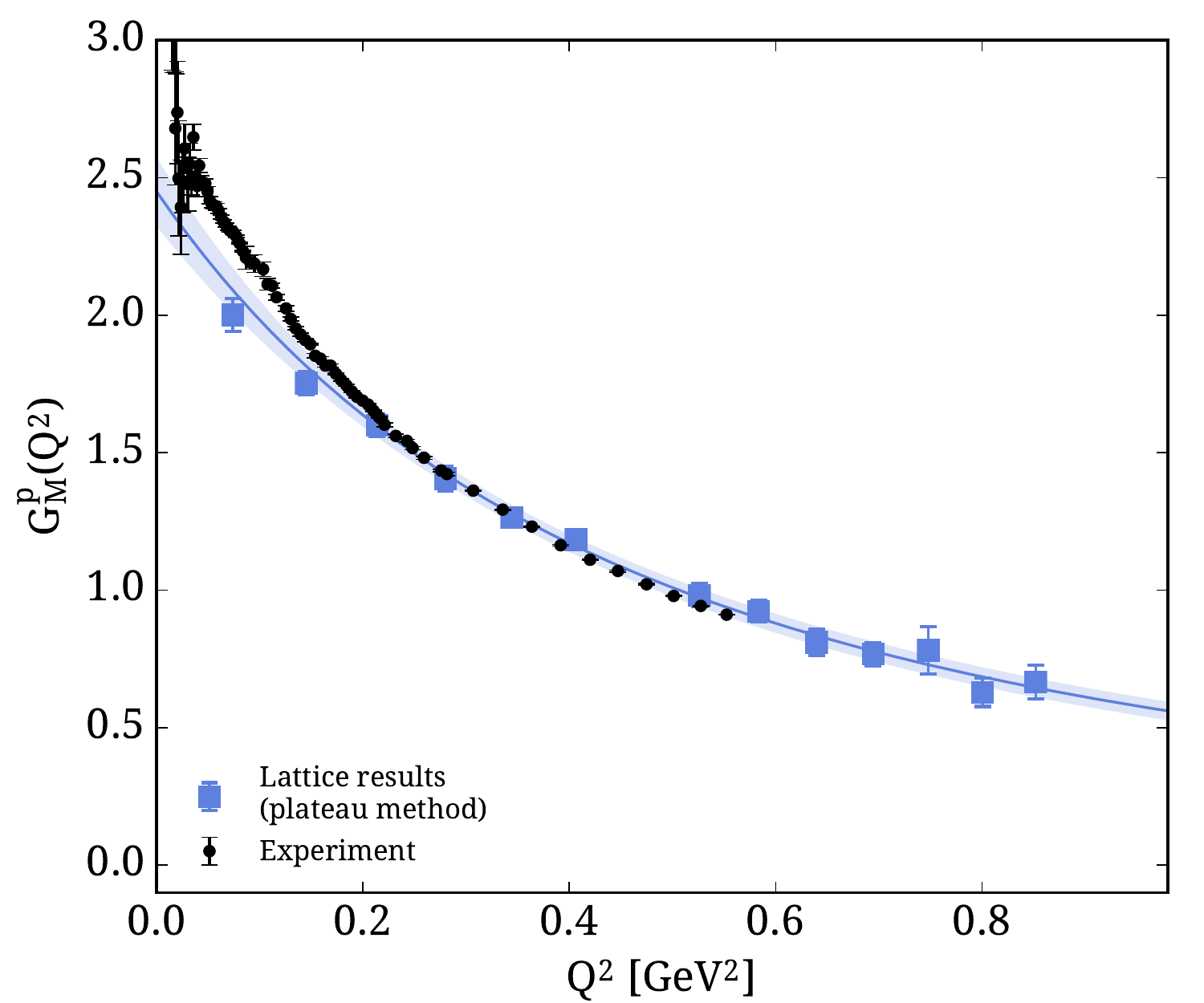}
    \caption{Proton magnetic Sachs form factor as a function of the
      momentum transfer. We show with squares the sum of connected and
      disconnected contributions, with the plateau result for
      $t_s=14a=1.3$~fm for the connected and for $t_s=10a=0.9$~fm for
      the disconnected. The band is a fit to the dipole form. The
      black points show experimental data from
      Ref.~\cite{Bernauer:2013tpr}.}
    \label{fig:gmp ff fit}
\end{figure}

Having the isovector and isoscalar contributions to the form factors,
we can obtain the proton ($G^p(Q^2)$) and neutron ($G^n(Q^2)$) form
factors via linear combinations taken from Eqs.~(\ref{eq:isovector})
and~(\ref{eq:isoscalar}) assuming isospin symmetry between up and down
quarks and proton and neutron. Namely, we have:
\begin{align}
  G^p(Q^2) =& \frac{1}{2}[G^{u+d}(Q^2) + G^{u-d}(Q^2)] \nonumber\\
  G^n(Q^2) =& \frac{1}{2}[G^{u+d}(Q^2) - G^{u-d}(Q^2)]
\end{align}
where $G^p(Q^2)$ ($G^n(Q^2)$) is either the electric or magnetic
proton (neutron) form factor. In Figs.~\ref{fig:gep ff fit}
and~\ref{fig:gmp ff fit} we show results for the proton electric and
magnetic Sachs form factors respectively. As for the isoscalar case,
the disconnected contributions have been included. The bands are from
fits to the dipole form of Eq.~(\ref{eq:dipole fit}). In these plots
we compare to experimental results from the A1
collaboration~\cite{Bernauer:2013tpr}. We observe a similar behavior
when comparing to experiment as for the case of the isovector form
factors. Namely, the dipole fit to the lattice data has a smaller
slope for small values of $Q^2$ as compared to experiment, while
$G_M^p(Q^2)$ reproduces the experimental momentum dependence for
$Q^2>0.2$~GeV$^2$.

\begin{figure}
    \includegraphics[width=\linewidth]{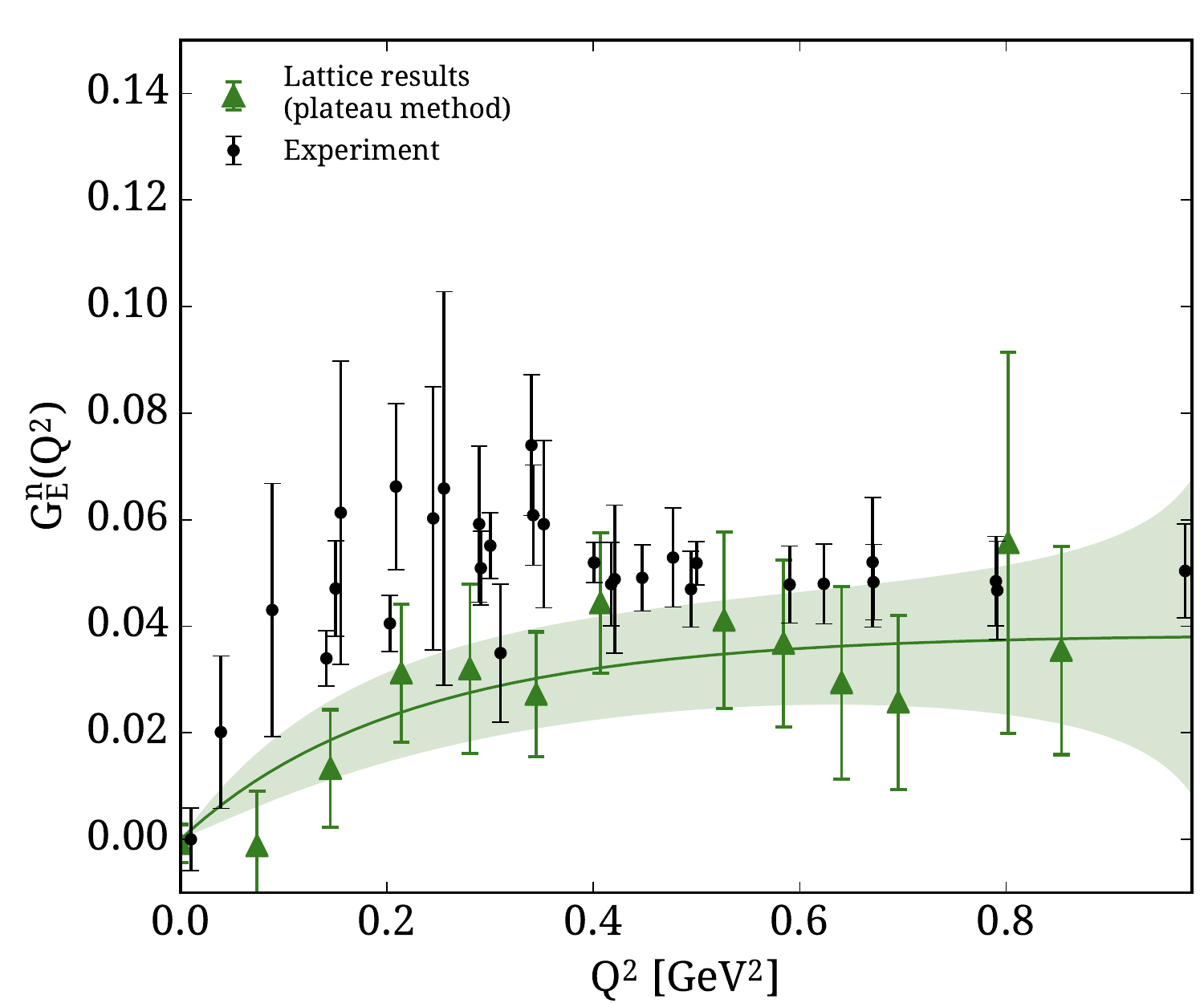}
    \caption{Neutron electric Sachs form factor as a function of the
      momentum transfer. Triangles are from the sum of connected and
      disconnected contributions, with the plateau result for
      $t_s=18a=1.7$~fm for the connected and for $t_s=10a=0.9$~fm for
      the disconnected. The band is a fit to the form of
      Eq.~(\ref{eq:dipole neutron}). Experimental data are shown with
      the black points, obtained from Refs.~\cite{Golak:2000nt,
        Becker:1999tw, Eden:1994ji, Meyerhoff:1994ev,
        Passchier:1999cj, Warren:2003ma, Zhu:2001md, Plaster:2005cx,
        Madey:2003av, Rohe:1999sh, Bermuth:2003qh, Glazier:2004ny,
        Herberg:1999ud, Schiavilla:2001qe, Ostrick:1999xa}.}
      \label{fig:gen ff fit}
\end{figure}

\begin{figure}
    \includegraphics[width=\linewidth]{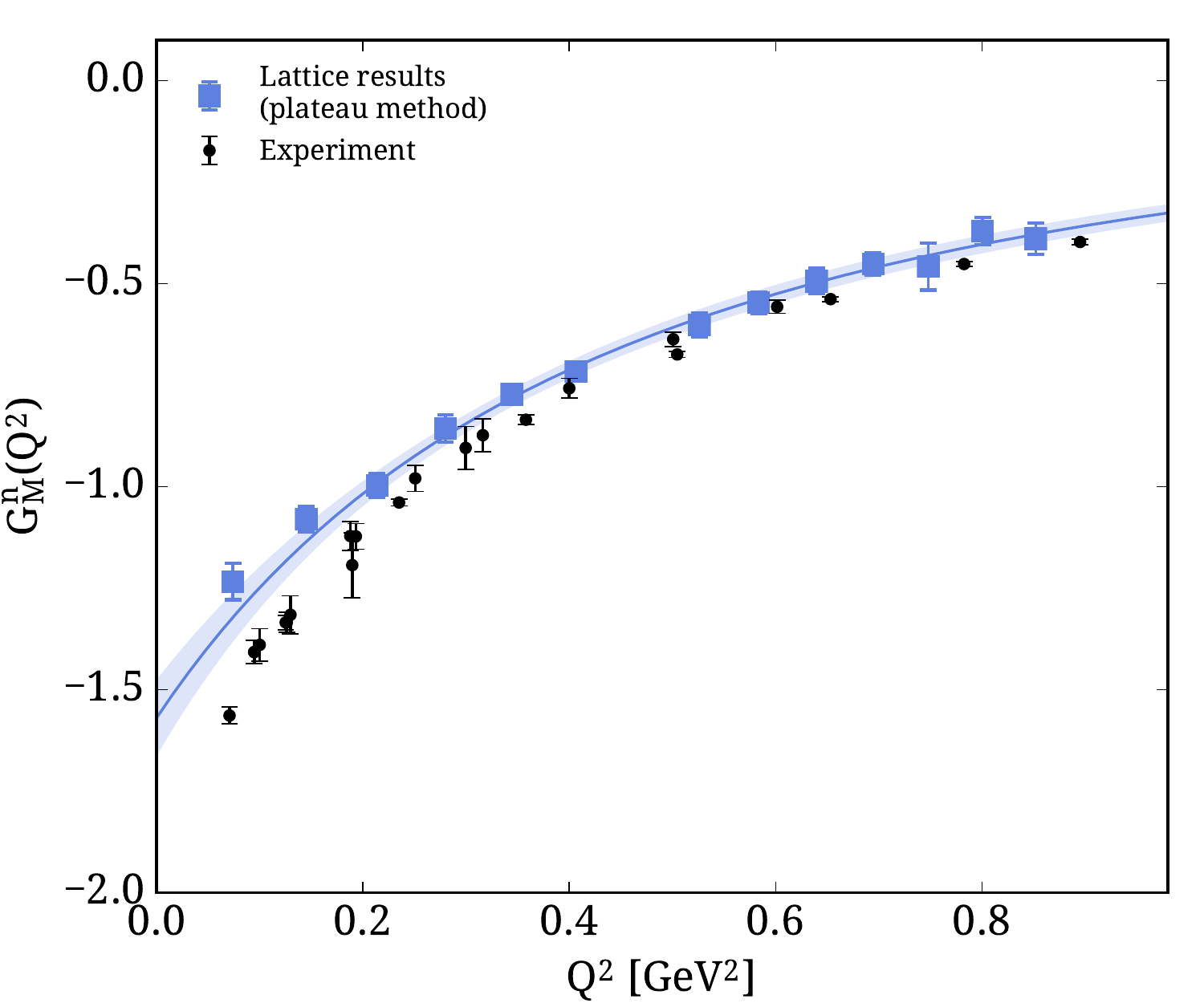}
    \caption{Neutron magnetic Sachs form factor as a function of the
      momentum transfer. We show with squares the sum of connected and
      disconnected contributions, with the plateau result for
      $t_s=14a=1.3$~fm for the connected and for $t_s=10a=0.9$~fm for
      the disconnected. The band is a fit to the dipole form. The
      black points show experimental data from
      Refs.~\cite{Anderson:2006jp, Gao:1994ud, Anklin:1994ae,
        Anklin:1998ae, Kubon:2001rj, Alarcon:2007zza}.}
      \label{fig:gmn ff fit}
\end{figure}

In Figs.~\ref{fig:gen ff fit} and~\ref{fig:gmn ff fit} we show the
same for the neutron form factors. For the neutron electric form
factor we fit to the form~\cite{Kelly:2004hm}:
\begin{equation}
  G^{n}_E(Q^2) = \frac{\tau A}{1 + \tau B}\frac{1}{(1+\frac{Q^2}{\Lambda^2})^2}
  \label{eq:dipole neutron}
\end{equation}
with $\tau=Q^2/(2m_N)^2$ and $\Lambda^2=0.71$~GeV$^2$ and allow $A$
and $B$ to vary. This Ansatz reproduces our data well. We compare to a
collection of experimental data from Refs.~\cite{Golak:2000nt,
  Becker:1999tw, Eden:1994ji, Meyerhoff:1994ev, Passchier:1999cj,
  Warren:2003ma, Zhu:2001md, Plaster:2005cx, Madey:2003av,
  Rohe:1999sh, Bermuth:2003qh, Glazier:2004ny, Herberg:1999ud,
  Schiavilla:2001qe, Ostrick:1999xa}. For $G_M^n(Q^2)$, we agree with
the experimental data for $Q^2>0.2$~GeV$^2$, however we underestimate
the magnetic moment by about 20\%. Experimental data for $G_M^n(Q^2)$
shown in Fig.~\ref{fig:gmn ff fit} are taken from
Refs.~\cite{Anderson:2006jp, Gao:1994ud, Anklin:1994ae, Anklin:1998ae,
  Kubon:2001rj, Alarcon:2007zza}.

We use Eq.~(\ref{eq:ff deriv}) to obtain the radii using the dipole
fits. For the case of $G_E^n(Q^2)$, the neutron electric radius is
obtained via: $\langle r^2_E\rangle^n = -\frac{3A}{2m_N^2}$, where $A$
is the parameter of Eq.~(\ref{eq:dipole neutron}). In all cases we
have combined connected and disconnected. We obtain:
\begin{align}
  \langle r_E^2\rangle^{p} &= 0.589(39)(33)~\textrm{fm}^2, \nonumber\\
  \langle r_M^2\rangle^{p} &= 0.506(51)(42)~\textrm{fm}^2,\,\textrm{and} \nonumber\\
  \mu_{p}                  &= 2.44(13)(14),
\end{align}
for the proton, and:
\begin{align}
  \langle r_E^2\rangle^{n} &= -0.038(34)(6)~\textrm{fm}^2, \nonumber\\
  \langle r_M^2\rangle^{n} &= \phantom{-}0.586(58)(75)~\textrm{fm}^2,\,\textrm{and} \nonumber\\
  \mu_{n}                  &= -1.58(9)(12),
\end{align}
for the neutron, where as in the case of the isoscalar and isovector,
the first error is statistical and the second is a systematic obtained
when comparing the plateau method to the two-state fit method as a
measure of excited state effects.

\section{Comparison with other results}
\label{sec:others}

\subsection{Comparison of isovector and isoscalar form factors}
Recent lattice calculations for the electromagnetic form factors of
the nucleon include an analysis from the Mainz
group~\cite{Capitani:2015sba} using $N_\textrm{f}=2$ clover fermions
down to a pion mass of 193~MeV, results from the PNDME
collaboration~\cite{Bhattacharya:2013ehc} using clover valence
fermions on $N_\textrm{f}=2+1+1$ HISQ sea quarks down to pion mass of
$\sim$220~MeV and $N_\textrm{f}=2+1+1$ results from the ETM
collaboration down to 213~MeV pion
mass~\cite{Alexandrou:2013joa}. Simulations directly at the physical
point have only been possible recently. The LHPC has published results
in Ref.~\cite{Green:2014xba} using $N_\textrm{f}=2+1$ HEX smeared
clover fermions, which include an ensemble with
$m_\pi=$149~MeV. Preliminary results for electromagnetic nucleon form
factors at physical or near physical pion masses have also been
reported by the PNDME collaboration in Ref.~\cite{Jang:2016kch} using
clover valence quarks on HISQ sea quarks at a pion mass of $130$~MeV
and by the RBC/UKQCD collaboration using Domain Wall fermions at
$m_\pi=172$~MeV in Ref.~\cite{Abramczyk:2016ziv}.

\begin{figure}
    \includegraphics[width=\linewidth]{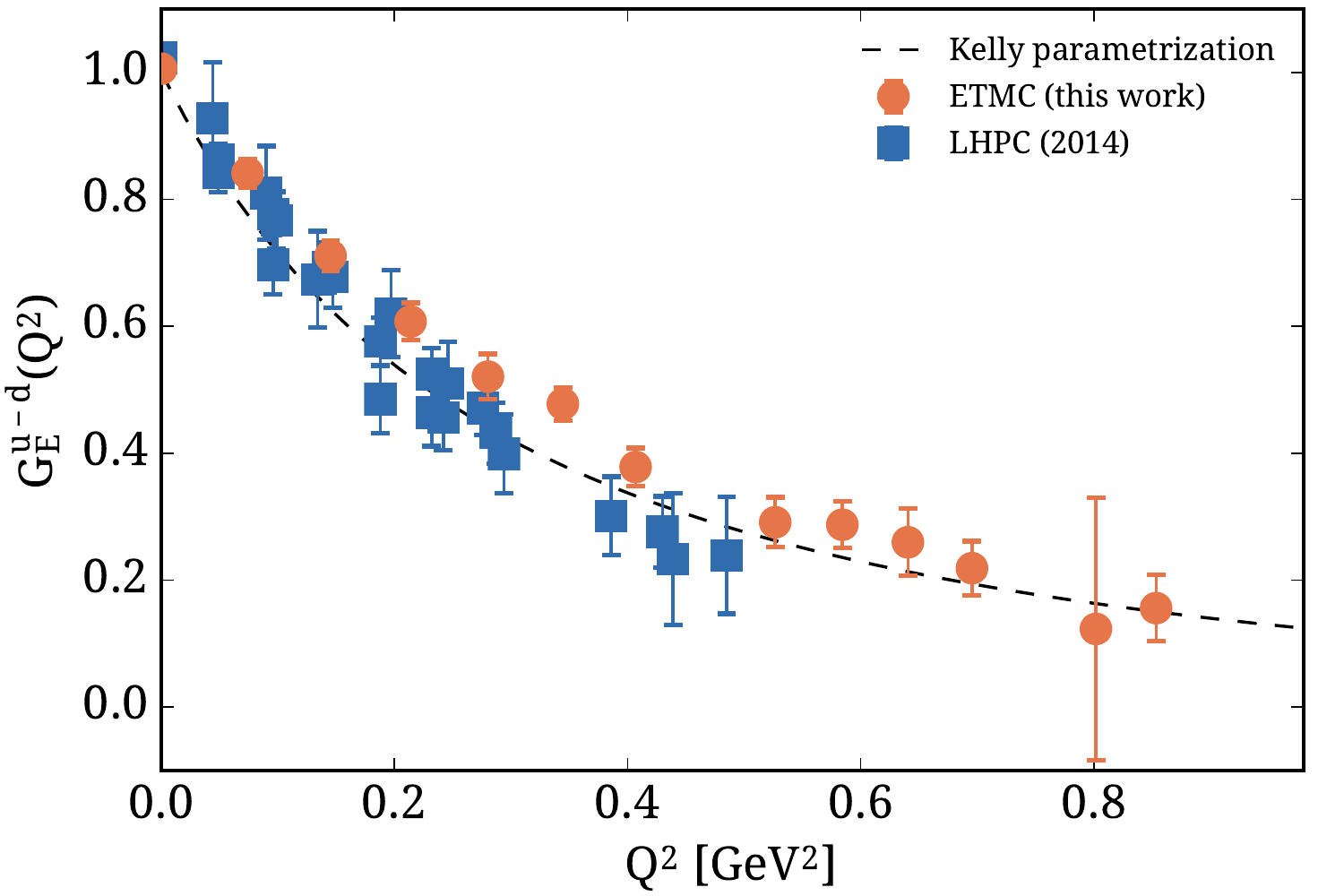}
    \caption{Comparison of $G^{u-d}_E(Q^2)$ between results from this
      work (circles) denoted by ETMC and from the LHPC taken from
      Ref.~\cite{Green:2014xba} (squares). The dashed line shows the
      parameterization of the experimental data.}
    \label{fig:gev comparison}
\end{figure}

In Fig.~\ref{fig:gev comparison} we compare our results for
$G^{u-d}_E(Q^2)$ from the plateau method using $t_s=18a=1.7$~fm to
published results.   We show
results from Ref.~\cite{Green:2014xba} extracted from the
summation method using three sink-source separations from 0.93 to
1.39~fm for their ensemble at the near-physical pion mass of
$m_\pi=$149~MeV.  We note that their statistics of 7752 are about six times
less than ours at the sink-source separation we use in this plot (see
Table~\ref{table:statistics}).

\begin{figure}
    \includegraphics[width=\linewidth]{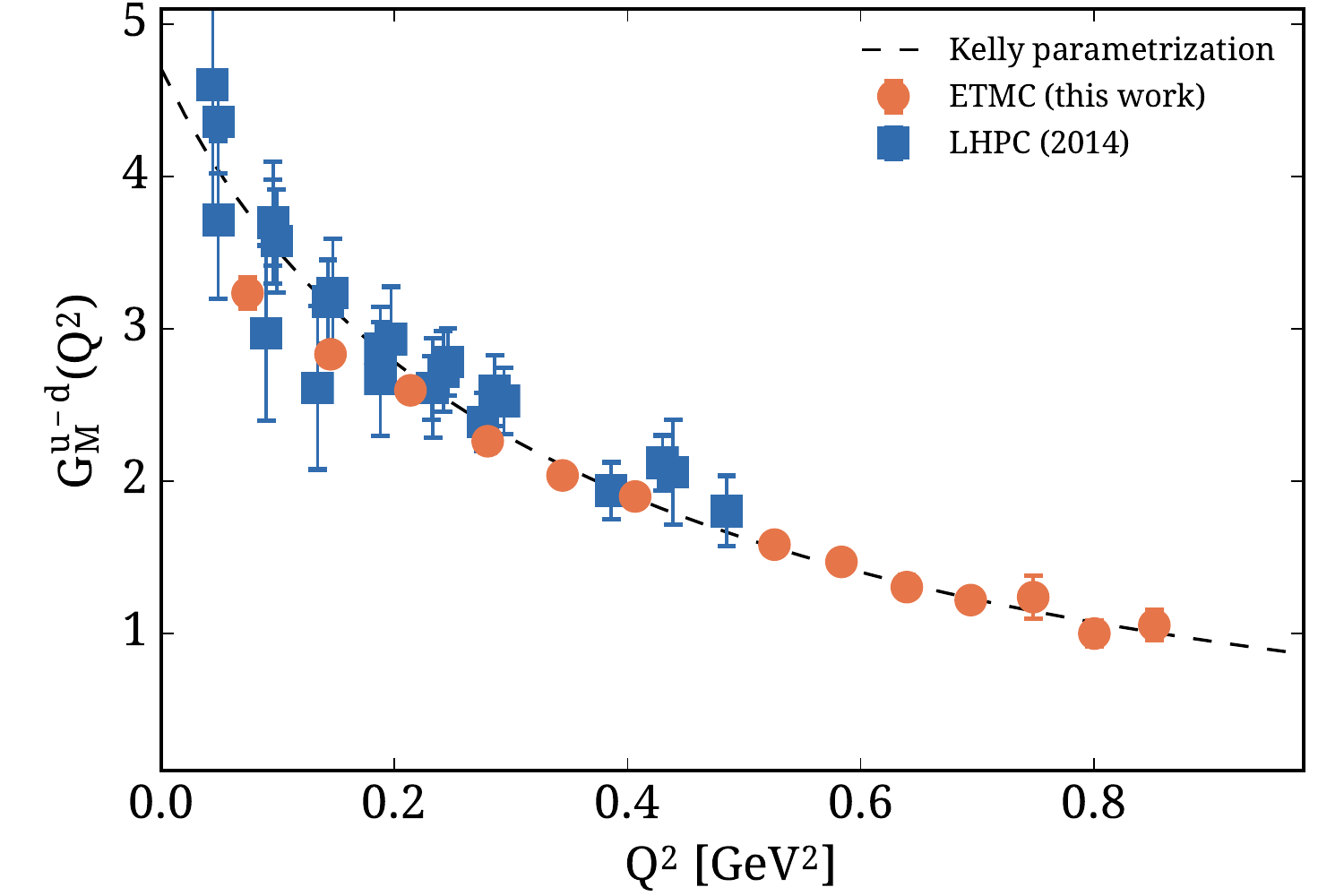}
    \caption{Comparison of $G^{u-d}_M(Q^2)$ between results from this
      work (circles) and Ref.~\cite{Green:2014xba} (squares). The
      dashed line shows the parameterization of the experimental data.}
    \label{fig:gmv comparison}
\end{figure}

In Fig.~\ref{fig:gmv comparison} we plot our results for
$G^{u-d}_M(Q^2)$ from the plateau method using $t_s=14a=1.3$~fm and
compare to those from LHPC. At this sink-source separation the
statistics are similar, namely 7752 for the LHPC data and 9248 for the
results from this work, however their errors are larger, possibly due
to the fact that the summation method is used for their final quoted
results. Within errors, we see consistent results at all $Q^2$ values.

\begin{figure}
    \includegraphics[width=\linewidth]{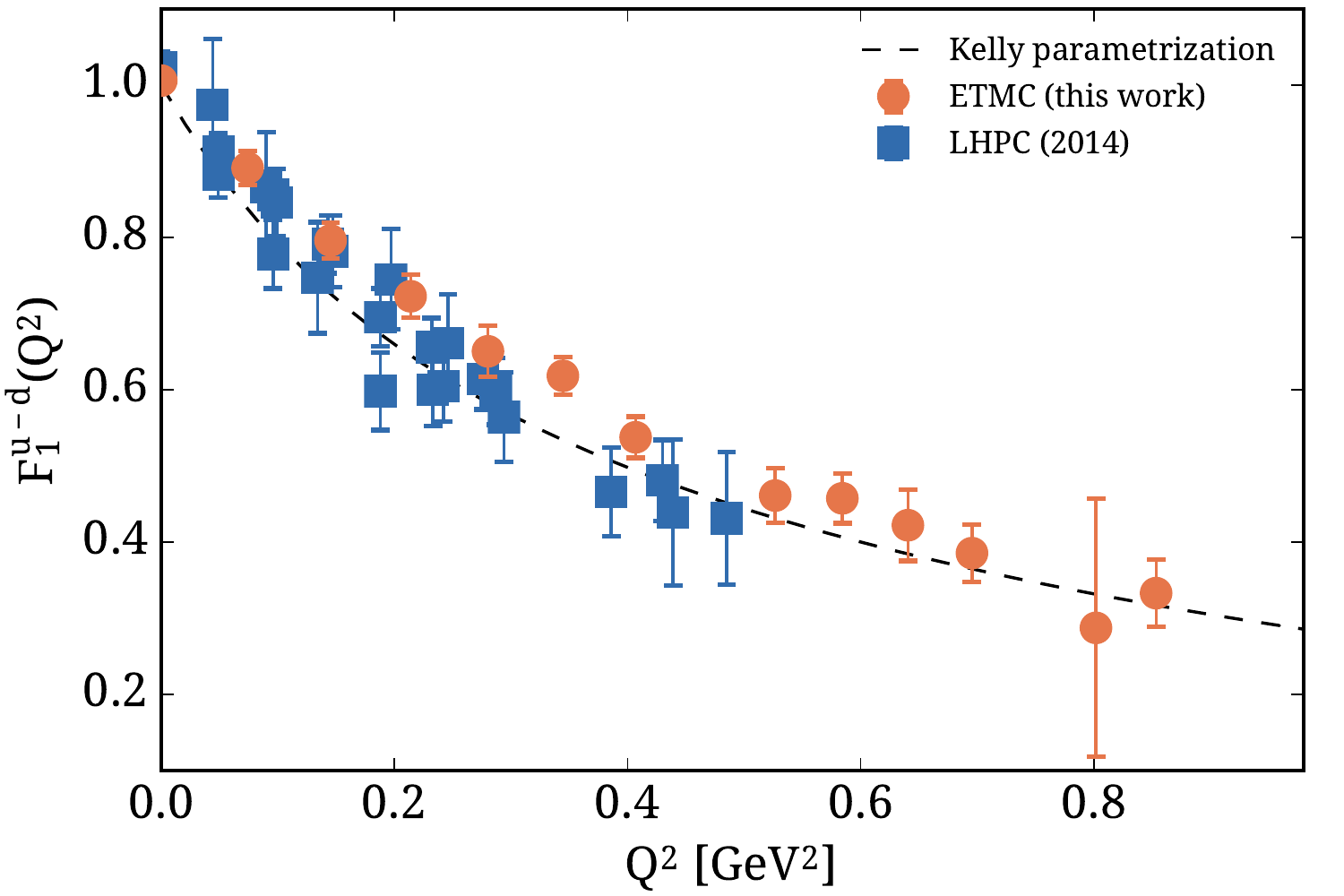}
    \caption{Comparison of $F^{u-d}_1(Q^2)$ between results from this
      work (circles) and Ref.~\cite{Green:2014xba} (squares). The
      dashed line shows the parameterization of the experimental
      data.}
    \label{fig:f1v comparison}
\end{figure}

\begin{figure}
    \includegraphics[width=\linewidth]{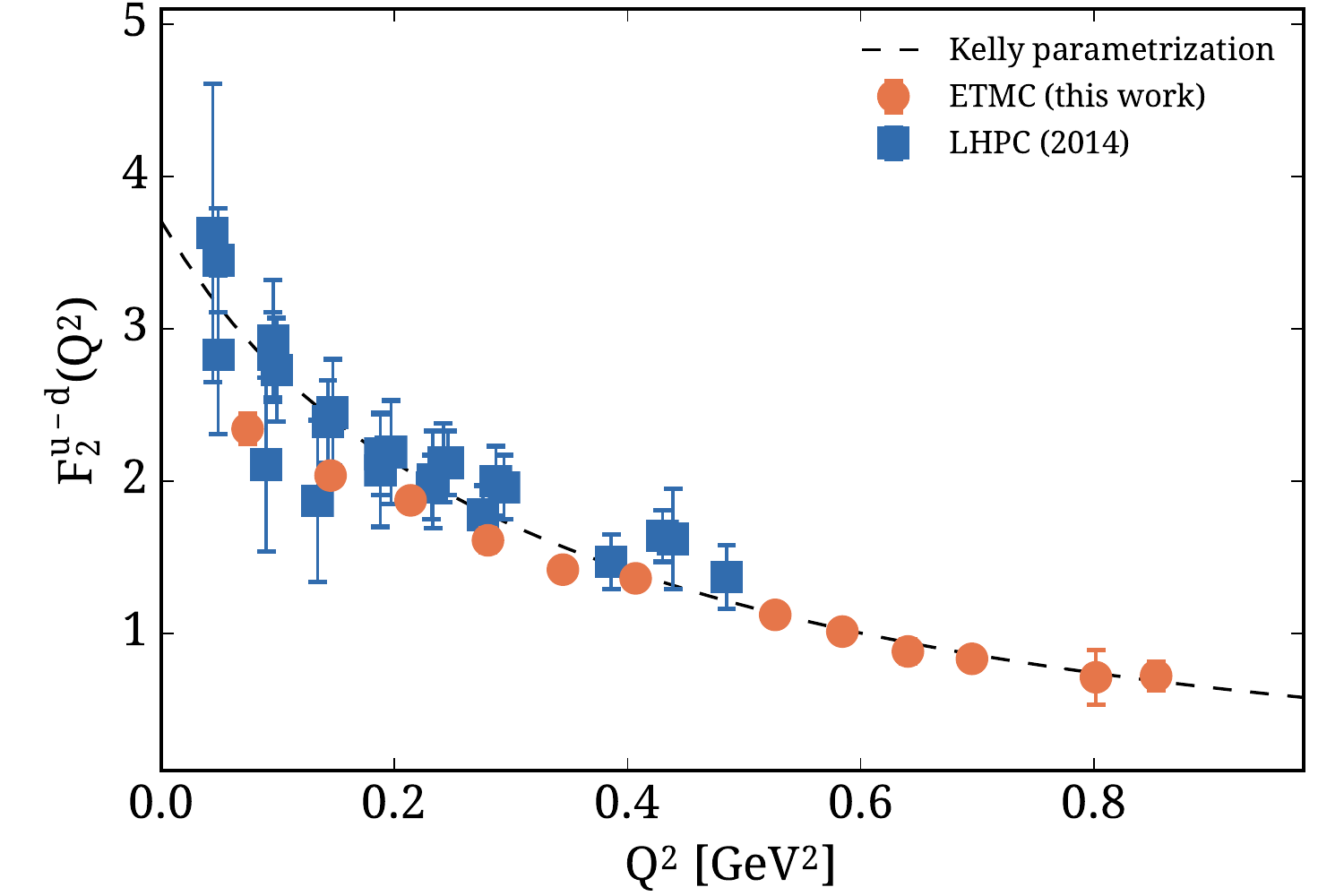}
    \caption{Comparison of $F^{u-d}_2(Q^2)$ between results from this
      work (circles) and Ref.~\cite{Green:2014xba} (squares). The
      dashed line shows the parameterization of the experimental
      data.}
    \label{fig:f2v comparison}
\end{figure}

In Figs.~\ref{fig:f1v comparison} and~\ref{fig:f2v comparison} we
compare our results for the isovector Dirac and Pauli form factors
$F^{u-d}_1(Q^2)$ and $F^{u-d}_2(Q^2)$ with those from
Ref.~\cite{Green:2014xba}. We use Eq.~(\ref{eq:f1f2}) to obtain
$F_1^{u-d}(Q^2)$ and $F^{u-d}_2(Q^2)$ from $G^{u-d}_E(Q^2)$ and
$G^{u-d}_M(Q^2)$ extracted from the plateau method at the same
sink-source separations used in Figs.~\ref{fig:gev comparison}
and~\ref{fig:gmv comparison}. As in the case of $G^{u-d}_E(Q^2)$ and
$G^{u-d}_M(Q^2)$ we see agreement between these two calculations. We
also note that the discrepancy with experiment of $G^{u-d}_M(Q^2)$ at
low $Q^2$ values carries over to $F^{u-d}_2(Q^2)$.

\begin{figure}
    \includegraphics[width=\linewidth]{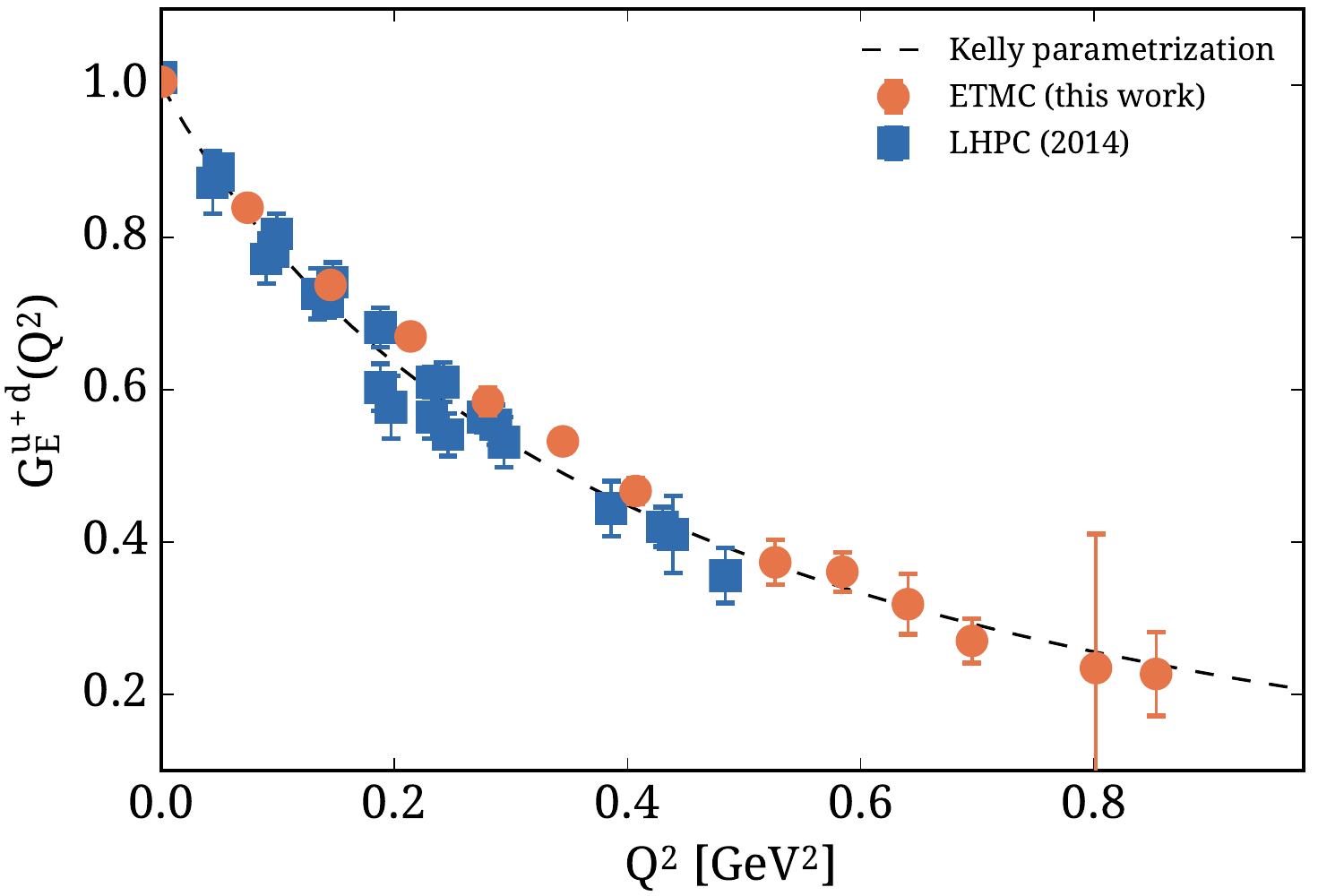}
    \caption{Comparison of $G^{u+d}_E(Q^2)$ between results from this
      work (circles) and Ref.~\cite{Green:2014xba} (squares). The
      dashed line shows the parameterization of the experimental
      data.}
    \label{fig:ges comparison}
\end{figure}

\begin{figure}
    \includegraphics[width=\linewidth]{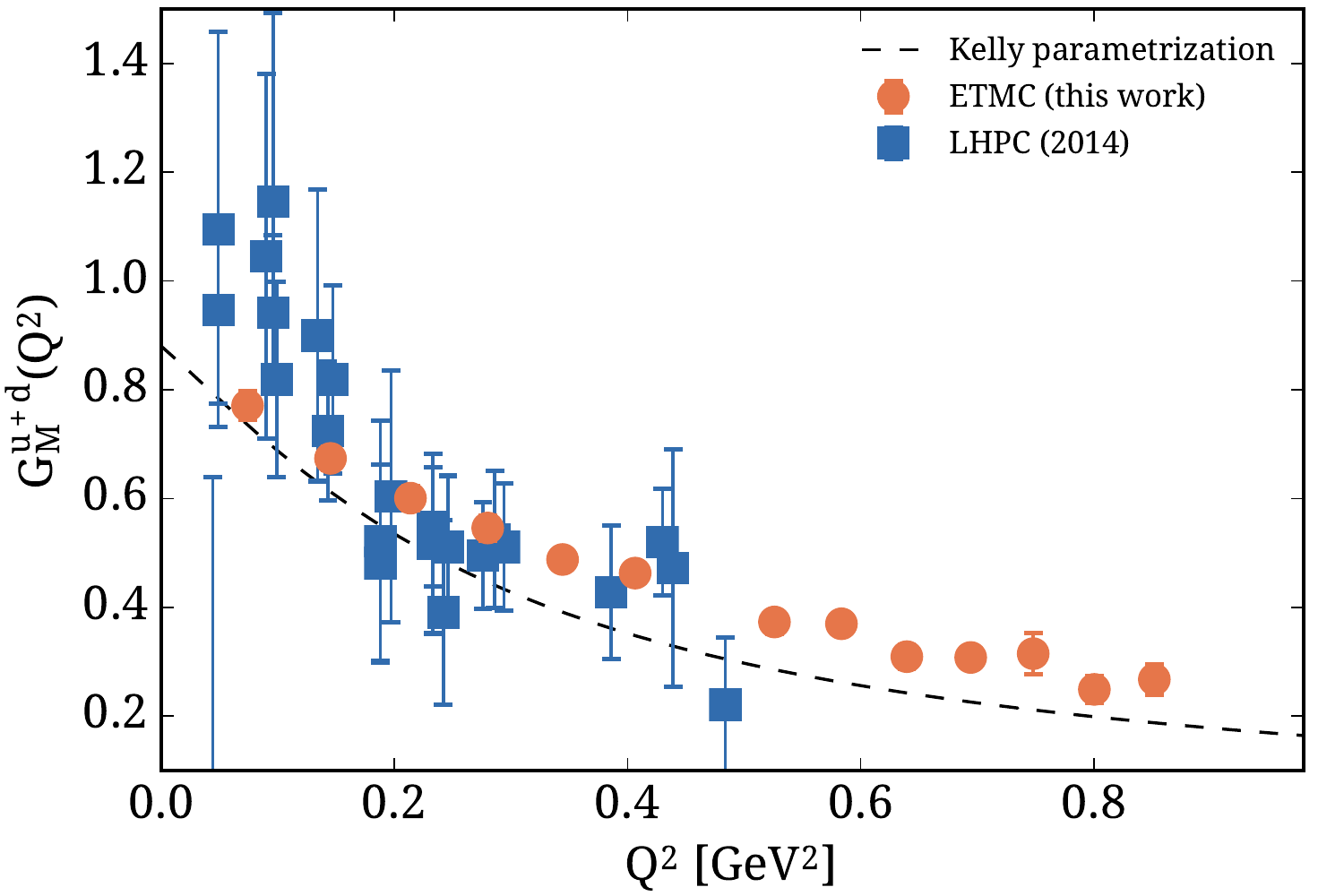}
    \caption{Comparison of $G^{u+d}_M(Q^2)$ between results from this
      work (circles) and Ref.~\cite{Green:2014xba} (squares). The
      dashed line shows the parameterization of the experimental
      data.}
    \label{fig:gms comparison}
\end{figure}

For the isoscalar case, we compare the connected contributions to the
Sachs form factors with Ref.~\cite{Green:2014xba} in
Figs.~\ref{fig:ges comparison} and~\ref{fig:gms comparison}. The
agreement between the two lattice formulations is remarkable given
that the results have not been corrected for finite volume or cut-off
effects. The gauge configurations used by LHPC were carried out using
the same spatial lattice size as ours but with a coarser lattice
spacing yielding $m_\pi L=4.2$ compared to ours of $m_\pi L=3$.
Although the LHPC results for the isovector magnetic form factor at
low $Q^2$ are in agreement with experiment, they carry large
statistical errors that do not allow us to draw any conclusion as to
whether the origin of the discrepancy in our much more accurate data
is due to the smaller $m_\pi L$ value.

For the radii and magnetic moment, we compare our result to recent
published results, which are available for the isovector case, from
Refs.~\cite{Green:2014xba,Bhattacharya:2013ehc,Capitani:2015sba,Alexandrou:2013joa}. We
quote their values obtained before extrapolation to the physical
point, using the smallest pion mass available. In Fig.~\ref{fig:rEv
  comparison} we see that the two results at physical or near-physical
pion mass, namely the result of this work and from LHPC, are within
one standard deviation from the spectroscopic determination of the
charged radius using muonic hydrogen~\cite{Pohl:2010zza}.

\begin{figure}
    \includegraphics[width=\linewidth]{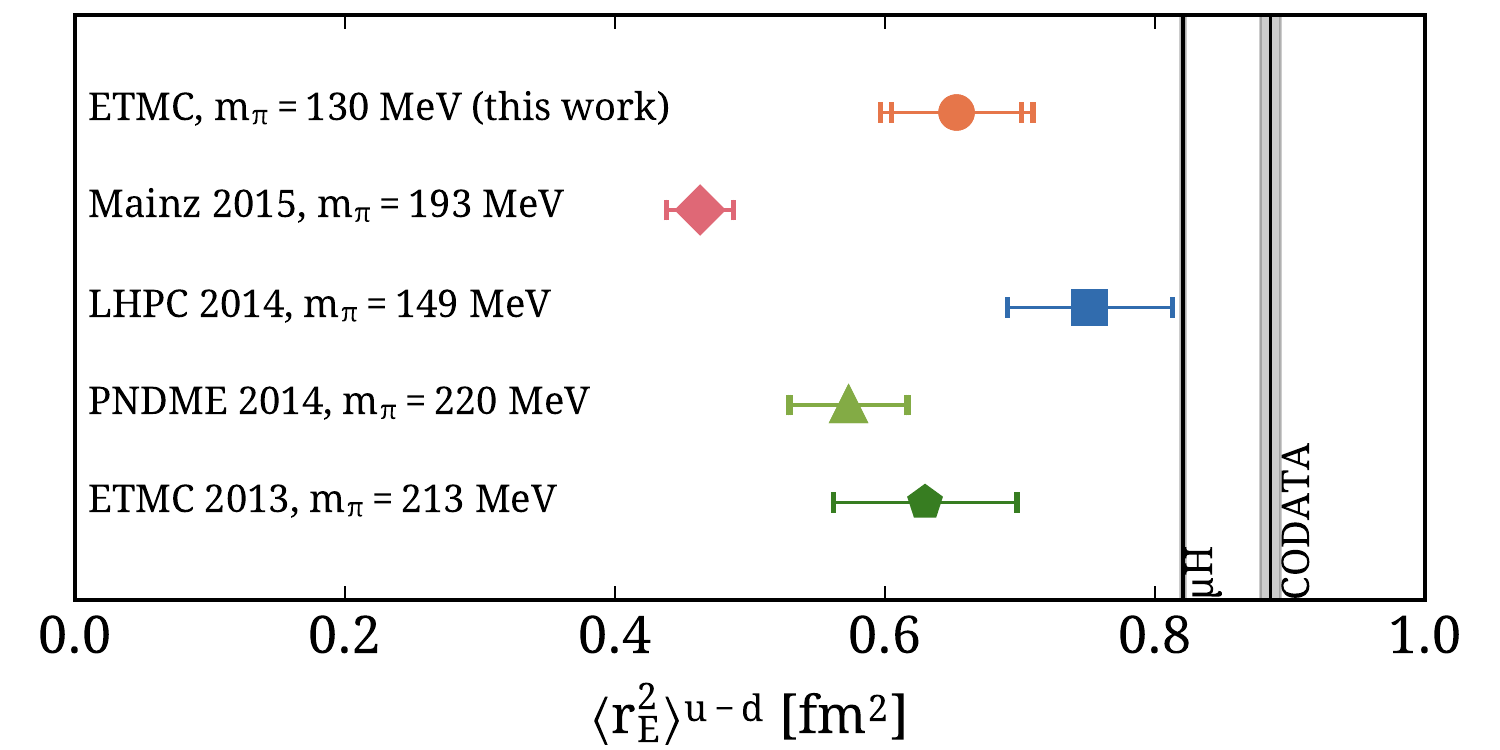}
    \caption{Our result for $\langle r^2_E\rangle^{u-d}$ at
      $m_\pi=$130~MeV (circle) compared to recent lattice results from
      LHPC~\cite{Green:2014xba} at $m_\pi=149$~MeV (square),
      PNDME~\cite{Bhattacharya:2013ehc} at $m_\pi=220$~MeV (triangle),
      the Mainz group~\cite{Capitani:2015sba} at $m_\pi=193$~MeV
      (diamond) and ETMC~\cite{Alexandrou:2013joa} (pentagon). We show
      two error bars when systematic errors are available, with the
      smaller denoting the statistical error and the larger denoting
      the combination of statistical and systematic errors added in
      quadrature. The vertical band denoted with $\mu$H is the
      experimental result using muonic hydrogen from
      Ref.~\cite{Pohl:2010zza} and the band denoted with CODATA is
      from Ref.~\cite{Mohr:2015ccw}.}
    \label{fig:rEv comparison}
\end{figure}
\begin{figure}
    \includegraphics[width=\linewidth]{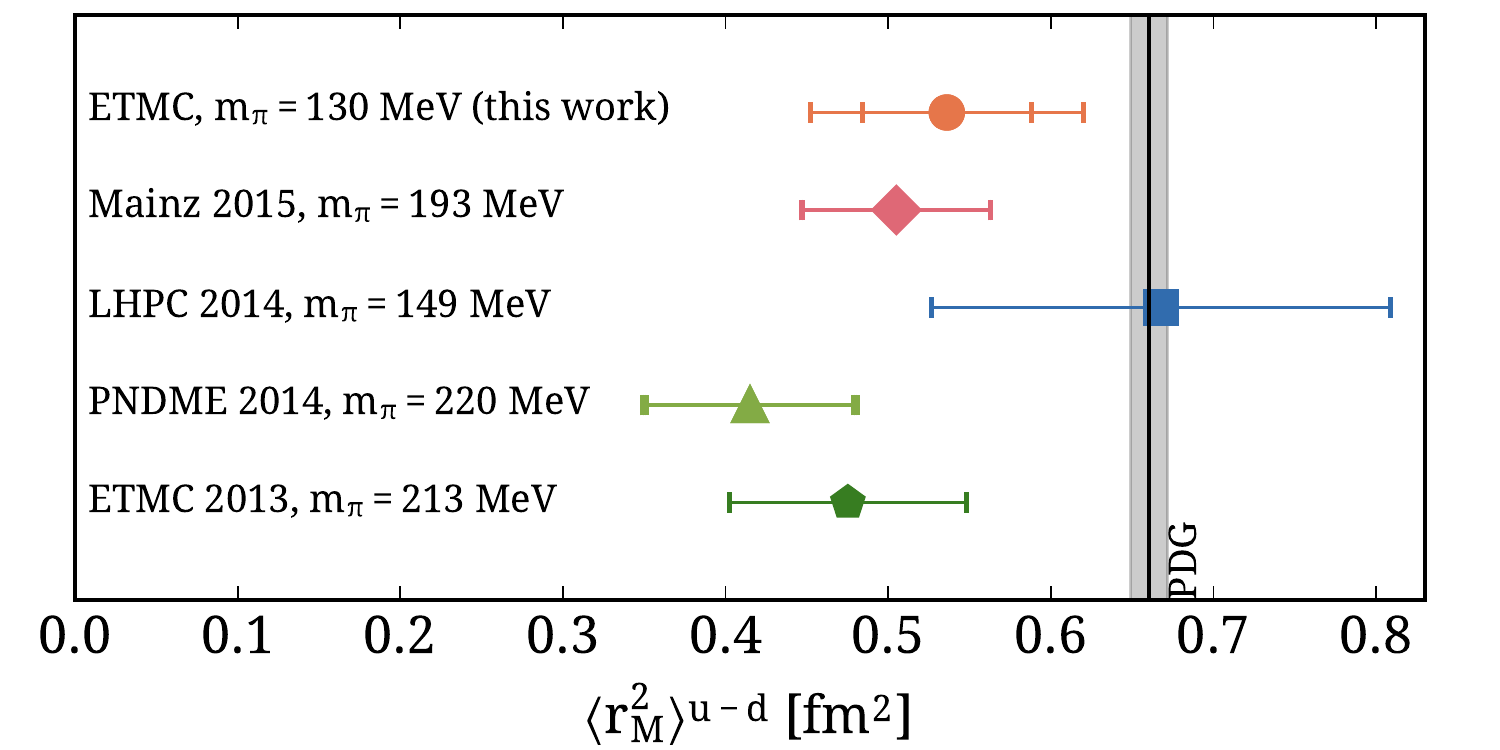}
    \caption{Comparison of results for $\langle r^2_M\rangle^{u-d}$
      with the notation of Fig.~\ref{fig:rEv comparison}. The
      experimental band is from Ref.~\cite{Mohr:2015ccw}.}
    \label{fig:rMv comparison}
\end{figure}
\begin{figure}
    \includegraphics[width=\linewidth]{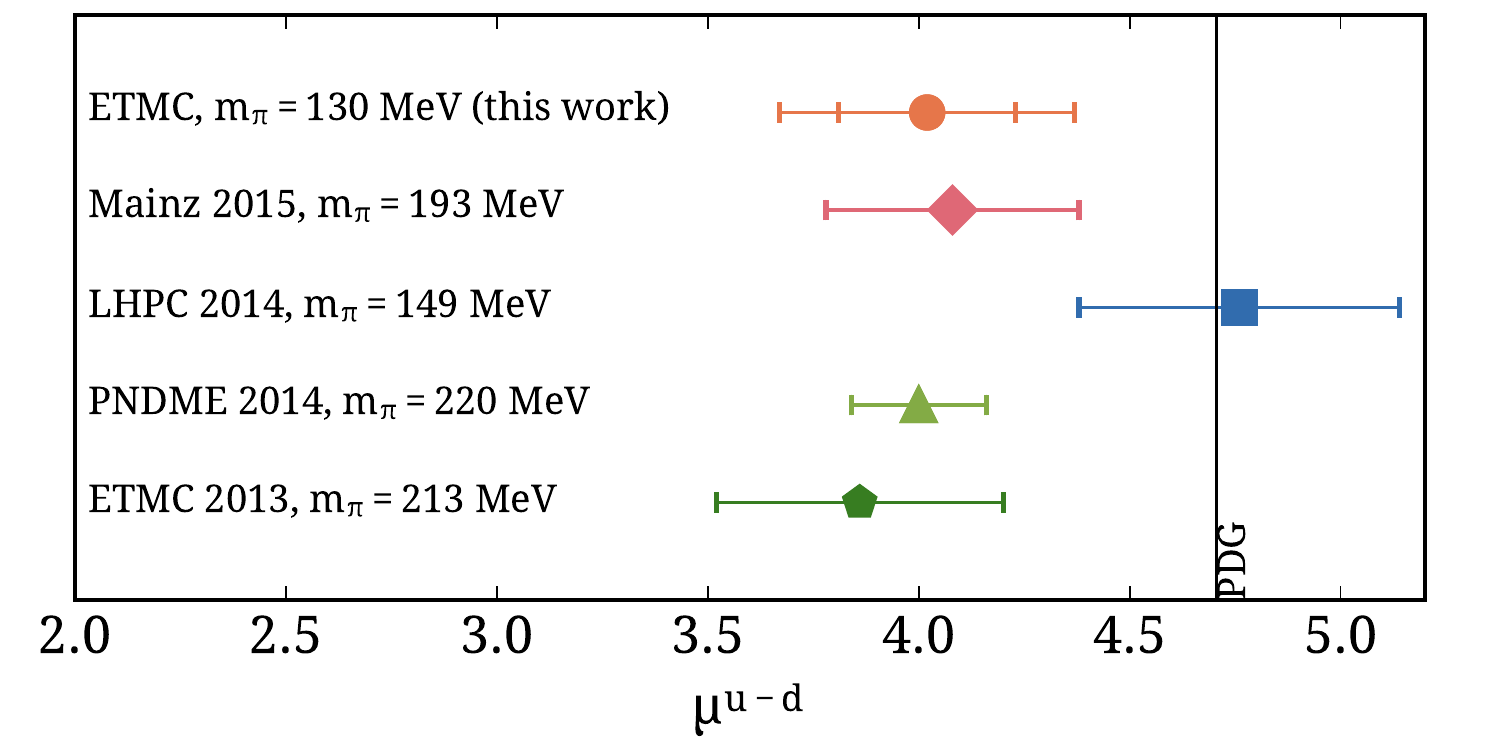}
    \caption{Comparison of results for the isovector nucleon magnetic
      moment $\mu^{u-d}$ with the notation of Fig.~\ref{fig:rEv
        comparison}.}
    \label{fig:muv comparison}
\end{figure}

A similar comparison is shown in Fig.~\ref{fig:rMv comparison} for the
magnetic radius. We see that all lattice results underestimate the
experimental band by at most 2$\sigma$, with the exception of the LHPC
value that used the summation method. Similar conclusions are drawn
for the isovector magnetic moment $G_M(0)=\mu^{u-d}=\mu_p-\mu_n$,
which we show in Fig.~\ref{fig:muv comparison}.

\subsection{Comparison of proton form factors}

\begin{figure}
    \includegraphics[width=\linewidth]{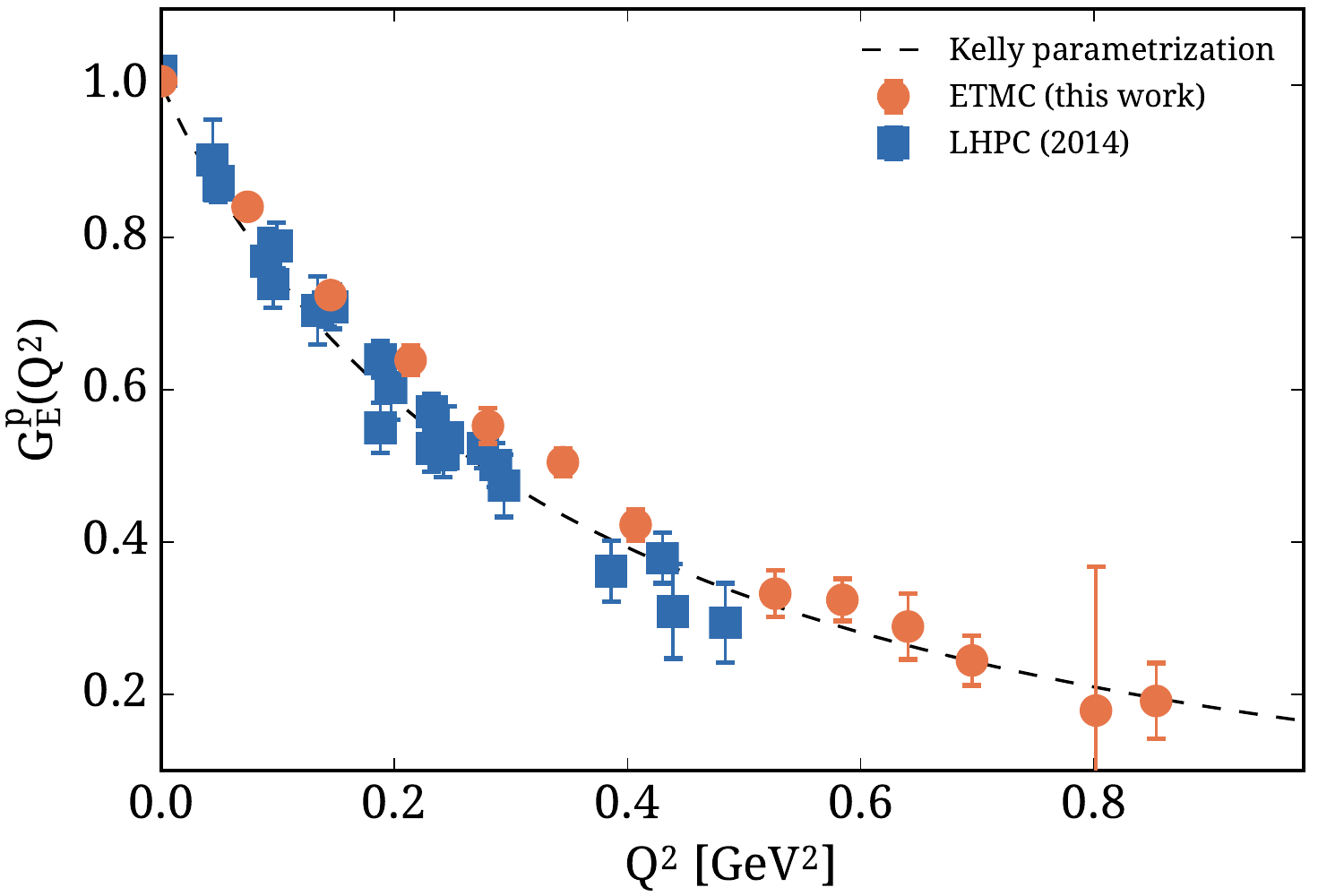}
    \caption{Comparison of $G^p_E(Q^2)$ between results from this work
      (circles) and Ref.~\cite{Green:2014xba} (squares). The dashed
      line shows the parameterization of the experimental data.}
    \label{fig:gep comparison}
\end{figure}

\begin{figure}
    \includegraphics[width=\linewidth]{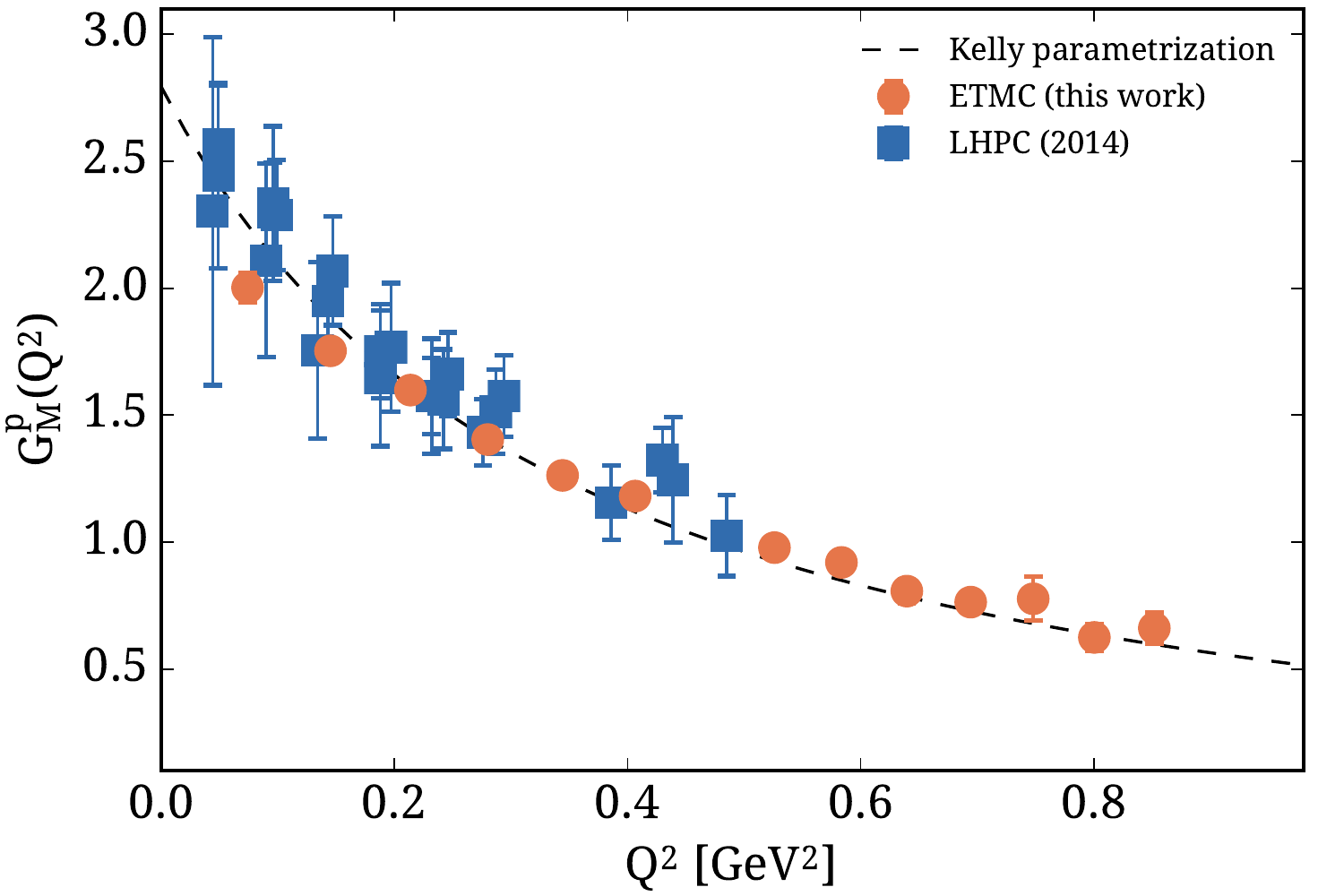}
    \caption{Comparison of $G^p_M(Q^2)$ between results from this work
      (circles) and Ref.~\cite{Green:2014xba} (squares). The dashed
      line shows the parameterization of experimental data.}
    \label{fig:gmp comparison}
\end{figure}

Published lattice QCD results for the proton form factors at physical
or near-physical pion masses are available from
LHPC~\cite{Green:2014xba}. We compare our results in
Figs.~\ref{fig:gep comparison} and~\ref{fig:gmp comparison} for the
proton electric and magnetic Sachs form factors respectively. We see
agreement with their results and note that their relatively larger
errors at small $Q^2$ for the case of the magnetic form factor are
consistent with both the experimentally determined curve and our
results.

\section{Summary and conclusions}
\label{sec:conclusions}
A first calculation of the isovector and isoscalar electromagnetic
Sachs nucleon form factors including the disconnected contributions is
presented directly at the physical point using an ensemble of
$N_\textrm{f}=2$ twisted mass fermions at maximal twist at a volume
of $m_\pi L\simeq 3$. Using five sink-source separations for
$G_E(Q^2)$ between 0.94~fm and 1.69~fm, we confirm our previous
findings that excited state contributions require a separation larger
than $\sim$1.5~fm to be sufficiently suppressed. For the case of
$G_M(Q^2)$ we use three sink-source separations between 0.94~fm and
1.31~fm and observe that for the isovector no excited state effects
are present within statistical errors, while for the connected
isoscalar, the largest separation of $t_s=1.31$~fm is sufficient for
their suppression. Our results for both the isovector and isoscalar
$G_E(Q^2)$ lie higher than experiment by about a standard
deviation. This may be due to small residual excited state
contamination since this difference is found to decrease as the
sink-source separation increases. Our results for
$G^{u-d}_M(Q^2)$ at the two lowest $Q^2$ values underestimate the
experimental ones but are in agreement for $Q^2>0.2$~GeV$^2$. Volume
effects are being investigated to determine whether these could be
responsible for this discrepancy.

The isoscalar matrix element requires both connected and disconnected
contributions, the latter requiring an order of magnitude more
statistics. We have computed the disconnected contributions to
$G_E^{u+d}(Q^2)$ and $G_M^{u+d}(Q^2)$ for the first four non-zero
momentum transfers up to $Q^2= 0.28$~GeV$^2$ and find that their
magnitude is smaller or comparable to the statistical error of the
connected contribution. We include the disconnected contributions to
combine isovector and isoscalar matrix elements and obtain the proton
and neutron electromagnetic Sachs form factors at the physical point.

We have used two methods to fit the $Q^2$-dependence of our data, both
a dipole Ansatz and the z-expansion. These two methods yield
consistent results, however the latter method yields parameters with
larger statistical errors. Using the dipole fits to determine the
electric and magnetic radii, as well as the magnetic moment, we find
agreement with other recent lattice QCD results for the isovector
case, and are within 2$\sigma$ with the experimental
determinations. Our result for the proton electric charge radius
$\langle r^2_E\rangle^p = 0.589(39)(33)$~fm$^2$, is two sigmas
smaller than the muonic hydrogen determination~\cite{Antognini:1900ns}
of $\langle r^2_p\rangle = 0.7071(4)(5)$~fm$^2$, which may be due to
remaining excited state effects or volume effects, which will be
investigated further.

Our final results are collected in Table~\ref{table:intro results}.
\begin{table}
  \caption{Our final results for the isovector ($p-n$), isoscalar
    ($p+n$), proton ($p$) and neutron ($n$) electric radius ($\langle
    r_E^2\rangle$), magnetic radius ($\langle r_M^2\rangle$) and
    magnetic moment ($\mu$). The first error is statistical and the
    second a systematic due to excited state contamination.}
  \label{table:intro results}
  \begin{tabular}{lr@{.}lr@{.}lr@{.}l}
    \hline\hline
    &\multicolumn{2}{c}{$\langle r_E^2\rangle$ [fm$^2$]} &
    \multicolumn{2}{c}{$\langle r_M^2\rangle$ [fm$^2$]} &
    \multicolumn{2}{c}{$\mu$} \\
    \hline
    $p$-$n$& 0&653(48)(30)  & 0&536(52)(66) & 4&02(21)(28) \\
    $p$+$n$& 0&537(53)(38)  & 0&394(82)(42) & 0&870(60)(39)\\
    $p$    & 0&589(39)(33)  & 0&506(51)(42) & 2&44(13)(14) \\
    $n$    &-0&038(34)(6)   & 0&586(58)(75) &-1&58(9)(12)  \\
    \hline\hline
  \end{tabular}
\end{table}
We plan to analyze the electromagnetic form factors using both an
ensemble of $N_\textrm{f}=2$ twisted mass clover-improved fermions
simulated at the same pion mass and lattice spacing as the ensemble
analyzed in this work but with a lattice size of $64^3\times 128$,
yielding $m_\pi L=4$ as well as with an $N_\textrm{f}=2+1+1$ ensemble
of finer lattice spacing. In addition, we are investigating improved
techniques for the computation of the disconnected quark loops at the
physical point. These future calculations will allow for further
checks of lattice artifacts and resolve the remaining small tension
between lattice QCD and experimental results for these important
benchmark quantities.

\textit{Acknowledgments:} We would like to thank the members of the
ETM Collaboration for a most enjoyable collaboration. We acknowledge
funding from the European Union's Horizon 2020 research and innovation
program under the Marie Sklodowska-Curie Grant Agreement
No. 642069. Results were obtained using Jureca, via a
John-von-Neumann-Institut für Computing (NIC) allocation ECY00,
HazelHen at H\"ochstleistungsrechenzentrum Stuttgart (HLRS) and SuperMUC
at the Leibniz-Rechenzentrum (LRZ), via Gauss allocations with ids
44066 and 10862, Piz Daint at Centro Svizzero di Calcolo Scientifico
(CSCS), via projects with ids s540, s625 and s702, and resources at
Centre Informatique National de l’Enseignement Supérieur (CINES) and
Institute for Development and Resources in Intensive Scientific
Computing (IDRIS) under allocation 52271. We thank the staff of these
centers for access to the computational resources and for their
support.

\bibliographystyle{unsrtnat}
\bibliography{refs.bib}

\clearpage

\appendix
\widetext
\section{Tables of Results}
\label{sec:appendix results}
\begin{table*}[!h]
  \caption{Results for the isovector $G_E(Q^2)$ using the plateau
    method for five sink-source separations and the summation and
    two-state fit methods fitted to all separations. Results where the
    operand of the square root in Eq.~\ref{eq:ratio} becomes negative
    are denoted with ``NA''.}
  \label{table:results gev}
  \begin{tabular}{crrrrrrr}
    \hline\hline
    \multirow{2}{*}{$Q^2$ [GeV$^2$]} &\multicolumn{5}{c}{Plateau, $t_s$ [fm]} & Summation & Two-state \\
    & 0.94 & 1.13 & 1.31 & 1.50 & 1.69 & $[$0.94, 1.69$]$ & $[$1.13, 1.69$]$ \\    
    \hline
  0.000 &         0.9982(08) &        0.998(2)  &        0.996(41) &        1.003(4) &           1.006(8) &           1.000(6) &          - \\
  0.074 &         0.8460(31) &        0.832(6)  &        0.826(11) &        0.819(15) &         0.841(23) &          0.798(20) &          0.849(18) \\
  0.145 &         0.7337(34) &        0.713(6)  &        0.703(12) &        0.701(16) &         0.711(24) &          0.664(21) &          0.717(17) \\
  0.214 &         0.6423(45) &        0.618(8)  &        0.598(15) &        0.615(21) &         0.608(29) &          0.556(26) &          0.617(19) \\
  0.280 &         0.5753(54) &        0.553(10) &        0.549(19) &        0.514(24) &         0.521(36) &          0.483(35) &          0.535(22) \\
  0.345 &         0.5222(43) &        0.503(7)  &        0.497(15) &        0.461(19) &         0.478(26) &          0.435(26) &          0.474(16) \\
  0.407 &         0.4761(49) &        0.456(8)  &        0.450(17) &        0.391(20) &         0.378(30) &          0.357(30) &          0.407(19) \\
  0.527 &         0.4000(62) &        0.380(12) &        0.379(22) &        0.326(31) &         0.291(39) &          0.283(49) &          0.334(24) \\
  0.584 &         0.3676(57) &        0.353(10) &        0.365(22) &        0.265(27) &         0.287(37) &          0.269(42) &          0.296(22) \\
  0.640 &         0.3500(67) &        0.338(13) &        0.339(25) &        0.256(35) &         0.260(53) &          0.229(52) &          0.292(25) \\
  0.695 &         0.3273(72) &        0.320(13) &        0.303(26) &        0.236(31) &         0.219(43) &          0.208(54) &          0.279(26) \\
  0.749 &         0.284(11)  &        0.282(20) &        0.343(47) &        0.138(68) &               NA &                NA  &          0.181(46) \\
  0.802 &         0.2847(85) &        0.262(16) &        0.215(28) &        0.196(49) &          0.12(21) &          0.058(79) &          0.203(35) \\
  0.853 &         0.2707(81) &        0.273(15) &        0.257(34) &        0.144(33) &         0.156(52) &          0.160(74) &          0.186(35) \\
    \hline\hline
  \end{tabular}
\end{table*}

\begin{table*}
  \caption{Results for the isovector $G_M(Q^2)$ using the plateau
    method for three sink-source separations and the summation and
    two-state fit method fitted to all separations.}
  \label{table:results gmv}
  \begin{tabular}{crrrrr}
    \hline\hline
    \multirow{2}{*}{$Q^2$ [GeV$^2$]} &\multicolumn{3}{c}{Plateau, $t_s$ [fm]} & Summation & Two-state \\
    & 0.94 & 1.13 & 1.31 & $[$0.94, 1.31$]$ & $[$0.94, 1.31$]$ \\
    \hline
  0.074 &   3.225(36) &   3.220(54) &   3.230(99) &    3.18(18) &   3.292(82) \\
  0.145 &   2.841(28) &   2.807(38) &   2.832(70) &    2.73(13) &   2.847(54) \\
  0.214 &   2.538(26) &   2.505(38) &   2.596(67) &    2.53(13) &   2.546(54) \\
  0.280 &   2.288(26) &   2.281(39) &   2.262(75) &    2.23(13) &   2.294(59) \\
  0.344 &   2.098(21) &   2.042(31) &   2.037(55) &    1.85(11) &   2.033(46) \\
  0.407 &   1.941(19) &   1.873(32) &   1.899(56) &    1.67(12) &   1.863(45) \\
  0.526 &   1.665(20) &   1.611(35) &   1.583(70) &    1.37(15) &   1.593(50) \\
  0.583 &   1.565(17) &   1.515(31) &   1.469(64) &    1.29(14) &   1.483(40) \\
  0.640 &   1.481(22) &   1.420(39) &   1.304(77) &    1.14(17) &   1.354(53) \\
  0.694 &   1.387(20) &   1.339(37) &   1.219(68) &    1.12(17) &   1.299(52) \\
  0.748 &   1.330(25) &   1.275(54) &    1.23(14) &    0.99(29) &   1.247(58) \\
  0.800 &   1.218(21) &   1.128(44) &   0.999(83) &    0.86(22) &   1.063(77) \\
  0.852 &   1.173(20) &   1.140(46) &   1.054(98) &    0.98(23) &   1.116(46) \\
    \hline\hline
  \end{tabular}
\end{table*}

\begin{table*}[!h]
  \caption{Results for the connected contribution to the isoscalar
    $G_E(Q^2)$ using the plateau method for five sink-source
    separations and the summation and two-state fit methods fitted to
    all separations. Results where the operand of the square root in
    Eq.~\ref{eq:ratio} becomes negative are denoted with ``NA''.}
  \label{table:results ges}
  \begin{tabular}{crrrrrrr}
    \hline\hline
    \multirow{2}{*}{$Q^2$ [GeV$^2$]} &\multicolumn{5}{c}{Plateau, $t_s$ [fm]} & Summation & Two-state \\
    & 0.94 & 1.13 & 1.31 & 1.50 & 1.69 & $[$0.94, 1.69$]$ & $[$1.13, 1.69$]$ \\    
    \hline
  0.000 &           0.999(0) &           1.000(1) &           0.999(1) &           1.000(2) &           1.004(3) &           1.000(2) &          - \\
  0.074 &           0.870(1) &           0.863(2) &           0.855(4) &           0.852(5) &           0.839(9) &           0.834(7) &          0.874(16) \\
  0.145 &           0.768(2) &           0.755(3) &           0.746(5) &           0.746(6) &          0.738(10) &           0.721(9) &          0.756(16) \\
  0.214 &           0.688(2) &           0.671(4) &           0.657(7) &           0.665(9) &          0.670(15) &          0.638(11) &          0.672(12) \\
  0.280 &           0.624(2) &           0.609(4) &           0.600(9) &          0.588(12) &          0.585(18) &          0.570(15) &          0.603(12) \\
  0.345 &           0.567(2) &           0.552(3) &           0.544(7) &          0.535(10) &          0.532(16) &          0.508(12) &          0.537(12) \\
  0.407 &           0.524(2) &           0.506(4) &           0.499(9) &          0.476(10) &          0.467(17) &          0.446(14) &          0.487(11) \\
  0.527 &           0.450(3) &           0.429(6) &          0.413(13) &          0.430(21) &          0.373(29) &          0.362(24) &          0.412(11) \\
  0.584 &           0.421(3) &           0.408(6) &          0.398(14) &          0.387(19) &          0.361(26) &          0.356(23) &          0.381(10) \\
  0.640 &           0.399(4) &           0.385(7) &          0.357(16) &          0.369(26) &          0.318(40) &          0.308(29) &          0.357(10) \\
  0.695 &           0.375(4) &           0.358(8) &          0.322(15) &          0.321(20) &          0.270(29) &          0.247(26) &          0.338(10) \\
  0.749 &           0.352(5) &          0.339(12) &          0.328(31) &          0.336(93) &                NA  &                NA  &          0.314(16) \\
  0.802 &           0.331(5) &          0.302(10) &          0.261(20) &          0.306(41) &         0.235(176) &          0.189(40) &          0.286(12) \\
  0.853 &           0.315(5) &          0.304(11) &          0.275(24) &          0.265(31) &          0.227(55) &          0.239(42) &          0.278(11) \\
    \hline\hline
  \end{tabular}
\end{table*}

\begin{table*}
  \caption{Results for the connected contribution to the isoscalar
    $G_M(Q^2)$ using the plateau method for three sink-source
    separations and the summation and two-state fit method fitted to
    all separations.}
  \label{table:results gms}
  \begin{tabular}{crrrrr}
    \hline\hline
    \multirow{2}{*}{$Q^2$ [GeV$^2$]} &\multicolumn{3}{c}{Plateau, $t_s$ [fm]} & Summation & Two-state \\
    & 0.94 & 1.13 & 1.31 & $[$0.94, 1.31$]$ & $[$0.94, 1.31$]$ \\
    \hline
  0.074 &          0.756(10) &          0.760(17) &          0.771(26) &          0.777(54) &          0.781(25) \\
  0.145 &           0.669(8) &          0.660(13) &          0.674(21) &          0.644(40) &          0.671(19) \\
  0.214 &           0.602(8) &          0.593(13) &          0.601(22) &          0.584(44) &          0.597(19) \\
  0.280 &           0.551(8) &          0.544(12) &          0.546(23) &          0.547(46) &          0.544(19) \\
  0.344 &           0.501(6) &          0.484(10) &          0.488(18) &          0.448(36) &          0.483(15) \\
  0.407 &           0.465(6) &          0.451(10) &          0.463(18) &          0.412(38) &          0.455(15) \\
  0.526 &           0.402(6) &          0.386(11) &          0.373(21) &          0.339(50) &          0.380(18) \\
  0.583 &           0.377(5) &          0.360(10) &          0.370(20) &          0.334(47) &          0.357(14) \\
  0.640 &           0.360(6) &          0.339(12) &          0.309(22) &          0.293(52) &          0.325(17) \\
  0.694 &           0.334(6) &          0.323(11) &          0.308(21) &          0.291(55) &          0.324(16) \\
  0.748 &           0.317(8) &          0.304(16) &          0.315(38) &          0.332(90) &          0.302(22) \\
  0.800 &           0.299(6) &          0.280(13) &          0.249(25) &          0.258(74) &          0.269(22) \\
  0.852 &           0.280(6) &          0.275(13) &          0.268(28) &          0.284(73) &          0.270(17) \\
  \hline\hline
  \end{tabular}
\end{table*}

\end{document}